\documentclass[onecolumn,sort&compress,numbers]{els-mrw} 
\usepackage{bm}
\usepackage{braket}
\usepackage{amsmath}
\usepackage{amssymb}
\usepackage{amsfonts}
\usepackage{amsthm}
\usepackage{makeidx}
\usepackage{mathrsfs}
\usepackage{graphicx}
\usepackage{txfonts}
\usepackage{helvet}
\usepackage{ulem}

\usepackage{tikz, tikz-feynhand}
\usetikzlibrary{intersections, calc, arrows.meta}
\usetikzlibrary{fadings, patterns}
\tikzfading[name=fade top,
            top color=transparent!0,
            bottom color=transparent!100]
\tikzfading[name=fade out,
            inner color=transparent!20,
            outer color=transparent!80]
            
\usepackage[colorlinks=true,pdfstartview=FitV,linkcolor=blue,citecolor=magenta,urlcolor=blue,bookmarks=true,bookmarksnumbered=true]{hyperref}

\newcommand{\red}[1]{\color{red}#1\color{black}}
\newcommand{\blue}[1]{\color{blue}#1\color{black}}
\newcommand{\green}[1]{\color{green}#1\color{black}}

\newcommand{\inlinephoton}{%
\begin{tikzpicture}[baseline=-0.6ex]
    \begin{feynhand}
        \vertex (a) at (0,0); \vertex (b) at (0.42,0);
        \propag[bos] (a) to (b);
    \end{feynhand}
\end{tikzpicture}%
}
\newcommand{\inlinegluon}{%
\begin{tikzpicture}[baseline=-0.6ex]
    \begin{feynhand}
        \vertex (a) at (0,0); \vertex (b) at (0.63,0);
        \propag[gluon] (a) to (b);
    \end{feynhand}
\end{tikzpicture}%
}
\newcommand{\inlineghost}{%
\begin{tikzpicture}[baseline=-0.6ex]
    \begin{feynhand}
        \vertex (a) at (0,0); \vertex (b) at (0.4,0);
        \propag[ghost] (a) to (b);
    \end{feynhand}
\end{tikzpicture}%
}

\begin{document}

\chapter{Schwinger effect in QCD and nuclear physics}

\author[1,2]{Hidetoshi~Taya}
\address[1]{\orgname{Keio University}, \orgdiv{Hiyoshi Department of Physics}, \orgaddress{Hiyoshi, Kanagawa 223-8521, Japan}}
\address[2]{\orgname{RIKEN}, \orgdiv{iTHEMS, TRIP Headquarters}, \orgaddress{Saitama, Wako 351-0198, Japan}}

\maketitle

\begin{abstract}[Abstract]
We provide a pedagogical review of the Schwinger effect, i.e., the non-perturbative production of particle and anti-particle pairs from the vacuum by strong fields, as well as related strong-field phenomena.  Beginning with an overview of the Schwinger effect in quantum electrodynamics, we discuss its extensions to quantum chromodynamics and its applications in nuclear physics, including high-$Z$ nuclei, string breaking, relativistic heavy-ion collisions, and the chiral anomaly.
\end{abstract}

\section{Introduction} \label{sec:intro}

Strong electromagnetic and color fields appear ubiquitously in nuclear physics under extreme conditions.  These fields can be so intense that their strengths reach, or even exceed, the characteristic energy scales of the underlying quantum theories, such as the electron mass $m_e \approx 511\;{\rm keV}$ in quantum electrodynamics (QED) and the pion mass $m_\pi \approx 140\;{\rm MeV}$ in quantum chromodynamics (QCD).  Such strong fields can drastically affect the system and induce nontrivial phenomena that would never occur otherwise.  These phenomena are collectively referred to as {\it strong-field physics} and are actively investigated not only in nuclear physics~\cite{Greiner:1985ce, Tuchin:2013ie, Hattori:2016emy, Fukushima:2018grm, Hattori:2023egw} but also across a wide range of disciplines, including particle physics~\cite{Hartin:2018egj}, laser and plasma physics~\cite{Marklund:2006my, DiPiazza:2011tq, Gonoskov:2021hwf, Fedotov:2022ely}, solid-state physics~\cite{Oka:2011kf, Ghimire_2014, Cavaletto2025, Stammer:2025ekh}, and astrophysics~\cite{Harding:2006qn, Ruffini:2009hg, Turolla:2015mwa}.

One of the most remarkable predictions of strong-field physics is that the vacuum becomes unstable in the presence of strong fields due to non-perturbative pair production.  This phenomenon is known as {\it the (Sauter-)Schwinger effect}, named after the pioneering works of Sauter in 1931~\cite{Sauter:1931zz, Sauter:1932gsa} and Schwinger in 1951~\cite{Schwinger:1951nm}\footnote{We use the term {\it Schwinger effect} to refer to {\it a non-perturbative pair-production mechanism from the vacuum by slowly varying, strong fields}.  As we review later (see Sec.~\ref{sec:2.3..1}), genuine non-perturbativity is ensured only for slowly-varying fields, whereas {\it perturbative pair production} occurs for rapidly-varying fields.  We distinguish the Schwinger effect from such a perturbative process.  Although this usage is widely accepted, it is broader than Schwinger’s original formulation for constant fields~\cite{Schwinger:1951nm}, and some works use the term in this original sense.  }. Since its prediction, the Schwinger effect has drawn sustained attention from researchers across many disciplines and continues to be at the forefront of contemporary physics, not only because of the intrinsic interest of the phenomenon itself but also due to its potential phenomenological applications, particularly in nuclear physics, as we review later.   

This chapter is intended as a primer on the Schwinger effect.  Our aim is to review the basic concepts, without delving into detailed theoretical derivations or overly technical discussions.  The primary targets are non-experts on the Schwinger effect, including motivated undergraduates and researchers from other fields.  A number of excellent reviews on the Schwinger effect and related topics already exist, e.g., Refs.~\cite{Dunne:2004nc, Ruffini:2009hg, Gelis:2015kya, Fedotov:2022ely, Hattori:2023egw}, and these provide valuable resources for further study.  Compared with the existing literature, the present chapter places greater emphasis on extensions to QCD and nuclear physics.

The remainder of this chapter is organized into four sections.  In Sec.~\ref{sec:2}, we review the Schwinger effect in QED, explaining not only its essential ideas but also covering recent developments.  Readers who are unfamiliar with the Schwinger effect are strongly encouraged to first go through this section (particularly, Sec.~\ref{sec:2.1}) to grasp the core concepts before proceeding to the discussions of QCD and nuclear-physics applications.  Section~\ref{sec:3} presents the extension of the Schwinger effect to QCD, with an emphasis on the formal aspects and on clarifying the similarities and differences between QED and QCD.  In Sec.~\ref{sec:4}, we highlight several phenomenological applications of the Schwinger effect to nuclear physics.  Finally, Sec.~\ref{sec:5} provides a summary, in which we reiterate several important ideas that we wish to emphasize.

{\it Notation}: We work in natural units with $\hbar = c = 1$ and do not explicitly display the Planck constant $\hbar$ or the speed of light $c$, unless necessary.  We adopt the mostly-minus convention for the Minkowski metric $g^{\mu\nu} = {\rm diag}(+1,-1,-1,-1)$ and normalize the Levi-Civita tensor as $\epsilon^{0123} = +1$.  Lorentz indices are denoted by Greek letters and run from $0$ to $3$.  Repeated indices are summed implicitly, otherwise stated.  Three-vectors are indicated by boldface letters ${\bm X}$.  The sign of the elementary electric charge $e$ is taken to be positive $e=|e|$.

\section{Schwinger effect in QED} \label{sec:2}

We begin by reviewing the Schwinger effect in QED, since the history of the Schwinger effect originates from QED and it provides a clear setting for understanding the essence.  The Schwinger effect in QED is of interest not only as a prototype for the extension to QCD but also in its own right, especially given the recent availability of strong electromagnetic fields from intense lasers~\cite{Danson:2019qlu, Fedotov:2022ely}.  

\subsection{Intuitive picture} \label{sec:2.1}

\setlength{\feynhandarrowsize}{3pt}
\newcommand{\vacuumloop}[6]{
	\pgfmathsetmacro{\radius}{#1}
	\pgfmathsetmacro{\radiusB}{#2}
	\pgfmathsetmacro{\ang}{#3}
	\pgfmathsetmacro{\offsetX}{#4}
	\pgfmathsetmacro{\offsetY}{#5}
	\pgfmathsetmacro{\loose}{#6}
  	\vertex            (s) at ({\offsetX+\radius*cos(\ang+180)}, {\offsetY+\radius*sin(\ang+180)}); 
	\vertex[dot, red]  (a) at ({\offsetX+\radiusB*cos(\ang+90)}, {\offsetY+\radiusB*sin(\ang+90)}) {}; 
	\vertex[dot, blue] (b) at ({\offsetX+\radiusB*cos(\ang-90)}, {\offsetY+\radiusB*sin(\ang-90)}) {}; 
	\vertex            (g) at ({\offsetX+\radius*cos(\ang)}, {\offsetY+\radius*sin(\ang)}); 
	\pgfmathparse{\ang+180}\edef\angA{\pgfmathresult}
	\pgfmathparse{\ang+90}\edef\angB{\pgfmathresult}
	\pgfmathparse{\ang-90}\edef\angC{\pgfmathresult}
	\propag [chagho, densely dotted, red, opacity=0.3]  (s) to [in=\angA, out=\angB, looseness=\loose] (a);
	\propag [chagho, densely dotted, red, opacity=0.3]  (a) to [in=\angB, out=\ang,  looseness=\loose] (g);
	\propag [chagho, densely dotted, blue, opacity=0.3] (s) to [in=\angA, out=\angC, looseness=\loose] (b);
	\propag [chagho, densely dotted, blue, opacity=0.3] (b) to [in=\angC, out=\ang,  looseness=\loose] (g);
}
\newcommand{\paircrea}[5]{
	\pgfmathsetmacro{\dist}{#1}
	\pgfmathsetmacro{\acc}{#2}
	\pgfmathsetmacro{\ang}{#3}
	\pgfmathsetmacro{\offsetX}{#4}
	\pgfmathsetmacro{\offsetY}{#5}	
  	\vertex (o)              at ({\offsetX}, {\offsetY}); 
	\vertex (a1) [dot, red]  at ({\offsetX+\dist*cos(\ang)},      {\offsetY+\dist*sin(\ang)}) {}; 
  	\vertex (a2)             at ({\offsetX+\dist*cos(\ang)+\acc}, {\offsetY+\dist*sin(\ang)}); 
	\vertex (b1) [dot, blue] at ({\offsetX-\dist*cos(\ang)},      {\offsetY-\dist*sin(\ang)}) {}; 
  	\vertex (b2)             at ({\offsetX-\dist*cos(\ang)-\acc}, {\offsetY-\dist*sin(\ang)}); 
	\propag [fer, red,  opacity=0.3, with arrow=1]  (a1) to (a2);
	\propag [fer, blue, opacity=0.3, with arrow=1]  (b1) to (b2);
	\propag [gho, densely dotted, red,  opacity=0.3] (o) to (a1);
	\propag [gho, densely dotted, blue, opacity=0.3] (o) to (b1);
}
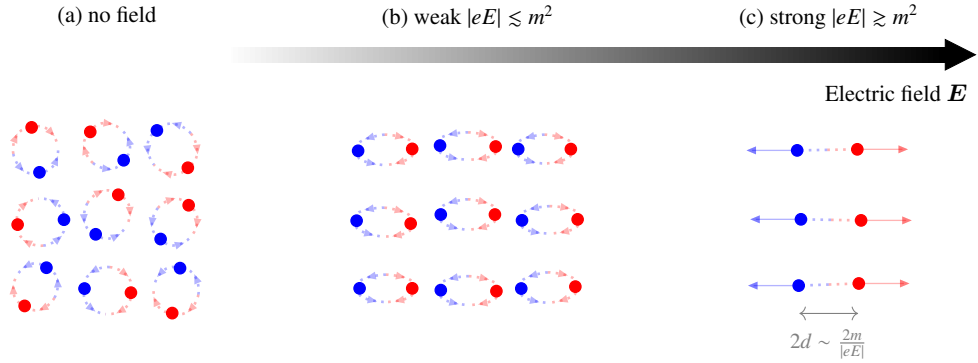
\begin{figure}[t]
\centering
\begin{tikzpicture}
\begin{feynhand}
 	\shade[left color=white, right color=black] (2.5,1.2) rectangle +(9.5,0.2);
 	\draw[fill=black] (12.0,1.1) -- (12.4,1.3) -- (12.0,1.5) -- cycle;
	\vertex[particle] (l0) at (11.35, 0.8) {\rm Electric field ${\bm E}$}; 
 
	\vertex[particle] (l1) at (0.85, 1.8) {\rm (a) no field};
	\vacuumloop{0.30}{0.30}{+10}{-0.10}{+0.05}{0.9}
	\vacuumloop{0.29}{0.29}{+50}{+0.85}{+0.10}{0.9}
	\vacuumloop{0.32}{0.32}{-140}{+1.72}{+0.06}{0.9}
	\vacuumloop{0.31}{0.31}{+96}{-0.03}{-0.90}{0.9}
	\vacuumloop{0.30}{0.30}{-30}{+0.85}{-0.80}{0.9}
	\vacuumloop{0.29}{0.29}{-40}{+1.75}{-0.90}{0.9}
	\vacuumloop{0.29}{0.29}{+150}{-0.10}{-1.75}{0.9}
	\vacuumloop{0.31}{0.31}{-95}{+0.86}{-1.80}{0.9}
	\vacuumloop{0.30}{0.30}{+170}{+1.77}{-1.80}{0.9}
	
	\vertex[particle] (l2) at (5.65, 1.8) {\rm (b) weak $|eE|\lesssim m^2$};
	\vacuumloop{0.20}{0.36}{-88}{4.55}{+0.05}{0.5}
	\vacuumloop{0.18}{0.37}{-91}{5.65}{+0.10}{0.5}
	\vacuumloop{0.19}{0.35}{-90}{6.67}{+0.06}{0.5}
	\vacuumloop{0.19}{0.36}{-93}{4.53}{-0.90}{0.5}
	\vacuumloop{0.20}{0.36}{-90}{5.65}{-0.80}{0.5}
	\vacuumloop{0.18}{0.37}{-89}{6.74}{-0.87}{0.5}
	\vacuumloop{0.18}{0.35}{-90}{4.56}{-1.77}{0.5}
	\vacuumloop{0.19}{0.37}{-91}{5.66}{-1.80}{0.5}
	\vacuumloop{0.20}{0.36}{-88}{6.72}{-1.76}{0.5}
	
	\vertex[particle] (l) at (10.45, 1.8) {\rm (c) strong $|eE|\gtrsim m^2$};
	\paircrea{0.4}{0.6}{+1}{10.43}{+0.05}
	\paircrea{0.4}{0.6}{-1}{10.48}{-0.87}
	\paircrea{0.4}{0.6}{+2}{10.45}{-1.72}
	\vertex[particle] (d0) at (10.45, -2.5) {\textcolor{gray}{\small $2d \sim \frac{2m}{|eE|}$}}; 
	\vertex[particle] (d1) at (9.95, -2.1) {}; 
	\vertex[particle] (d2) at (10.95, -2.1) {}; 
	\draw[<->, gray] (d1) to (d2);
\end{feynhand}
\end{tikzpicture}
\vspace*{3mm}
\caption{Schematic illustration of the quantum vacuum and its response to an external electric field ${\bm E}$.  (a) In the absence of an electric field, the spontaneous creation and annihilation of virtual particle (\red{$\bullet$}) and anti-particle (\blue{$\bullet$}) pairs occur randomly in the vacuum.  (b) Once an electric field is applied, the virtual pairs tend to align along the field direction (provided they are electrically charged), leading to vacuum polarization.  (c) When the field strength goes beyond the critical strength $|eE| \gtrsim m^2$, the vacuum decays: the field supplies sufficient energy to tear the virtual pairs apart, allowing them to materialize as real particles.  The same picture applies equally to QCD, where the vacuum fluctuations of colorful quarks and gluons are polarized and eventually materialized by the presence of a color electric field.  
}
\label{fig:1}
\end{figure}

The essence of the Schwinger effect (more broadly, strong-field physics) lies in the recognition that the vacuum is not empty space.  It is instead filled with virtual particle and anti-particle pairs that ceaselessly pop in and out of existence due to quantum fluctuations, as illustrated in Fig.~\ref{fig:1}~(a).  The vacuum can therefore exhibit nontrivial responses against external fields through coupling to these fluctuations.  For example, the QED vacuum can become electrically polarized along the direction of an electric field, much like dielectric polarization in materials; see Fig.~\ref{fig:1}~(b).  The vacuum polarization leads to intriguing strong-field QED phenomena such as photon-photon scattering~\cite{Euler1935}, photon splitting~\cite{Adler:1971wn}, vacuum birefringence~\cite{Toll:1952rq}, and vacuum dichroism~\cite{Hattori:2020htm}.  As the field strength increases, the vacuum polarization becomes stronger, and eventually the vacuum ``decays," in a manner analogous to dielectric breakdown in materials (the Landau-Zener transition~\cite{Landau:1932vnv, 10.1098/rspa.1932.0165}).  In this process, virtual particle and anti-particle pairs are torn apart and materialize as real pairs due to the large energy supplied by the strong electric field, as in Fig.~\ref{fig:1}~(c).  The vacuum, which initially contains no real particles, thus decays into a state populated by a finite number of real particles.  This provides an intuitive picture of the Schwinger effect, i.e., how the vacuum pair production by a strong electric field occurs.  

It should be noted that the Schwinger effect is intrinsically an {\it electric} phenomenon: pair production does not occur in fields that cannot exert work on vacuum fluctuations to promote them into real particles; for example, a pure magnetic field or a plane electromagnetic wave~\cite{Schwinger:1951nm}.  This observation implies that Schwinger-like pair production can arise in other contexts as well, provided that the background field is capable of doing work on the vacuum (or on the ground state of the system).  Examples include Hawking radiation generated by a strong gravitational field~\cite{Hawking:1975vcx}, particle production in curved spacetime~\cite{Birrell:1982ix, Parker:2009uva}, and (p)reheating in the early Universe driven by a coherent inflaton field~\cite{Kofman:1997yn, Taya:2022gzp}.

The Schwinger effect can also be viewed as a quantum-tunneling process.  Based on this picture, let us provide a simple order-of-magnitude estimate of the pair-production probability~\cite{Casher:1978wy, Glendenning:1983qq}.  We apply a constant electric field ${\bm E} = E {\bf e}_\parallel$ (with $E>0$ and ${\bf e}_\parallel$ being a unit vector) to the vacuum and suppose that a virtual electron-positron pair that momentarily appears at $x_\parallel = 0$.  Since the pair originates from the vacuum, its total energy and momentum must vanish.  Thus, for the electron and positron with four-momenta $(\epsilon, {\bm p})$ and $(\epsilon',{\bm p}')$, respectively, we have $0 = \epsilon + \epsilon' = \sqrt{m^2 + {\bm p}^{2}} + \sqrt{m^2 + {\bm p}^{\prime 2}}$ and ${\bm 0} = {\bm p} + {\bm p}'$, where $m$ is the electron mass.  These relations yield $\epsilon = \epsilon' = 0$ and that the longitudinal momenta must be purely imaginary, $p_\parallel = -{\rm i}\sqrt{m^2 + {\bm p}_\perp^{2}}$ and $p'_\parallel = -p_\parallel$, where ${\bm p}_\perp$ is the electron momentum transverse to the electric field.  Being imaginary, these momenta correspond to virtual states that cannot be observed as real particles.  However, now we have the electric field.  The electron (positron) is, thus, accelerated in the $-x_\parallel$ ($+x_\parallel$) direction, and the energy evolves as $\epsilon = -eE x_\parallel$ ($\epsilon' = +eE x_\parallel$), with $e>0$ being the elementary electric charge.  Consequently, the imaginary parts of the longitudinal momenta shrink, $p_\parallel = -{\rm i}\sqrt{m^2 + {\bm p}_\perp^{2} - (-eE x_\parallel)^2}$ and $p'_\parallel = +{\rm i}\sqrt{m^2 + {\bm p}_\perp^{2} - (+eE x_\parallel)^2}$, and they become real at $x_\parallel = -\sqrt{m^2 + {\bm p}_\perp^{2}}/eE =: -d$ and $x_\parallel = +d$ for the electron and positron, respectively.  In the language of quantum tunneling, $|x_\parallel|<d$ corresponds to the classically-forbidden region, and the points $\pm d$ are the classical turning points where the virtual particles emerge as real ones.  According to the undergraduate quantum-mechanics course, the tunneling probability can be estimated using Gamow's method~\cite{Gamow:1928zz}, i.e., the probability is given by the exponential of the integral of the imaginary momentum over the classically-forbidden region.  The total pair-production probability $P_{\rm pair}$ is the product of the tunneling probabilities for the electron (traveling from $0$ to $-d$) and for the
positron ($0$ to $+d$), which we write $P$ and $P'$, respectively, and hence
\begin{align}
	P_{\rm pair}
	\sim \underbrace{ \left| \exp \left[ +{\rm i} \int_0^{-d} {\rm d}x_\parallel \; p_\parallel \right] \right|^2 }_{=\,P} \times \underbrace{ \left| \exp \left[ +{\rm i} \int_0^{+d} {\rm d}x_\parallel \; p'_\parallel \right] \right|^2 }_{=\,P'}
	= \exp \left[ -2 \int_{-d}^{+d} {\rm d}x_\parallel \sqrt{ m^2 + {\bm p}_\perp^2 - (eEx_\parallel)^2 } \right]
	= \exp\left[ -\pi \frac{m^2 + {\bm p}_\perp^2}{|eE|} \right] \;. \label{eqwsd1}
\end{align} 

The order estimate~(\ref{eqwsd1}) captures the essential feature of the Schwinger effect; namely, the exponential factor ${\rm e}^{-\pi m^2/|eE|}$.  Several important observations follow from this factor:
\begin{itemize}
\item {\it Non-perturbativity}: The Schwinger effect is a genuinely non-perturbative phenomenon.  The characteristic exponential factor~(\ref{eqwsd1}) cannot be expanded in powers of the QED coupling constant $e$ or the field strength $E$.  Thus, no finite-order perturbative calculation can reproduce the effect.  While the perturbative regime of QED has been explored in great depth---most notably the extremely high-precision computation of the anomalous magnetic moment~\cite{Aoyama:2019ryr, Aoyama:2020ynm}---non-perturbative QED phenomena remain poorly understood.  This is one of the reasons why the Schwinger effect is intriguing in QED.  Physically, the non-perturbativity reflects the fact that a constant electromagnetic field consists of zero-energy photons: absorbing an {\it infinite} number of such photons is necessary to supply the finite excitation energy $\sim 2m$.  

\item {\it Schwinger limit}: The exponential factor shows that the Schwinger effect is strongly suppressed by the ratio $m^{2}/eE$.  To make the effect observable, the field strength should thus satisfy $m^2/eE \lesssim 1$, or 
\begin{align}
	E \gtrsim E_{\rm cr} := \frac{m^{2}}{e} \;, \label{eq::::2}
\end{align}
which is known as the Schwinger limit.  Note that the Schwinger effect is not a strict threshold phenomenon: it occurs at any electric-field strength in principle, but the strong exponential suppression makes it observable effectively only above the Schwinger limit.  The Schwinger limit arises naturally from the vacuum-polarization picture in Fig.~\ref{fig:1}.  A vacuum fluctuation typically persists for a time $\Delta t \sim 1/m$ because of the uncertainty principle.  For the electric field to supply the excitation energy $\sim m$ within this time, we must have $eE \times \Delta t \gtrsim m\ \Rightarrow\ E \gtrsim E_{\rm cr}$, since the power (work per unit time) exerted by the field is $eE$.

\item {\it Observability}: For the electron mass $m\approx 511\;{\rm keV}$, the Schwinger limit is $E_{\rm cr}\approx 1.3\times 10^{18}\;{\rm V}/{\rm m}$.  For the $u$ quark, the corresponding values are $E_{\rm cr}\approx 7\times10^{19}\;{\rm V}/{\rm m}$ ($7\times10^{23}\;{\rm V}/{\rm m}$) for the current (constituent) mass $m\approx 3\;{\rm MeV}$ ($m\approx 300\;{\rm MeV}$)\footnote{Among all quarks, the $u$ quark has the smallest Schwinger limit essentially because it is the lightest.  Although the $d$ quark has a comparable mass, $m_d\approx m_u$, it carries a smaller electric charge, $|q_d| = e/3 = |q_u|/2$, which increases its Schwinger limit by a factor of two.}.  These field strengths are many orders of magnitude beyond what is currently achievable in controlled laboratory experiments.  For example, the current Guinness World Record was achieved by the HERCULES laser in 2008~\cite{Yanovsky:08}, which reached an intensity of $I \approx 2 \times 10^{22}\;{\rm W}/{\rm cm}^2$, corresponding to $E = {\mathcal O}(10^{14}\;{\rm V}/{\rm m})$.  This record was recently surpassed by the CoReLS laser, which achieved $I \approx 1.1 \times 10^{23}\;{\rm W}/{\rm cm}^2$, corresponding to $E = {\mathcal O}(10^{15}\;{\rm V}/{\rm m})$~\cite{2021Optic...8..630Y}.  Consequently, the Schwinger effect has not yet been experimentally verified (although analogous mechanisms have been observed in other systems, such as a cold-atomic setup~\cite{Pineiro:2019uzb}).  
\end{itemize}

\subsection{Basics and early development} \label{sec:2.2}

The foundational concept of the Schwinger effect dates back to 1931, when Sauter first pointed out the possibility of vacuum pair production by high-energy potential barriers in relativistic quantum mechanics~\cite{Sauter:1931zz, Sauter:1932gsa}.  It was originally proposed as a resolution to the Klein paradox~\cite{Klein:1929zz}, which is a paradoxical result of relativistic quantum mechanics that the number of reflected particles by a sufficiently high potential-barrier becomes larger than that of the injected ones, implying an inadequacy of single-particle frameworks in the presence of a strong potential.  

Shortly thereafter, in 1936, Heisenberg and Euler~\cite{Heisenberg:1936nmg} analyzed the vacuum polarization in the presence of a constant classical electromagnetic background field\footnote{\label{foot:2}The ``classical'' electromagnetic field $A^\mu$ is defined as the vacuum expectation value of the photon field operator $\hat{A}^\mu$, i.e., $A^\mu := \braket{ \hat{A}^\mu }$.  When the classical field is sufficiently strong that quantum fluctuations about it, $\hat{a}^\mu := \hat{A}^\mu - A^\mu$, can be neglected (i.e., no radiation), the electron dynamics is fully determined by the Dirac equation in the presence of a classical four-vector potential $[ {\rm i}\gamma^\mu ( \partial_\mu + {\rm i}e A_\mu ) - m ] \hat{\psi} = e \gamma^\mu \hat{a}_\mu \hat{\psi} \approx 0$, where $\gamma^\mu$ are the Dirac gamma matrices and $\hat{\psi}$ denotes the electron field operator.   The effects of $\hat{a}^\mu$ can be incorporated perturbatively by expanding the full electron field operator $\hat{\psi}$ around the electron basis $\hat{\psi}_{(0)}$ determined by the leading-order equation $[ {\rm i}\gamma^\mu ( \partial_\mu + {\rm i}e A_\mu ) - m ] \hat{\psi}_{(0)} = 0$.  This framework is known as the Furry-picture perturbation theory~\cite{Furry:1951bef, Fradkin:1981sc, Fradkin:1991zq, Fukushima:2025eyt} and forms the basis for strong-field processes (see Ref.~\cite{Fedotov:2022ely} for review), including the advanced treatment of radiative corrections to the Schwinger effect discussed in Sec.~\ref{sec::2.3.2}.  } based on relativistic quantum mechanics.  They carried out a non-perturbative calculation with respect to the electromagnetic field and found that the classical Maxwell Lagrangian ${\mathscr L}_{\rm M} = -{\mathscr F}$ is modified by the vacuum polarization as 
\begin{align}
    {\mathscr L}_{\rm HE}
        &= {\mathscr L}_{\rm M} - \frac{1}{8\pi^2} \int_0^\infty \frac{{\rm d}s}{s^3} {\rm e}^{-m^2s} \left[ (es)^2 \frac{{\rm Re}\,\cosh\left(es\sqrt{2({\mathscr F}+{\rm i}{\mathscr G})}\right)}{{\rm Im}\,\cosh\left(es\sqrt{2({\mathscr F}+{\rm i}{\mathscr G})}\right)}{\mathscr G} - \frac{2}{3}(es)^2 {\mathscr F} - 1 \right] \;. \label{eq:1} 
\end{align}
Here, ${\mathscr F}$ and ${\mathscr G}$ are the electromagnetic Lorentz-invariants,
\begin{align}
	{\mathscr F} := \frac{1}{4}F^{\mu\nu}F_{\mu\nu} = \frac{1}{2} \left( {\bm B}^2 - {\bm E}^2 \right) 
    \ \ {\rm and}\ \ 
	{\mathscr G} := -\frac{1}{4} F^{\mu\nu}\tilde{F}_{\mu\nu} = {\bm E}\cdot{\bm B} \;, \label{eq:2} 
\end{align}
where $F^{\mu\nu} := \partial^\mu A^\nu - \partial^\nu A^\mu$ is the electromagnetic field-strength tensor and $\tilde{F}^{\mu\nu} := (1/2)\epsilon^{\mu\nu\rho\sigma}F_{\rho\sigma}$ is its dual, and the electric and magnetic fields, respectively, are defined as ${\bm E}:=(F^{10},F^{20},F^{30})$ and ${\bm B}:=(F^{32},F^{13},F^{21})$.  The modified Lagrangian (\ref{eq:1}) is what is known as the Heisenberg-Euler effective Lagrangian and is one of the most basic tools to analyze strong-field QED (see Ref.~\cite{Dunne:2004nc} for review).  It is important for the Schwinger effect that the Heisenberg-Euler effective Lagrangian~(\ref{eq:2}) can have a non-vanishing imaginary part ${\rm Im}\,{\mathscr L}_{\rm HE} \neq 0$ because the integrand has a sequence of poles on the integration contour if ${\rm Im}\sqrt{ {\mathscr F} + {\rm i}{\mathscr G}} \neq 0$, or
\begin{align}
	{\rm Im}\,{\mathscr L}_{\rm HE} \neq 0
	\ \ \ {\rm if}\ \ \ {\mathscr F} < 0 \ \ {\rm or}\ \ {\mathscr G}\neq 0 
	\ \ \Leftrightarrow\ \ |{\bm B}| < |{\bm E}| \ \ {\rm or}\ \ {\bm E}\cdot{\bm B}\neq 0 \;. \label{eq::2}
\end{align}
Thus, an imaginary part appears if an electric field is applied.  Conversely, electromagnetic fields that do not satisfy the condition~(\ref{eq::2}), such as pure magnetic fields and plane waves, cannot induce any imaginary part.  Heisenberg and Euler noted that ``the imaginary part cannot be interpreted physically” and refrained from analyzing it in detail, but they briefly speculated about a possible connection to Sauter’s vacuum pair production.

The result of Heisenberg and Euler was re-examined by Schwinger in 1951~\cite{Schwinger:1951nm}.  Schwinger reproduced the Heisenberg-Euler Lagrangian~(\ref{eq:1}) in a fully quantum-field theoretic manner based on QED by developing the so-called proper-time method.  In the modern Feynman-diagram language, Schwinger’s calculation amounts to evaluating the following one-loop diagram responsible for the vacuum polarization:
\begin{align}
	\vcenter{\hbox{
	\begin{tikzpicture}
		\begin{feynhand}
				\vertex[particle] (a1) at (-0.5, +0.0);
				\vertex[particle] (a2) at (+0.0, +0.5);
				\vertex[particle] (a3) at (+0.5, +0.0);
				\vertex[particle] (a4) at (+0.0, -0.5);
				\propag [plain, line width=3pt]  (a1) to [in=180, out=90, looseness=1.0] (a2);
				\propag [plain, line width=3pt]  (a2) to [in=90, out=0, looseness=1.0] (a3);
				\propag [plain, line width=3pt]  (a1) to [in=180, out=-90, looseness=1.0] (a4);
				\propag [plain, line width=3pt]  (a4) to [in=-90, out=0, looseness=1.0] (a3);
			\vertex[particle] (eq) at (+1.0, +0.0) {$=$};
				\vertex[particle] (a1) at (+1.5, +0.0);
				\vertex[particle] (a2) at (+2.0, +0.5);
				\vertex[particle] (a3) at (+2.5, +0.0);
				\vertex[particle] (a4) at (+2.0, -0.5);
				\propag [plain, line width=0.8pt]  (a1) to [in=180, out=90, looseness=1.0] (a2);
				\propag [plain, line width=0.8pt]  (a2) to [in=90, out=0, looseness=1.0] (a3);
				\propag [plain, line width=0.8pt]  (a1) to [in=180, out=-90, looseness=1.0] (a4);
				\propag [plain, line width=0.8pt]  (a4) to [in=-90, out=0, looseness=1.0] (a3);
			\vertex[particle] (eq) at (+3.0, +0.0) {$+$};
				\vertex[particle] (a1) at (+3.8, +0.0);
				\vertex[particle] (a2) at (+4.3, +0.5);
				\vertex[particle] (a3) at (+4.8, +0.0);
				\vertex[particle] (a4) at (+4.3, -0.5);
				\vertex[crossdot, minimum size=5pt, line width=0.5pt] (b1) at (+5.1, +0.0) {};
				\vertex[crossdot, minimum size=5pt, line width=0.5pt] (b2) at (+3.5, +0.0) {};
				\propag [plain, line width=0.8pt]  (a1) to [in=180, out=90, looseness=1.0] (a2);
				\propag [plain, line width=0.8pt]  (a2) to [in=90, out=0, looseness=1.0] (a3);
				\propag [plain, line width=0.8pt]  (a1) to [in=180, out=-90, looseness=1.0] (a4);
				\propag [plain, line width=0.8pt]  (a4) to [in=-90, out=0, looseness=1.0] (a3);
				\propag [bos]  (a1) to (b2);
				\propag [bos]  (a3) to (b1);
			\vertex[particle] (eq) at (+5.6, +0.0) {$+$};
				\vertex[particle] (a1) at (+6.4, +0.0);
				\vertex[particle] (a2) at (+6.9, +0.5);
				\vertex[particle] (a3) at (+7.4, +0.0);
				\vertex[particle] (a4) at (+6.9, -0.5);
				\vertex[crossdot, minimum size=5pt, line width=0.5pt] (b1) at (+7.7, +0.0) {};
				\vertex[crossdot, minimum size=5pt, line width=0.5pt] (b2) at (+6.1, +0.0) {};
				\vertex[crossdot, minimum size=5pt, line width=0.5pt] (b3) at (+6.9, +0.8) {};
				\vertex[crossdot, minimum size=5pt, line width=0.5pt] (b4) at (+6.9, -0.8) {};
				\propag [plain, line width=0.8pt]  (a1) to [in=180, out=90, looseness=1.0] (a2);
				\propag [plain, line width=0.8pt]  (a2) to [in=90, out=0, looseness=1.0] (a3);
				\propag [plain, line width=0.8pt]  (a1) to [in=180, out=-90, looseness=1.0] (a4);
				\propag [plain, line width=0.8pt]  (a4) to [in=-90, out=0, looseness=1.0] (a3);
				\propag [bos]  (a1) to (b2);
				\propag [bos]  (a3) to (b1);
				\propag [bos]  (a4) to (b4);
				\propag [bos]  (a2) to (b3);
			\vertex[particle] (eq) at (+8.2, +0.0) {$+$};
			\vertex[particle] (eq) at (+8.7, -0.03) {$\cdots$};
			\vertex[particle] (eq) at (+9.3, +0.0) {$=$};
			\vertex[particle] (eq) at (+9.9, -0.1) {$\displaystyle \sum_{{\rm\ all}\;\otimes}$};
				\vertex[particle] (a1) at (+10.7, +0.0);
				\vertex[particle] (a2) at (+11.2, +0.5);
				\vertex[particle] (a3) at (+11.7, +0.0);
				\vertex[particle] (a4) at (+11.2, -0.5);
				\propag [plain, line width=0.8pt]  (a1) to [in=180, out=90, looseness=1.0] (a2);
				\propag [plain, line width=0.8pt]  (a2) to [in=90, out=0, looseness=1.0] (a3);
				\propag [plain, line width=0.8pt]  (a1) to [in=180, out=-90, looseness=1.0] (a4);
				\propag [plain, line width=0.8pt]  (a4) to [in=-90, out=0, looseness=1.0] (a3);
			\pgfmathsetmacro{\offs}{11.2}		
			\pgfmathsetmacro{\radius}{0.5}
			\pgfmathsetmacro{\prop}{0.3}		
			\pgfmathsetmacro{\ang}{0}
				\vertex[particle] (b1) at ({\offs+\radius*cos(\ang)}, {\radius*sin(\ang)});				
				\vertex[crossdot, minimum size=5pt, line width=0.5pt] (b2) at ({\offs+\radius*cos(\ang)+\prop*cos(\ang)}, {\radius*sin(\ang)+\prop*sin(\ang)}) {};
				\propag [bos]  (b1) to (b2);
			\pgfmathsetmacro{\ang}{45}
				\vertex[particle] (b1) at ({\offs+\radius*cos(\ang)}, {\radius*sin(\ang)});				
				\vertex[crossdot, minimum size=5pt, line width=0.5pt] (b2) at ({\offs+\radius*cos(\ang)+\prop*cos(\ang)}, {\radius*sin(\ang)+\prop*sin(\ang)}) {};
				\propag [bos]  (b1) to (b2);
			\pgfmathsetmacro{\ang}{90}
				\vertex[particle] (b1) at ({\offs+\radius*cos(\ang)}, {\radius*sin(\ang)});				
				\vertex[crossdot, minimum size=5pt, line width=0.5pt] (b2) at ({\offs+\radius*cos(\ang)+\prop*cos(\ang)}, {\radius*sin(\ang)+\prop*sin(\ang)}) {};
				\propag [bos]  (b1) to (b2);
			\pgfmathsetmacro{\ang}{135}
				\vertex[particle] (b1) at ({\offs+\radius*cos(\ang)}, {\radius*sin(\ang)});				
				\vertex[crossdot, minimum size=5pt, line width=0.5pt] (b2) at ({\offs+\radius*cos(\ang)+\prop*cos(\ang)}, {\radius*sin(\ang)+\prop*sin(\ang)}) {};
				\propag [bos]  (b1) to (b2);
			\pgfmathsetmacro{\ang}{180}
				\vertex[particle] (b1) at ({\offs+\radius*cos(\ang)}, {\radius*sin(\ang)});				
				\vertex[crossdot, minimum size=5pt, line width=0.5pt] (b2) at ({\offs+\radius*cos(\ang)+\prop*cos(\ang)}, {\radius*sin(\ang)+\prop*sin(\ang)}) {};
				\propag [bos]  (b1) to (b2);
			\pgfmathsetmacro{\ang}{225}
				\vertex[particle] (b1) at ({\offs+\radius*cos(\ang)}, {\radius*sin(\ang)});				
				\vertex[crossdot, minimum size=5pt, line width=0.5pt] (b2) at ({\offs+\radius*cos(\ang)+\prop*cos(\ang)}, {\radius*sin(\ang)+\prop*sin(\ang)}) {};
				\propag [bos]  (b1) to (b2);
			\pgfmathsetmacro{\ang}{255}
			\vertex[dot, minimum size=1.5pt] (v2) at ({\offs+\radius*cos(\ang)+\prop*cos(\ang)}, {\radius*sin(\ang)+\prop*sin(\ang)}) {};
			\pgfmathsetmacro{\ang}{270}
			\vertex[dot, minimum size=1.5pt] (v2) at ({\offs+\radius*cos(\ang)+\prop*cos(\ang)}, {\radius*sin(\ang)+\prop*sin(\ang)}) {};	
			\pgfmathsetmacro{\ang}{285}
			\vertex[dot, minimum size=1.5pt] (v2) at ({\offs+\radius*cos(\ang)+\prop*cos(\ang)}, {\radius*sin(\ang)+\prop*sin(\ang)}) {};
			\pgfmathsetmacro{\ang}{300}
			\vertex[dot, minimum size=1.5pt] (v2) at ({\offs+\radius*cos(\ang)+\prop*cos(\ang)}, {\radius*sin(\ang)+\prop*sin(\ang)}) {};
			\pgfmathsetmacro{\ang}{315}
			\vertex[dot, minimum size=1.5pt] (v2) at ({\offs+\radius*cos(\ang)+\prop*cos(\ang)}, {\radius*sin(\ang)+\prop*sin(\ang)}) {};
			\pgfmathsetmacro{\ang}{330}
			\vertex[dot, minimum size=1.5pt] (v2) at ({\offs+\radius*cos(\ang)+\prop*cos(\ang)}, {\radius*sin(\ang)+\prop*sin(\ang)}) {};
		\end{feynhand}
	\end{tikzpicture}
	}} \;, \label{eq:3}
\end{align}
where the thin lines (\hspace*{0.15mm}\rule[0.5ex]{1.1em}{0.8pt}\hspace*{0.15mm}) represent the bare electron propagators, while the thick lines (\hspace*{0.15mm}\rule[0.4ex]{1.1em}{2pt}\hspace*{0.15mm}) denote the dressed electron propagators that non-perturbatively incorporate the effects of the background electromagnetic field indicated by the blobs ($\sim\hspace*{-1.2mm}\otimes$), i.e., 
\begin{align}
	\vcenter{\hbox{
	\begin{tikzpicture}
		\begin{feynhand}
				\vertex[particle] (a1) at (-0.5, +0.0);
				\vertex[particle] (a4) at (+0.5, +0.0);
				\propag [plain, line width=3pt]  (a1) to (a4);
			\vertex[particle] (eq) at (+1.0, +0.0) {$=$};
				\vertex[particle] (a1) at (+1.5, +0.0);
				\vertex[particle] (a4) at (+2.5, +0.0);
				\propag [plain, line width=0.8pt]  (a1) to (a4);
			\vertex[particle] (eq) at (+3.0, +0.0) {$+$};
				\vertex[particle] (a1) at (+3.5, +0.0);
				\vertex[particle] (a2) at (+4.0, +0.0);			
				\vertex[particle] (a4) at (+4.5, +0.0);
				\vertex[crossdot, minimum size=5pt, line width=0.5pt] (b1) at (+4.0, -0.3) {};
				\propag [plain, line width=0.8pt]  (a1) to (a4);
				\propag [bos]  (a2) to (b1);
			\vertex[particle] (eq) at (+5.0, +0.0) {$+$};
				\vertex[particle] (a1) at (+5.5, +0.0);
				\vertex[particle] (a2) at (+5.83, +0.0);			
				\vertex[particle] (a3) at (+6.17, +0.0);			
				\vertex[particle] (a4) at (+6.5, +0.0);
				\vertex[crossdot, minimum size=5pt, line width=0.5pt] (b1) at (+5.83, -0.3) {};
				\vertex[crossdot, minimum size=5pt, line width=0.5pt] (b2) at (+6.17, -0.3) {};
				\propag [plain, line width=0.8pt]  (a1) to (a4);
				\propag [bos]  (a2) to (b1);				
				\propag [bos]  (a3) to (b2);				
			\vertex[particle] (eq) at (+7.0, +0.0) {$+$};
			\vertex[particle] (eq) at (+7.5, -0.03) {$\cdots$};
		\end{feynhand}
	\end{tikzpicture}
	}} \;.  
\end{align}
Note that contributions from loop diagrams containing an odd number of the blobs vanish by virtue of Furry's theorem (i.e., charge-conjugation symmetry forbids an electron loop from coupling to the electromagnetic field an odd number of times, since such loops are odd in charge conjugation)~\cite{PhysRev.51.125} and thus are omitted in Eq.~(\ref{eq:3}).  

Schwinger then performed a careful analysis of the imaginary part of the Heisenberg-Euler effective Lagrangian ${\rm Im}\,{\mathscr L}_{\rm HE}$.  He pointed out that ${\rm Im}\,{\mathscr L}_{\rm HE}$ is directly related to the vacuum-decay rate $w$.  In fact, the vacuum persistence probability---the probability that the initial vacuum in the infinite past, $\ket{{\rm vac;in}}$, remains the vacuum in the infinite future, $\ket{{\rm vac;out}}$---is given by
\begin{align}
	\left| \braket{{\rm vac;out} | {\rm vac;in}} \right|^2
	= \exp \left[ -\int {\rm d}^4x\, 2\,{\rm Im}\,{\mathscr L}_{\rm HE} \right]
	= 1 - \int {\rm d}^4x\, 2\,{\rm Im}\,{\mathscr L}_{\rm HE} + {\mathcal O}\left( |{\rm Im}\,{\mathscr L}_{\rm HE}|^2 \right) \;, \label{eq:4}
\end{align}
which suggests that the vacuum decays with the rate per unit spacetime volume,
\begin{align}
	w := 2\,{\rm Im}\,{\mathscr L}_{\rm HE} \; .
\end{align}
Schwinger also explicitly evaluated the imaginary part for the case of a purely electric field (${\bm E} \neq {\bm 0}$ and ${\bm B} = {\bm 0}$) and derived a compact expression for the vacuum-decay rate,
\begin{align}
	w = 2\,\frac{(eE)^2}{(2\pi)^3} \sum_{k=1}^{\infty} \frac{1}{k^2} \exp\left[ -k\pi \frac{m^2}{|eE|} \right] \;, \label{eq:6}
\end{align}
where the sum over $k$ accounts for the pole contributions (sometimes called ``instantons") in Eq.~(\ref{eq:1}).  This is the famous Schwinger formula.  Note that the $k$ summation can be done analytically, and the Schwinger formula can also be expressed in terms of the dilogarithm function $w = 2((eE)^2/(2\pi)^3) {\rm Li}_2({\rm e}^{-\pi m^2/|eE|})$~\cite{Dunne:2025cyo}.  We emphasize that the characteristic exponential factor in the intuitive order estimate~(\ref{eqwsd1}) is reproduced in the Schwinger formula.  The non-vanishing vacuum-decay rate $w \neq 0$ means that the initial vacuum evolves into a non-vacuum, which should be a state containing a finite number of particles.  In other words, Sauter’s vacuum pair production indeed occurs if ${\rm Im}\,{\mathscr L}_{\rm HE} \neq 0$.  This interpretation is also supported by the optical theorem (or Cutkosky's rule in quantum-field theory~\cite{Cutkosky:1960sp}), which tells us that the loop diagram~(\ref{eq:3}) can have an imaginary part when the electron becomes on-shell.  In this way, Sauter’s idea was fully formulated in QED by Schwinger.  

However, this is not the end of the story: it is impossible to extract detailed information about the produced pairs, such as the number of pairs and the momentum spectrum, from the Heisenberg-Euler effective Lagrangian~(\ref{eq:1}), where the electron degrees of freedom have already been integrated out.  An alternative approach is necessary to obtain such information.  It was Nikishov~\cite{Nikishov:1969tt}, who first resolved this issue in 1969 by using what is known today as the Bogoliubov-transformation approach.  Nikishov obtained the phase-space distribution of the produced electrons per spin at the infinite future, 
\begin{align}
	f
	:= (2\pi)^3 \frac{{\rm d}^6N}{{\rm d}{\bm p}^3{\rm d}{\bm x}^3} 
	= \exp\left[ - \pi \frac{m^2 + {\bm p}_\perp^2}{|eE|} \right] \; , \label{eq:7}
\end{align}
where ${\bm p}_\perp$ is the momentum transverse to the electric field.  Note that the distribution of the pair-produced positrons, $\bar{f}$, is obtained by flipping the sign of momentum, $\bar{f}({\bm p}) = f(-{\bm p})$, because they are produced as a pair and the total momentum of the pair should be zero because there is no momentum supply from the constant field.  We refer Eq.~(\ref{eq:7}) as Nikishov's formula\footnote{Equation~(\ref{eq:7}) is often referred as Schwinger formula, which is however inappropriate from a historical viewpoint, since what Schwinger derived is the formula for the vacuum decay rate $w$~(\ref{eq:6}) and did not obtain that for the spectrum $f$~(\ref{eq:7}) nor for the production rate $\Gamma$~(\ref{eq:8}).   }.  Integrating the spectrum~(\ref{eq:7}) over momentum and spin yields the total number of pairs produced, $N$, and the rate of the pair production per unit spacetime volume, $\Gamma$, as
\begin{align}
	\Gamma
	:= \frac{N}{V_4}
	:= \frac{1}{T} \sum_{\rm spin} \int \frac{{\rm d}^3{\bm p}}{(2\pi)^3} f
	= 2 \frac{(eE)^2}{(2\pi)^3} \exp\left[ - \pi \frac{m^2}{|eE|} \right] \; , \label{eq:8}
\end{align}
where $V_4 := \int {\rm d}^4x = \int {\rm d}^3{\bm x} \times \int {\rm d}x^0 =: V \times T$ is the spacetime volume of the system and $\sum_{\rm spin} = 2$ accounts for the spin degeneracy.  

We remark that the rates for the number of pairs $\Gamma$~(\ref{eq:8}) and the vacuum decay $w$~(\ref{eq:6}) do not coincide with each other.  They are different quantities~\cite{Cohen:2008wz, Fukushima:2009er, Nishida:2021qta}.  The difference may become evident, as done by Nikishov, by expressing the vacuum-decay rate $w$~(\ref{eq:6}) in terms of the phase-space density $f$~(\ref{eq:7}), 
\begin{align}
    w 
    = \frac{1}{T} \sum_{\rm spin} \sum_{k=1}^\infty \int \frac{{\rm d}^3{\bm p}}{(2\pi)^3}\frac{f^{k}}{k} 
    = \Gamma + \frac{1}{T} \sum_{\rm spin} \sum_{k=2}^\infty \int \frac{{\rm d}^3{\bm p}}{(2\pi)^3}\frac{f^{k}}{k} \; . \label{eq:9}
\end{align}
Thus, they agree with each other only in the weak-field limit $|eE| \ll m^2$, where the lowest $k=1$ contribution dominates.  The distinction between $w$ and $\Gamma$ become important for strong fields $|eE| \gtrsim m^2$ or for massless particles like gluon [see Eq.~(\ref{eq:---10})].   

The classic results by Sauter, Heisenberg-Euler, Schwinger, and Nikishov for the Schwinger effect for the electron were extended to charged particles with other spin statistics: to scalars by Weisskopf in 1936~\cite{Weisskopf:406571}, to vectors by Vanyashin and Terentev in 1965~\cite{Vanyashin:1965ple}, and to arbitrary spin by Marinov and Popov in 1972~\cite{Marinov:1972nx}.  They found that the pair-production rate $\Gamma$ is unchanged from the fermionic one~(\ref{eq:8}), except for an overall spin degeneracy factor $2s+1$ (with $s=0, 1/2, 1, \cdots$), while the vacuum decay rate $w$ acquires a correction due to spin statistics (i.e., fermion or boson) compared to Eq.~(\ref{eq:9}): 
\begin{align}
\begin{split}
	&\Gamma = \frac{1}{T} \sum_{\rm spin} \int \frac{{\rm d}^3{\bm p}}{(2\pi)^3} f 
	= (2s+1)\frac{(eE)^2}{(2\pi)^3} \exp\left[ -\pi\frac{m^2}{|eE|} \right] \;, \\
	&w = \frac{1}{T} \sum_{\rm spin} \sum_{k=1}^\infty \int \frac{{\rm d}^3{\bm p}}{(2\pi)^3} \beta_k \frac{f^{k}}{k} 
	= (2s+1) \frac{(eE)^2}{(2\pi)^3} \sum_{k=1}^{\infty} \frac{\beta_k}{k^2} \exp\left[ -k\pi\frac{m^2}{|eE|} \right] \;, \label{eq:10} 
\end{split}
\end{align}
where the phase-space distribution $f$ remains the same as Eq.~(\ref{eq:7}) and $\beta_k$ is the statistical factor such that $\beta_k = (-1)^{k-1}$ for bosons and $=1$ for fermions.  It is reasonable that $f$ is unchanged because the work exerted by an electric field do not depend on spin and hence the pair-production mechanism itself does not change.  

These results for a purely electric field were extended to configurations including a magnetic field.  This was first done by Nikishov~\cite{Nikishov:1969tt} and independently by Bunkin and Tugov~\cite{Bunkin:1969if} for spin one-half in 1969, and by Marinov and Popov~\cite{Marinov:1972nx} for arbitrary spin in 1972.  The essence of the magnetic-field effect is that it modifies the dispersion relation through the Landau quantization (i.e., a charged particle in a magnetic field undergoes a circular motion, which is first-quantized and hence the transverse motion is discretized)~\cite{Landau:1930fja}.  Meanwhile, it does not change the production mechanism itself, since a magnetic field does no work.  For definiteness, let us consider a parallel electromagnetic field, ${\bm E}\parallel{\bm B}$ (i.e., ${\bm E}=E{\bf e}_\parallel$ and ${\bm B}=B{\bf e}_\parallel$), which is in fact the most general case because any electromagnetic-field configuration can be made parallel by a Lorentz boost.  Due to the Landau quantization by the magnetic field, the orbital motion is discretized as 
\begin{align}
	\sqrt{m^2 + {\bm p}_{\perp}^2} \to \sqrt{m^2 + (2n + 1)|eB| - {\mathfrak g}s_\parallel eB}
	\ \ \ \ {\rm and}\ \ 
	\int \frac{{\rm d}^2{\bm p}_\perp}{(2\pi)^2} \to \frac{|eB|}{2\pi} \sum_{n=0}^\infty \;, \label{eq:11}
\end{align}
where $n = 0,1,2,\ldots$ labels the Landau level, ${\mathfrak g}$ is the gyromagnetic ratio, and $s_\parallel$ is the longitudinal spin component, $s_\parallel = -s, -s+1, \ldots, s$.  Accordingly, the phase-space distribution of the produced particles $f$~(\ref{eq:7}) is modified to
\begin{align}
	f
	\to \exp\left[ - \pi \frac{m^2 + (2n +1)|eB| - {\mathfrak g}s_{\parallel} eB}{|eE|} \right] \; .  \label{eq--14}
\end{align}
It is clear that particles with spin aligned with the magnetic field, $s_{\parallel} eB>0$, are produced more efficiently because they are energetically favorable due to the Landau quantization (\ref{eq:11}).  The production rate $\Gamma$ and the vacuum decay rate $w$ are then modified from their purely electric-field forms~(\ref{eq:10}) as
\begin{align}
\begin{split}
	&\Gamma 
	\to \frac{1}{T} \sum_{\rm spin} \frac{|eB|}{2\pi} \sum_{n=0}^\infty \int \frac{{\rm d}p_\parallel}{2\pi} f 
	= \Gamma(B=0) \times R \left( \pi \left|\frac{B}{E}\right|  \right) \;, \\
	&w 
	\to \frac{1}{T} \sum_{\rm spin} \sum_{k=1}^\infty \frac{|eB|}{2\pi} \sum_{n=0}^\infty \int \frac{{\rm d}p_\parallel}{2\pi} \beta_k \frac{f^{k}}{k} 
	= \sum_{k=1}^\infty w_k(B=0) \times R \left( k \pi \left|\frac{B}{E}\right|  \right)  \;, \label{eq:13} 
\end{split}
\end{align}
where the modification factor $R$ is given by
\begin{align}
	R(x) := \frac{x}{2s+1} \frac{\sinh\left( (2s+1) \frac{{\mathfrak g}}{2}x \right)}{ \sinh\left( \frac{{\mathfrak g}}{2}x \right) \sinh\left( x \right) }
	= \left\{ \begin{array}{ll} \displaystyle 1 + \frac{{\mathfrak g}^2s(1+s)-1}{6} x^2 + {\mathcal O}(x^4) & (x\ll 1) \\[8pt] \displaystyle \frac{2x}{2s+1} {\rm e}^{(s{\mathfrak g}-1)x} & (x \gg 1)\end{array} \right. \;.
\end{align}
Thus, the Schwinger effect is suppressed ($R\leq 1$) for scalars $s=0$.  For higher spin $s \geq 1/2$, the value of ${\mathfrak g}$ becomes relevant, but for usual particles with ${\mathfrak g}=2$, the magnetic field enhances the Schwinger effect ($R\geq 1$).

\subsection{Going beyond the early treatment} \label{sec:2.3}

The early developments reviewed in Sec.~\ref{sec:2.2} form the foundation of Schwinger-effect studies.  However, it should be emphasized that they rely on a number of idealized assumptions.  Relaxing these idealizations remains an active area of research and is crucial for realistic phenomenological applications.  

\subsubsection{Inhomogeneous fields} \label{sec:2.3..1}

In the early treatments, the electromagnetic field was assumed to be homogeneous in spacetime.  Consequently, the resulting equations are strictly valid only for homogeneous fields, or for fields that vary so slowly that the characteristic length scale of the inhomogeneity is much larger than the microscopic length scales relevant to the Schwinger effect.  Realistic electromagnetic fields, however, are not necessarily slowly varying, which underscores the need to go beyond the constant-field approximation.  For example, although the electromagnetic fields generated in heavy-ion collisions can be extremely strong, their spacetime extent is typically very small~\cite{Skokov:2009qp, Voronyuk:2011jd, Bzdak:2011yy, Deng:2012pc, Taya:2024wrm, Taya:2025utb} (see Refs.~\cite{Hattori:2016emy, Shen:2025unr} for reviews).  In laser experiments, the field profile can be engineered, suggesting the possibility of optimizing it to enhance characteristic signatures of the Schwinger effect, provided that inhomogeneity effects are properly controlled~\cite{Hebenstreit:2014lra, Hebenstreit:2015jaa}.

The easiest extension is the so-called locally-constant-field approximation (LCFA)~\cite{Bulanov:2004de}.  The idea is to naively extend the homogeneous formulas, Eqs.~(\ref{eq:6}) and~(\ref{eq:7}), by promoting the fields appearing in them to spacetime-dependent ones by hand, ${\bm E} \to {\bm E}(t,{\bm x})$ and ${\bm B} \to {\bm B}(t,{\bm x})$.  This is a crude approximation (see Refs.~\cite{Aleksandrov:2018zso, Aleksandrov:2019ddt} for validity), but due to its simpleness and convenience, it is widely used for Schwinger-effect studies with inhomogeneous fields.  

A more sophisticated treatment was done by Brezin and Itzykson~\cite{1970PhRvD...2.1191B} and, independently, by Popov~\cite{Popov:1971, Popov:1971iga} around 1970 based on semi-classical methods (i.e., the $\hbar$ expansion)\footnote{Brezin and Itzykson employed a steepest-descent analysis, similar to the Dykhne-Davis-Pechukas method in solid-state physics~\cite{dykhne1962adiabatic, doi:10.1063/1.432648, Fukushima:2019iiq}, and Popov developed the imaginary-time method~\cite{Popov:2005rp}.  Since these pioneering works, a number of semi-classical methods have been invented to analyze the Schwinger effect for inhomogeneous fields; e.g., the worldline instanton method~\cite{Dunne:2005sx, Dunne:2006st, Dunne:2006ur} and the WKB method~\cite{Taya:2020dco}.   It is beyond our scope to detail these semi-classical analyses, and we instead refer the reader to Ref.~\cite{Fedotov:2022ely} and references therein.  }.  They considered a purely time-dependent electric field, with peak strength $E_0$ and frequency $\omega$, and showed that the pair-production rate $\Gamma$ scales as 
\begin{align}
	\Gamma
	\propto 
	\left\{  \begin{array}{ll} 
		{\rm e}^{-\pi \frac{m^2}{eE_0}} & (\gamma_{\rm K} \ll 1) \\[5pt]
		\left( \frac{eE_0}{m^2} \right)^{4m/\omega} & (\gamma_{\rm K} \gg 1) 
	\end{array} \right.
	\ {\rm with}\ \ \gamma_{\rm K} := \frac{m\omega}{eE_0} \;. \label{eq:::17}
\end{align}
The dimensionless parameter $\gamma_{\rm K}$ is known as the Keldysh parameter, reflecting its historical origin in the study of strong-field atomic ionization, where it was first introduced by Keldysh~\cite{Keldysh:1965ojf}.  Intuitively, the Keldysh parameter can be interpreted as the ratio of the tunneling time $\tau = d/c = m/eE_0$ (see Fig.~\ref{fig:1}) to the characteristic time scale of the electric field, $1/\omega$, as $\gamma_{\rm K} = \tau/\omega^{-1}$.  In other words, when $\omega$ is sufficiently large, so is $\gamma_{\rm K}$, the electric field varies before the tunneling can be completed, and the non-perturbative Schwinger effect cannot take place.  For such rapidly varying fields, as is evident from Eq.~(\ref{eq:::17}), the dependence on $eE_0$ becomes at most polynomial, indicating that the vacuum pair production is governed by a {\it perturbative} mechanism.  Physically, when the field varies rapidly, it interacts with the electron incoherently; namely, not as a coherent classical field but as individual photons constituting the field, which is analogous to the photoelectric effect in materials.  Each such incoherent photon scattering transfers an energy $\omega$ to a virtual pair in the vacuum, and the energy threshold for a pair production, $2m$, is overcome after approximately $2m/\omega$ multi-photon scatterings.  The corresponding scattering amplitude therefore scales as $e^{2m/\omega}$, and squaring this amplitude yields the pair-production probability $|e^{2m/\omega}|^2$, reproducing the exponent $4m/\omega$ appearing in Eq.~(\ref{eq:::17}).

Note that the semi-classical analysis is valid only when $\omega/m \ll 1$.  Roughly speaking, this is because $\hbar$ always appears together with derivatives in the Dirac (or Klein-Gordon) equation, $\partial_t \to \hbar \partial_t$, so changing the value of $\hbar$ is equivalent to rescaling the time variable, or equivalently the frequency $\omega$~\cite{Taya:2020dco}.  Since $\omega$ has mass dimension, whether it is large or small should be assessed relative to the characteristic energy scale of the problem, namely, the electron mass $m$.  Thus, the condition $\omega/m \ll 1$ is required to justify the semi-classical $\hbar$ expansion.  Conversely, this means that the semi-classical result~(\ref{eq:::17}) is not justified for very large $\gamma_{\rm K}$, where $\omega/m \not\ll 1$.  Possible approaches to discussing pair production in such a large-$\omega$ regime are to use exactly solvable field configurations such as the Sauter pulse (see also Refs.~\cite{Bagrov:2014rss,Fedotov:2022ely} for other solvable fields) and low-order perturbation theories with respect to the field~\cite{Fukushima:2009er, Taya:2014taa, Gelis:2015kya}.  In the extreme limit $\gamma_{\rm K} \gg 1$ and $\nu := eE_0/\omega^2 \ll 1$, pair production has been shown to reduce to the lowest-order perturbative process (one-photon pair production).

Temporal inhomogeneities generally enhance pair production.  This is evident from Eq.~(\ref{eq:::17}), which shows that the perturbative pair production for $\gamma_{\rm K} \gg 1$ is only weakly suppressed by a power of $eE_0/m^2$, whereas the non-perturbative one for $\gamma_{\rm K} \ll 1$ is strongly suppressed by an exponential factor.  Intuitively, this enhancement arises because the time dependence supplies additional energy to virtual pairs, thereby reducing the tunneling barrier.  Consequently, controlling temporal inhomogeneities provides an efficient means of enhancing vacuum pair production and is exploited, for example, in the dynamically assisted Schwinger effect~\cite{Schutzhold:2008pz, Dunne:2009gi} (an analog of the Franz-Keldysh effect in materials~\cite{1958ZNatA..13..484F, keldysh1958effect, Taya:2018eng, Huang:2019uhf, Taya:2023ltd}).  Note that the time dependence for the enhancement is not necessarily supplied by the electromagnetic field; e.g., the Schwigner effect for a constant electric field can be enhanced by time-dependent geometries~\cite{Taya:2020pkm}.  

The impact of spatial inhomogeneity is characterized by a spatial analog of the Keldysh parameter, $\tilde{\gamma}_{\rm K}$, and is given by~\cite{Nikishov:1970br, Gies:2005bz, Dunne:2005sx, Dunne:2006st, Gies:2015hia, Gies:2016coz}
\begin{align}
	\Gamma
	\propto 
	\left\{  \begin{array}{ll} 
		{\rm e}^{-\pi \frac{m^2}{eE_0}} & (\tilde{\gamma}_{\rm K} \ll 1) \\[5pt]
		0 & (\tilde{\gamma}_{\rm K} \to 1) 
	\end{array} \right.
	\ {\rm with}\ \ \tilde{\gamma}_{\rm K} := \frac{m}{eE_0 \lambda} \;, \label{eq::::17}
\end{align}
where $\lambda$ denotes the characteristic length scale of the spatial inhomogeneity.   This result means that vacuum pair production via the Schwinger effect occurs only when $\lambda$ is sufficiently large.  As $\lambda$ decreases, the pair production is suppressed and eventually ceases at $\lambda = m/eE_0$.  The absence of the perturbative multi-photon pair production in this case can be understood from the fact that spatial inhomogeneities cannot supply energy to the vacuum.  Instead, they impart momentum $\Delta p$ to virtual pairs, which, in contrast to the temporal case, increases the effective tunneling barrier as $m \to \sqrt{m^2 + (\Delta p)^2}$, thereby suppressing pair production.

When the field orientation varies in space and time (i.e., beyond linear polarization), pair production can acquire a spin dependence even in the absence of magnetic fields~\cite{Strobel:2014tha, Ebihara:2015aca, Wollert:2015kra, Blinne:2015zpa, Huang:2019uhf, Kohlfurst:2018kxg, Li:2019rex, Takayoshi:2020afs, Aleksandrov:2024cqh}, and may induce a spin current from the vacuum~\cite{Huang:2019szw, Chu:2021eae}.  This additional spin dependence originates from the electron spin-orbit coupling, ${\bm s} \cdot ({\bm p} \times {\bm E})$.  Intuitively, an electron moving in an electric field experiences an effective magnetic field in its rest frame through the Lorentz transformation, ${\bm B} \propto {\bm p} \times {\bm E}$, which tends to polarize the spin along the direction of this effective magnetic field.

\subsubsection{Radiative corrections} \label{sec::2.3.2}

The early development is based on the one-loop treatment~(\ref{eq:3}) (see also footnote~\ref{foot:2}), meaning that higher-loop contributions, arising from quantum photons on top of the classical electromagnetic field, are neglected.  For example, two-loop contributions can be depicted as
\begin{align}
	\vcenter{\hbox{
	\begin{tikzpicture}
		\begin{feynhand}
				\vertex[particle] (a1) at (-0.5, +0.0);
				\vertex[particle] (a2) at (+0.0, +0.5);
				\vertex[particle] (a3) at (+0.5, +0.0);
				\vertex[particle] (a4) at (+0.0, -0.5);
				\propag [plain, line width=3pt]  (a1) to [in=180, out=90, looseness=1.0] (a2);
				\propag [plain, line width=3pt]  (a2) to [in=90, out=0, looseness=1.0] (a3);
				\propag [plain, line width=3pt]  (a1) to [in=180, out=-90, looseness=1.0] (a4);
				\propag [plain, line width=3pt]  (a4) to [in=-90, out=0, looseness=1.0] (a3);
				\propag [bos]  (a1) to (a3);
				\vertex[particle] (o) at (1.2, +0.0) {${\rm and}$};
				\vertex[particle] (a1) at (+1.9, +0.0);
				\vertex[particle] (a2) at (+2.4, +0.5);
				\vertex[particle] (a3) at (+2.9, +0.0);
				\vertex[particle] (a4) at (+2.4, -0.5);
				\propag [plain, line width=3pt]  (a1) to [in=180, out=90, looseness=1.0] (a2);
				\propag [plain, line width=3pt]  (a2) to [in=90, out=0, looseness=1.0] (a3);
				\propag [plain, line width=3pt]  (a1) to [in=180, out=-90, looseness=1.0] (a4);
				\propag [plain, line width=3pt]  (a4) to [in=-90, out=0, looseness=1.0] (a3);
				\vertex[particle] (b1) at (+3.3, +0.0);
				\vertex[particle] (b2) at (+3.8, +0.5);
				\vertex[particle] (b3) at (+4.3, +0.0);
				\vertex[particle] (b4) at (+3.8, -0.5);
				\propag [plain, line width=3pt]  (b1) to [in=180, out=90, looseness=1.0] (b2);
				\propag [plain, line width=3pt]  (b2) to [in=90, out=0, looseness=1.0] (b3);
				\propag [plain, line width=3pt]  (b1) to [in=180, out=-90, looseness=1.0] (b4);
				\propag [plain, line width=3pt]  (b4) to [in=-90, out=0, looseness=1.0] (b3);
				\propag [bos]  (a3) to (b1);
		\end{feynhand}
	\end{tikzpicture}
	}} \ \ , \label{eq:::3}
\end{align}
where the wavy line (\inlinephoton) represents the photon propagator.  While this neglect may be a reasonable approximation in QED, where the coupling constant $e$ is small, it is no longer justified in strongly-coupled theories such as QCD.  Moreover, it has been argued that strong background fields can effectively enhance the coupling constant, known as Ritus-Narozhny conjecture or $\alpha \chi^{2/3}$ problem~\cite{ritus1970radiative, Narozhnyi:1980dc} (see Refs.~\cite{Fedotov:2016afw, Fedotov:2022ely} for review).  Higher-loop contributions are also essential for describing post-pair-production dynamics, such as scattering among the produced particles and radiation emitted from them (e.g., high-harmonics generation from the vacuum~\cite{PhysRevD.72.085005, Fedotov:2006ii, Kuchiev:2015qua, Aleksandrov:2021ylw, Taya:2021dcz}, similarly to that in materials~\cite{Corkum:1993zz, Lewenstein1994, Vampa2014, Vampa2015}).  

Ritus and Lebedev were the first to carry out the two-loop calculation~(\ref{eq:::3}) for a purely electric case and derived the corresponding effective Lagrangian in 1984~\cite{Ritus:1975pcc, Lebedev:1984mei}.  The result is given by a complicated double integral and cannot be simplified in a closed form for general field strength $E$ (cf. for the special case of Euclidean self-dual fields $F^{\mu\nu}=\tilde{F}^{\mu\nu}$, the result can be expressed in a simple form~\cite{Dunne:2001pp, Dunne:2002qf, Dunne:2002qg}).  In the limit of $|eE|/m^2 \ll 1$, the imaginary part of the effective Lagrangian is given by
\begin{align}
	w_{\rm two\mathchar`-loop}
	= 2\,{\rm Im}\,{\mathscr L}_{\rm two\mathchar`-loop}
	= 2\frac{(eE)^2}{(2\pi)^3}  \sum_{k=1}^\infty \left( \frac{1}{k^2} + \frac{e^2}{4} K_k \right) {\rm e}^{-k\frac{\pi m^2}{|eE|}}
	\xrightarrow{ |eE|/m^2 \ll 1 } 2\frac{(eE)^2}{(2\pi)^3} \left( 1 + \frac{e^2}{4} \right) {\rm e}^{-\frac{\pi m^2}{|eE|}} \;,  \label{eq::20}
\end{align}
where $K_k = - (m/\sqrt{|eE|}) c_k + 1 + {\mathcal O}(\sqrt{|eE|}/m)$ with $c_1=0$ and $c_k = \sum_{\ell=1}^{k-1} 1/(2\sqrt{k\ell(k-\ell)}) \sim \pi/(2\sqrt{l})$ for $k>1$.  Thus, the radiative corrections enhance the Schwinger effect.  Note that Eq.~(\ref{eq::20}) is for the fermionic case.  The coefficient $K_n$ is spin-independent up to corrections of order ${\mathcal O}( \sqrt{|eE|}/m )$.  In scalar QED, the sum obtains the additional statistical factor $\beta_k = (-1)^{k-1}$, and the spin degeneracy factor $2$ is replaced by $1$ [see Eq.~(\ref{eq:10})].

Having obtained Eq.~(\ref{eq::20}), Ritus and Lebedev conjectured that higher-loop contributions can be resummed into an exponential form, which is known as the Ritus exponentiation conjecture~\cite{Lebedev:1984mei}: 
\begin{align}
	2\,{\rm Im}\,{\mathscr L}_{\rm all\mathchar`-loop} 
	\overset{?}{=} 2\frac{(eE)^2}{(2\pi)^3} \sum_{k=1}^\infty \frac{1}{k^2} {\rm e}^{-k\frac{\pi m^2_{\rm eff}}{|eE|}}
	\ \ {\rm with}\ \ 
	m_{\rm eff} = m + \frac{e^2}{4\pi} \frac{k c_k}{2}\sqrt{|eE|} - \frac{e^2}{4\pi} \frac{k |eE|}{2m} + {\mathcal O} \left( \left| \sqrt{|eE|}/m \right|^3 \right)
	\xrightarrow{k=0} m - \frac{e^2}{4\pi}\frac{|eE|}{2m} \;. \label{eq::21}
\end{align}
The negative mass shift in the effective mass $m_{\rm eff}$ can naturally be interpreted as the effect of attractive Coulomb force.  Namely, when a pair production occurs, the electron-positron pair is typically separated by the distance $2d \sim 2m/|eE|$ (see Fig.~\ref{fig:1}).  The attractive Coulomb potential between the pair then induces a mass shift $\Delta m = -(e^2/4\pi)/(2d)$, which reproduces the negative mass correction appearing in the expression for $m_{\rm eff}$~(\ref{eq::21}).  

To date, the Ritus exponentiation conjecture has not been proven nor denied.  This is simply because higher-loop calculations are technically difficult, and even the three-loop contribution has not yet been completed in the $(3+1)$-dimensional QED (see Refs.~\cite{Huet:2018ksz, Huet:2020awq} for the recent progress in ${\rm QED}_{1+1}$).  Nonetheless, it is notable that a scalar-QED calculation based on the worldline instanton method in the weak-field limit at $k=0$ reproduces the effective mass expression~(\ref{eq::21})~\cite{Affleck:1981bma}.  These results are for the weak-coupling limit $e \ll 1$.  The strong-coupling limit $e \gg 1$ has been analyzed in the 2010s with the holography technique based on gauge/gravity duality, which suggests a different form of the imaginary part to Eq.~(\ref{eq::21}) and the existence of the upper limit for the electric field~\cite{Gorsky:2001up, Semenoff:2011ng, Sato:2013iua, Sato:2013hyw}.

\subsubsection{Realtime dynamics} \label{sec:2.3.3}

The formulas reviewed in Sec.~\ref{sec:2.2} describe observables only at the asymptotic times and therefore cannot capture their behaviors at intermediate times.  Access to such realtime information is desirable, since pair production and the accompanying vacuum decay are inherently non-equilibrium processes.

It is, however, unavoidable that realtime information about particle production is ambiguous~\cite{Dabrowski:2014ica, Dabrowski:2016tsx}, which is a well-known fact in the context of particle production in curved spacetime~\cite{Birrell:1982ix, Parker:2009uva}.  In fact, the notion of a particle is well-defined only at the asymptotic times $t \to \pm \infty$, where all interactions are switched off.  This is a general principle of quantum-field theory~\cite{Gell-Mann:1951ooy}.  At the asymptotic times, the plane wave $\propto {\rm e}^{-{\rm i}p_\mu x^\mu}$, i.e., the eigen-function of the translation operator $-{\rm i}\partial_\mu$, becomes the exact solution to the Dirac equation (or the Klein-Gordon equation for charged bosons).  The electron and positron states, respectively, are then identified unambiguously with the positive- ($p^0>0$) and negative-frequency ($p^0<0$) solutions.  In contrast, in the presence of a background electromagnetic field, spacetime-translation invariance is explicitly broken.  As a consequence, the plane wave is no longer a solution to the Dirac equation, and positive- and negative-frequency modes are mixed with each other during the time evolution.  It is, therefore, not justified to define ``a particle" using the plane wave in such a background.  Instead, one must introduce a {\it physically-natural ansatz} to define a particle at intermediate times.  Any such intermediate particle picture is thus inherently ambiguous and depends sensitively on the chosen ansatz~\cite{Dabrowski:2014ica, Dabrowski:2016tsx}.  It is also impossible to justify/deny any ansatz by experiments, since we by no means observe interacting states at the intermediate times.

A commonly adopted ansatz for defining particles at intermediate times is the so-called adiabatic basis~\cite{Tanji:2008ku, Dabrowski:2014ica, Dabrowski:2016tsx}.  Note that the same ansatz is implicitly assumed in the Dirac-Heisenberg-Wigner formalism~\cite{Hebenstreit:2010vz} and in the quantum Vlasov approach~\cite{Kluger:1998bm, Schmidt:1998vi}.  The basic idea is to expand the electron field operator in terms of adiabatic solutions to the Dirac equation, of the form ${\rm e}^{-{\rm i} \int {\rm d}x^\mu\, p_\mu(x)}$, where $p_\mu$ is the kinetic four-momentum in the background field.  The expansion coefficients are then promoted to annihilation and creation operators at intermediate times by following the standard canonical-quantization procedure of quantum-field theory.

Although the concept of an intermediate particle picture is ambiguous, it nevertheless provides a useful intuitive description of the realtime dynamics of pair production, and a number of numerical simulations have been done over the several decades:
\begin{itemize}
\item A pair production occurs around when the threshold energy is minimized.  The single-particle energy evolves in an electromagnetic field as $p^0 = \sqrt{m^2 + {\bm p}^2}$, where ${\bm p}$ is the kinetic momentum.  Since the threshold energy for a pair production is twice this quantity, $2p^0$, it is minimized when the kinetic momentum gets the smallest and at that instant pairs are preferably produced.

\item After being produced, electron-positron pairs are accelerated by the electromagnetic field according to the classical Lorentz equation ${\rm d}{\bm p}/{\rm d}t = e({\bm E} + {\bm v} \times {\bm B})$.  

\item If pair production occurs at more than one time, the contributions from the different production events can interfere with each other~\cite{Tanji:2008ku, Hebenstreit:2009km}.  This effect is analogous to the St\"uckelberg-phase interference in materials~\cite{stuckelberg1933theorie, Shevchenko:2010ms}.  To illustrate this, consider two production events occurring at times $t_1 < t_2$.  The wavefunction $\Psi_1$ of an electron with energy $p^0$ produced at $t_1$ accumulates a phase $\Psi_1 \propto \exp [ -{\rm i} \int_{t_1}^{t_2} {\rm d}t\, p^0 ]$ by the time $t_2$.  In contrast, the wavefunction $\Psi_2$ corresponding to production at $t_2$ was a vacuum (or a negative-energy electron filling the Dirac sea) during the time interval $[t_1,t_2]$ and hence acquires a different phase to $\Psi_1$.  Thus, at time $t_2$, $\Psi_1$ and $\Psi_2$ interfere with each other, producing characteristic constructive or destructive patterns.  Mathematically, this interference can be understood elegantly in terms of the Stokes phenomenon~\cite{Dumlu:2010ua, Dumlu:2010vv, Dumlu:2011rr, Taya:2020dco}.
\end{itemize}

\subsubsection{Backreaction} \label{sec:2.3.4}

So far, the electromagnetic field has been treated as fixed.  This treatment, however, violates energy conservation once pair production occurs.  Indeed, pairs are produced by drawing energy from the field, and the field should therefore decay as production proceeds.  Such backreaction effects become significant when the number of produced pairs gets large and the energy carried by the particles becomes comparable to the energy stored in the field.

To account for this, the electromagnetic field must be treated dynamically and evolved according to the Maxwell equation (assuming the weak-coupling limit, where radiative corrections can be neglected),
\begin{align}
	\partial_\nu F^{\nu\mu} = J^\mu \;, \label{eq::;22}
\end{align}
where $F^{\nu\mu}$ is the electromagnetic field-strength tensor and $J^\mu$ denotes the charge current generated by the Schwinger effect.  The first issue of the backreaction problem is, therefore, how to compute the current $J^\mu$.  Once it is computed, one then needs to solve the evolution equation for the electron (e.g., the Dirac equation) simultaneously with the Maxwell equation~(\ref{eq::;22}), so as to consistently describe the dynamics of the pair production.  Numerical methods are needed to complete these procedures in general (e.g., realtime lattice technique~\cite{Aarts:1998td, Hebenstreit:2013qxa, Kasper:2014uaa, Gelis:2015kya}), but there exist a few situations where analytical calculations are doable, such as massless QED in a strong magnetic-field limit~\cite{Iwazaki:2009bg}.  

The backreaction problem was first investigated by Matsui and his collaborators in the 1980s~\cite{Glendenning:1983qq, Kajantie:1985jh, Gatoff:1987uf}, motivated by the early-time  dynamics of relativistic heavy-ion collisions (see Sec.~\ref{sec:4.3}).  They pointed out that the current $J^\mu$ consists of two distinct contributions; namely, the conduction and polarization currents (referred to as intra- and inter-band currents, respectively, in solid-state physics~\cite{Ghimire_2014}),
\begin{align}
	J^\mu = J^\mu_{\rm cond} + J^\mu_{\rm pol} \;.
\end{align}
The conduction current $J^\mu_{\rm cond}$ is the ``usual'' current carried by real particles and takes the form ${\bm J}_{\rm cond} = \int \frac{{\rm d}^3{\bm p}}{(2\pi)^3}\,{\bm v}\, f$, where $f$ is the phase-space distribution of the produced particles and ${\bm v} = {\bm p}/p^0$ is velocity.  In addition to real particles, virtual particles can also contribute to the current, in a manner analogous to the dielectric current in materials.  This is the polarization current ${\bm J}_{\rm pol}$ and is induced by the polarization of the vacuum [see Fig.~\ref{fig:1}~(b)].  Without the polarization current, energy conservation is violated, since it costs energy to polarize the vacuum.

Matsui phenomenologically identified the form of the polarization current ${\bm J}_{\rm pol}$ from the energy-conservation argument and analyzed the backreaction problem using a Boltzmann equation [see Eq.~(\ref{eq::35})].  A field-theoretic approach was developed later by Kluger and his collaborators in the 1990s~\cite{Kluger:1991ib, Kluger:1992gb, Kluger:1992md, Cooper:1992hw, Kluger:1998bm}.  They computed the vacuum expectation value of the current operator $\braket{\hat{\bar{\psi}} \gamma^\mu \hat{\psi}}$ by employing the adiabatic regularization scheme based on the adiabatic particle picture to fix ${\bm J}_{\rm pol}$, and then self-consistently solved the Dirac equation.  

As a physical consequence, it was found that the backreaction induces plasma oscillation.  Initially, particles created by the Schwinger effect are accelerated by the electric field, generating a current in the direction of the field.  The electric field, then, begins to decay according to Amp\`ere's law, $\partial_t E = -J$, assuming spatial homogeneity so that the rotation term is dropped.  At a later time, the electric field reaches $E = 0$; however, the current $J$ is still flowing in the positive direction and continues to decrease $E$.  Consequently, the electric field becomes negative and starts to decelerate the particles.  Eventually, the current $J$ changes sign from positive to negative, after which the electric field begins to increase again.  This sequence of processes repeats, leading to oscillatory behavior of both the electric field $E$ and the current $J$ in time, before the electric field is finally damped away.  The oscillatory motion of the produced particles induces quantum interference (see Sec.~\ref{sec:2.3.3}), which distorts the momentum spectrum of the produced particles. 

Inclusion of radiative corrections to the backreaction formalism remains an open issue.  It is essential for the equilibration of the system, and hence is crucially important for the early-time dynamics of relativistic heavy-ion collisions (see Sec.~\ref{sec:4.3}).  Also, the classical treatment of the electromagnetic field is justified only when it is strong, implying that it needs to be treated as incoherent photons and properly taken into account the associated radiative effects when the field strength gets sufficiently weak~\cite{Mueller:2002gd, Jeon:2004dh}.  Another notable idea is the so-called QED cascade, in which the electromagnetic field decays rapidly due to energy dissipation through cascade; namely, successive processes of particle acceleration, pair production, and bremsstrahlung, seeded by the Schwinger effect~\cite{Bell:2008zzb, Fedotov:2010ja, Elkina:2010up}.  This is an analog of the avalanche breakdown in materials.  A promising framework for incorporating such radiative effects is the $n$-particle-irreducible effective action technique~\cite{Berges:2004pu, Berges:2004yj}, which has recently been applied to strong-field QED~\cite{Fauth:2021nwe}.

\section{Schwinger effect in QCD} \label{sec:3}

The idea of the Schwinger effect in QED can be naturally extended to QCD: if a color electromagnetic field becomes sufficiently strong, colored particles such as quarks and gluons tunnel out from the vacuum and the vacuum decays accordingly.  In this section, we review this extension to QCD, focusing on its formal aspects.  Phenomenological applications are discussed in Sec.~\ref{sec:4}.  

It was Batalin,  Matinyan, and Savvidi~\cite{Batalin:1976uv}\footnote{Savvidy used this one-loop result to discuss the vacuum structure of QCD, rather than the Schwinger effect.  He pointed out that the vacuum with non-zero color magnetic field can be energetically favored than the pure vacuum without color fields, which is known as the Savvidy vacuum~\cite{Savvidy:1977as}.  This idea has been investigated in more depth by the Copenhagen group around 1980s, suggesting a picture that the vacuum is a ``spaghetti" of color magnetic tubes (the Copenhagen vacuum)~\cite{etde_6248178, Ambjorn:1980ms}.  } and, independently, Brown, Duff, and Ram\'on-Medrano~\cite{Brown:1975bc, Duff:1975ue}, who first discussed the Yang-Mills (pure gluon) extension of the Heisenberg-Euler Lagrangian~(\ref{eq:1}) in the mid 1970s.  It was followed by Claudson, Cox, and Yildiz to include the quark contribution in 1979~\cite{Yildiz:1979vv, Claudson:1980yz}.  In a diagrammatic language, their calculations amount to evaluating the following one-loop diagrams [cf. compare the one-loop electron diagram in QED~(\ref{eq:3})], 
\begin{align}
	\vcenter{\hbox{
	\begin{tikzpicture}
		\begin{feynhand}
		\pgfmathsetmacro{\radius}{0.65}
		\pgfmathsetmacro{\prop}{0.65}
		\vertex (eq) at (-3.1,0) {${\mathscr L}_{\rm QCD}$};
		\vertex (eq) at (-2.45,0) {$=$};
		\vertex (eq) at (-1.9,-0.1) {$\displaystyle \sum_{{\rm\ all}\;\otimes}$};
		\pgfmathsetmacro{\ang}{0}
		\vertex (v1) at ({\radius*cos(\ang)}, {\radius*sin(\ang)});
		\vertex[crossdot, minimum size=5pt, line width=0.5pt] (v1-) at ({\radius*cos(\ang)+\prop*cos(\ang)}, {\radius*sin(\ang)+\prop*sin(\ang)}) {};
		\propag [gluon] (v1) to (v1-);
		\pgfmathsetmacro{\ang}{45}
		\vertex (v1) at ({\radius*cos(\ang)}, {\radius*sin(\ang)});
		\vertex[crossdot, minimum size=5pt, line width=0.5pt] (v1-) at ({\radius*cos(\ang)+\prop*cos(\ang-15)}, {\radius*sin(\ang)+\prop*sin(\ang-15)}) {};
		\propag [gluon] (v1) to (v1-);
		\vertex[crossdot, minimum size=5pt, line width=0.5pt] (v1-) at ({\radius*cos(\ang)+\prop*cos(\ang+15)}, {\radius*sin(\ang)+\prop*sin(\ang+15)}) {};
		\propag [gluon] (v1) to (v1-);
		\pgfmathsetmacro{\ang}{90}
		\vertex (v1) at ({\radius*cos(\ang)}, {\radius*sin(\ang)});
		\vertex[crossdot, minimum size=5pt, line width=0.5pt] (v1-) at ({\radius*cos(\ang)+\prop*cos(\ang)}, {\radius*sin(\ang)+\prop*sin(\ang)}) {};
		\propag [gluon] (v1) to (v1-);
		\pgfmathsetmacro{\ang}{135}
		\vertex (v1) at ({\radius*cos(\ang)}, {\radius*sin(\ang)});
		\vertex[crossdot, minimum size=5pt, line width=0.5pt] (v1-) at ({\radius*cos(\ang)+\prop*cos(\ang)}, {\radius*sin(\ang)+\prop*sin(\ang)}) {};
		\propag [gluon] (v1) to (v1-);
		\pgfmathsetmacro{\ang}{180}
		\vertex (v1) at ({\radius*cos(\ang)}, {\radius*sin(\ang)});
		\vertex[crossdot, minimum size=5pt, line width=0.5pt] (v1-) at ({\radius*cos(\ang)+\prop*cos(\ang-15)}, {\radius*sin(\ang)+\prop*sin(\ang-15)}) {};
		\propag [gluon] (v1) to (v1-);
		\vertex[crossdot, minimum size=5pt, line width=0.5pt] (v1-) at ({\radius*cos(\ang)+\prop*cos(\ang+15)}, {\radius*sin(\ang)+\prop*sin(\ang+15)}) {};
		\propag [gluon] (v1) to (v1-);
		\pgfmathsetmacro{\ang}{225}
		\vertex (v1) at ({\radius*cos(\ang)}, {\radius*sin(\ang)});
		\vertex[crossdot, minimum size=5pt, line width=0.5pt] (v1-) at ({\radius*cos(\ang)+\prop*cos(\ang)}, {\radius*sin(\ang)+\prop*sin(\ang)}) {};
		\propag [gluon] (v1) to (v1-);
		\pgfmathsetmacro{\ang}{255}
		\vertex[dot, minimum size=1.2pt] (v2) at ({\radius*cos(\ang)+\prop*cos(\ang)}, {\radius*sin(\ang)+\prop*sin(\ang)}) {};
		\pgfmathsetmacro{\ang}{270}
		\vertex[dot, minimum size=1.2pt] (v2) at ({\radius*cos(\ang)+\prop*cos(\ang)}, {\radius*sin(\ang)+\prop*sin(\ang)}) {};
		\pgfmathsetmacro{\ang}{285}
		\vertex[dot, minimum size=1.2pt] (v2) at ({\radius*cos(\ang)+\prop*cos(\ang)}, {\radius*sin(\ang)+\prop*sin(\ang)}) {};
		\pgfmathsetmacro{\ang}{300}
		\vertex[dot, minimum size=1.2pt] (v2) at ({\radius*cos(\ang)+\prop*cos(\ang)}, {\radius*sin(\ang)+\prop*sin(\ang)}) {};
		\pgfmathsetmacro{\ang}{315}
		\vertex[dot, minimum size=1.2pt] (v2) at ({\radius*cos(\ang)+\prop*cos(\ang)}, {\radius*sin(\ang)+\prop*sin(\ang)}) {};
		\pgfmathsetmacro{\ang}{330}
		\vertex[dot, minimum size=1.2pt] (v2) at ({\radius*cos(\ang)+\prop*cos(\ang)}, {\radius*sin(\ang)+\prop*sin(\ang)}) {};
		\vertex (vi) at ({\radius*cos(0)}, {\radius*sin(0)});
		\vertex (vf) at ({\radius*cos(180)}, {\radius*sin(180)});
		\propag [gluon] (vi) to [half right, looseness=1.68] (vf);
		\propag [gluon] (vf) to [half right, looseness=1.68] (vi);
		\vertex (eq) at (2.0,0) {$+$};
		\vertex (eq) at (2.65,-0.1) {$\displaystyle \sum_{{\rm\ all}\;\otimes}$};
		\pgfmathsetmacro{\offs}{4.5}
		\pgfmathsetmacro{\ang}{0}
		\vertex (v1) at ({\offs+\radius*cos(\ang)}, {\radius*sin(\ang)});
		\vertex[crossdot, minimum size=5pt, line width=0.5pt] (v1-) at ({\offs+\radius*cos(\ang)+\prop*cos(\ang)}, {\radius*sin(\ang)+\prop*sin(\ang)}) {};
		\propag [gluon] (v1) to (v1-);
		\pgfmathsetmacro{\ang}{45}
		\vertex (v1) at ({\offs+\radius*cos(\ang)}, {\radius*sin(\ang)});
		\vertex[crossdot, minimum size=5pt, line width=0.5pt] (v1-) at ({\offs+\radius*cos(\ang)+\prop*cos(\ang-15)}, {\radius*sin(\ang)+\prop*sin(\ang-15)}) {};
		\propag [gluon] (v1) to (v1-);
		\vertex[crossdot, minimum size=5pt, line width=0.5pt] (v1-) at ({\offs+\radius*cos(\ang)+\prop*cos(\ang+15)}, {\radius*sin(\ang)+\prop*sin(\ang+15)}) {};
		\propag [gluon] (v1) to (v1-);
		\pgfmathsetmacro{\ang}{90}
		\vertex (v1) at ({\offs+\radius*cos(\ang)}, {\radius*sin(\ang)});
		\vertex[crossdot, minimum size=5pt, line width=0.5pt] (v1-) at ({\offs+\radius*cos(\ang)+\prop*cos(\ang)}, {\radius*sin(\ang)+\prop*sin(\ang)}) {};
		\propag [gluon] (v1) to (v1-);
		\pgfmathsetmacro{\ang}{135}
		\vertex (v1) at ({\offs+\radius*cos(\ang)}, {\radius*sin(\ang)});
		\vertex[crossdot, minimum size=5pt, line width=0.5pt] (v1-) at ({\offs+\radius*cos(\ang)+\prop*cos(\ang)}, {\radius*sin(\ang)+\prop*sin(\ang)}) {};
		\propag [gluon] (v1) to (v1-);
		\pgfmathsetmacro{\ang}{180}
		\vertex (v1) at ({\offs+\radius*cos(\ang)}, {\radius*sin(\ang)});
		\vertex[crossdot, minimum size=5pt, line width=0.5pt] (v1-) at ({\offs+\radius*cos(\ang)+\prop*cos(\ang-15)}, {\radius*sin(\ang)+\prop*sin(\ang-15)}) {};
		\propag [gluon] (v1) to (v1-);
		\vertex[crossdot, minimum size=5pt, line width=0.5pt] (v1-) at ({\offs+\radius*cos(\ang)+\prop*cos(\ang+15)}, {\radius*sin(\ang)+\prop*sin(\ang+15)}) {};
		\propag [gluon] (v1) to (v1-);
		\pgfmathsetmacro{\ang}{225}
		\vertex (v1) at ({\offs+\radius*cos(\ang)}, {\radius*sin(\ang)});
		\vertex[crossdot, minimum size=5pt, line width=0.5pt] (v1-) at ({\offs+\radius*cos(\ang)+\prop*cos(\ang)}, {\radius*sin(\ang)+\prop*sin(\ang)}) {};
		\propag [gluon] (v1) to (v1-);
		\pgfmathsetmacro{\ang}{255}
		\vertex[dot, minimum size=1.2pt] (v2) at ({\offs+\radius*cos(\ang)+\prop*cos(\ang)}, {\radius*sin(\ang)+\prop*sin(\ang)}) {};
		\pgfmathsetmacro{\ang}{270}
		\vertex[dot, minimum size=1.2pt] (v2) at ({\offs+\radius*cos(\ang)+\prop*cos(\ang)}, {\radius*sin(\ang)+\prop*sin(\ang)}) {};
		\pgfmathsetmacro{\ang}{285}
		\vertex[dot, minimum size=1.2pt] (v2) at ({\offs+\radius*cos(\ang)+\prop*cos(\ang)}, {\radius*sin(\ang)+\prop*sin(\ang)}) {};
		\pgfmathsetmacro{\ang}{300}
		\vertex[dot, minimum size=1.2pt] (v2) at ({\offs+\radius*cos(\ang)+\prop*cos(\ang)}, {\radius*sin(\ang)+\prop*sin(\ang)}) {};
		\pgfmathsetmacro{\ang}{315}
		\vertex[dot, minimum size=1.2pt] (v2) at ({\offs+\radius*cos(\ang)+\prop*cos(\ang)}, {\radius*sin(\ang)+\prop*sin(\ang)}) {};
		\pgfmathsetmacro{\ang}{330}
		\vertex[dot, minimum size=1.2pt] (v2) at ({\offs+\radius*cos(\ang)+\prop*cos(\ang)}, {\radius*sin(\ang)+\prop*sin(\ang)}) {};
		\vertex (vi) at ({\offs+\radius*cos(0)}, {\radius*sin(0)});
		\vertex (vf) at ({\offs+\radius*cos(180)}, {\radius*sin(180)});
		\propag [ghost] (vi) to [half right, looseness=1.68] (vf);
		\propag [ghost] (vf) to [half right, looseness=1.68] (vi);
		\vertex (eq) at (6.5,0) {$+$};
		\vertex (eq) at (7.15,-0.1) {$\displaystyle \sum_{{\rm\ all}\;\otimes}$};
		\pgfmathsetmacro{\offs}{9}
		\pgfmathsetmacro{\ang}{0}
		\vertex (v1) at ({\offs+\radius*cos(\ang)}, {\radius*sin(\ang)});
		\vertex[crossdot, minimum size=5pt, line width=0.5pt] (v1-) at ({\offs+\radius*cos(\ang)+\prop*cos(\ang)}, {\radius*sin(\ang)+\prop*sin(\ang)}) {};
		\propag [gluon] (v1) to (v1-);
		\pgfmathsetmacro{\ang}{45}
		\vertex (v1) at ({\offs+\radius*cos(\ang)}, {\radius*sin(\ang)});
		\vertex[crossdot, minimum size=5pt, line width=0.5pt] (v1-) at ({\offs+\radius*cos(\ang)+\prop*cos(\ang)}, {\radius*sin(\ang)+\prop*sin(\ang)}) {};
		\propag [gluon] (v1) to (v1-);
		\pgfmathsetmacro{\ang}{90}
		\vertex (v1) at ({\offs+\radius*cos(\ang)}, {\radius*sin(\ang)});
		\vertex[crossdot, minimum size=5pt, line width=0.5pt] (v1-) at ({\offs+\radius*cos(\ang)+\prop*cos(\ang)}, {\radius*sin(\ang)+\prop*sin(\ang)}) {};
		\propag [gluon] (v1) to (v1-);
		\pgfmathsetmacro{\ang}{135}
		\vertex (v1) at ({\offs+\radius*cos(\ang)}, {\radius*sin(\ang)});
		\vertex[crossdot, minimum size=5pt, line width=0.5pt] (v1-) at ({\offs+\radius*cos(\ang)+\prop*cos(\ang)}, {\radius*sin(\ang)+\prop*sin(\ang)}) {};
		\propag [gluon] (v1) to (v1-);
		\pgfmathsetmacro{\ang}{180}
		\vertex (v1) at ({\offs+\radius*cos(\ang)}, {\radius*sin(\ang)});
		\vertex[crossdot, minimum size=5pt, line width=0.5pt] (v1-) at ({\offs+\radius*cos(\ang)+\prop*cos(\ang)}, {\radius*sin(\ang)+\prop*sin(\ang)}) {};
		\propag [gluon] (v1) to (v1-);
		\pgfmathsetmacro{\ang}{225}
		\vertex (v1) at ({\offs+\radius*cos(\ang)}, {\radius*sin(\ang)});
		\vertex[crossdot, minimum size=5pt, line width=0.5pt] (v1-) at ({\offs+\radius*cos(\ang)+\prop*cos(\ang)}, {\radius*sin(\ang)+\prop*sin(\ang)}) {};
		\propag [gluon] (v1) to (v1-);
		\pgfmathsetmacro{\ang}{255}
		\vertex[dot, minimum size=1.2pt] (v2) at ({\offs+\radius*cos(\ang)+\prop*cos(\ang)}, {\radius*sin(\ang)+\prop*sin(\ang)}) {};
		\pgfmathsetmacro{\ang}{270}
		\vertex[dot, minimum size=1.2pt] (v2) at ({\offs+\radius*cos(\ang)+\prop*cos(\ang)}, {\radius*sin(\ang)+\prop*sin(\ang)}) {};
		\pgfmathsetmacro{\ang}{285}
		\vertex[dot, minimum size=1.2pt] (v2) at ({\offs+\radius*cos(\ang)+\prop*cos(\ang)}, {\radius*sin(\ang)+\prop*sin(\ang)}) {};
		\pgfmathsetmacro{\ang}{300}
		\vertex[dot, minimum size=1.2pt] (v2) at ({\offs+\radius*cos(\ang)+\prop*cos(\ang)}, {\radius*sin(\ang)+\prop*sin(\ang)}) {};
		\pgfmathsetmacro{\ang}{315}
		\vertex[dot, minimum size=1.2pt] (v2) at ({\offs+\radius*cos(\ang)+\prop*cos(\ang)}, {\radius*sin(\ang)+\prop*sin(\ang)}) {};
		\pgfmathsetmacro{\ang}{330}
		\vertex[dot, minimum size=1.2pt] (v2) at ({\offs+\radius*cos(\ang)+\prop*cos(\ang)}, {\radius*sin(\ang)+\prop*sin(\ang)}) {};
		\vertex (vi) at ({\offs+\radius*cos(0)}, {\radius*sin(0)});
		\vertex (vf) at ({\offs+\radius*cos(180)}, {\radius*sin(180)});
		\propag [plain] (vi) to [half right, looseness=1.68] (vf);
		\propag [plain] (vf) to [half right, looseness=1.68] (vi);
		\end{feynhand}
	\end{tikzpicture}
	}} \ \ \ . \label{eq:-3}
\end{align}
The first and second diagrams represent the contributions from gluons ($\inlinegluon$) and ghosts ($\inlineghost$), respectively~\cite{Batalin:1976uv, Duff:1975ue, Yildiz:1979vv}, and the third from quarks (\hspace*{0.15mm}\rule[0.5ex]{1.1em}{0.8pt}\hspace*{0.15mm})~\cite{Claudson:1980yz}.  The blobs ($\sim\hspace*{-1.2mm}\otimes$) denote interactions with the background color electromagnetic field.  Unlike QED and the quark contribution, gluons and ghosts can interact with the background field not only once but also twice at a single vertex, reflecting the non-linear nature of the Yang-Mills theory (i.e., gluon is self-interacting). For example, the Yang-Mills theory contains a quartic interaction term, schematically $\propto \hat{{\mathcal A}}^4$, with $\hat{{\mathcal A}}$ being the gauge-field operator.  Upon decomposing the total gauge field $\hat{{\mathcal A}}$ into a classical component ${\mathcal A} := \braket{\hat{{\mathcal A}}}$ and quantum gluon fluctuations around it $\hat{a} := \hat{{\mathcal A}} - {\mathcal A}$ (cf. footnote~\ref{foot:2}), this term generates contributions quadratic in the classical field of the form $\propto \hat{a}^2 {\mathcal A}^2$, which is responsible for the double-interaction vertices in the diagram~(\ref{eq:-3}).  Note that a background electromagnetic field can be added straightforwardly to the quark-loop diagram~\cite{Ozaki:2015yja}, while such an electromagnetic coupling does not appear for gluons and ghosts at the one-loop order, since they are electrically neutral (they can be affected at higher orders through coupling to quark loops).  

They considered the so-called {\it covariantly-constant field} as a color background such that~\cite{Batalin:1976uv}
\begin{align}
	0 = {\mathcal D}_\lambda {\mathcal F}_{\mu\nu} := \partial_\lambda {\mathcal F}_{\mu\nu} + {\rm i}g [{\mathcal A}_\lambda , {\mathcal F}_{\mu\nu} ] \ \ {\rm for\ any}\ \lambda,\mu,\nu \;, \label{eq:29}
\end{align}
where $[A,B] := AB-BA$ denotes the commutator and ${\mathcal F}_{\mu\nu} := \partial_\mu {\mathcal A}_\nu - \partial_\nu {\mathcal A}_\mu + {\rm i}g[{\mathcal A}_\mu, {\mathcal A}_\nu]$ is the non-Abelian field-strength tensor.  This is a solution to the sourceless Yang-Mills equation ${\mathcal D}_\mu {\mathcal F}^{\mu\nu}=0$ and is a non-Abelian, gauge-covariant extension of a constant electromagnetic field in QED, $\partial_\lambda F_{\mu\nu}=0$.  A crucial property of this special background is that it is essentially Abelian.  As a result, the non-Abelian calculation can be reduced to an Abelian one, allowing the direct application of the QED knowledge reviewed in Sec.~\ref{sec:2}.  In fact, it can be shown that the general solution to the covariantly-constant-field condition~(\ref{eq:29}) takes the form (up to an unimportant gauge freedom)~\cite{Batalin:1976uv, Hattori:2023egw},
\begin{align}
	{\mathcal F}_{\mu\nu} = \bar{F}_{\mu\nu} n^a T^a \;, \label{eq:30}
\end{align}
where $n^a$ is a constant unit vector $n^a n^a=1$, specifying the color orientation of the field, and $\bar{F}_{\mu\nu}$ is an ``Abelian constant field-strength tensor" that satisfies the condition~$\partial_\lambda \bar{F}_{\mu\nu}=0$.  The color-matrix structure is completely factored out into $n^a T^a$.  Since $n^a T^a$ is a constant hermitian matrix in color space, there exists a global gauge transformation that diagonalizes it.  Consequently, a covariantly-constant field ${\mathcal F}_{\mu\nu}$ can always be brought to a diagonal form, effectively reducing the problem to that of an Abelian constant field $\bar{F}_{\mu\nu}$.  Note that a similar but a bit more general field configuration than the covariantly-constant field~(\ref{eq:29}) is a color field with fixed color orientation, ${\mathcal A}^\mu(x) = \bar{A}^\mu(x) n^a T^a$, where $\bar{A}^\mu(x)$ is an arbitrary scalar function (i.e., the Abelian part of the Cho-Faddeev-Niemi decomposition~\cite{Faddeev:1998eq, Cho:1979nv}).  The corresponding field-strength tensor ${\mathcal F}_{\mu\nu}$ takes the same form as Eq.~(\ref{eq:30}), and hence is Abelianizable, but the corresponding $\bar{F}_{\mu\nu}=\partial_\mu\bar{A}_\nu-\partial_\nu\bar{A}_\mu$ can be spacetime-dependent.  Such a field configuration is useful to analyze the effects of spacetime-dependence and realtime dynamics with backreaction~\cite{Tanji:2011di, Taya:2016ovo, Taya:2017pdp}.  

The resulting effective Lagrangian ${\mathscr L}_{\rm QCD}$~(\ref{eq:-3}) acquires an imaginary part, once a color electric field is turned on.  This imaginary part signals the Schwinger effect in QCD; namely, pairs of quarks and gluons are produced from the vacuum.  Note that the ghost loop also develops an imaginary part; however, this contribution exactly cancels that arising from the two unphysical gluon polarization modes.  As a result, only the $4-2=2$ physical gluons contribute to the spectrum, yielding no observable effect from the unphysical ghost and gluon modes~\cite{Ambjorn:1982bp}.  The corresponding decay rate of the QCD vacuum was derived explicitly in the late 1970s, providing a generalization of Schwinger's formula in QED~(\ref{eq:6})~\cite{Batalin:1976uv, Yildiz:1979vv, Claudson:1980yz}.  Likewise, Nikishov's formula for the momentum spectrum~(\ref{eq:7}) was extended to QCD by Ambjørn and Hughes in 1982 using the Bogoliubov-transformation technique~\cite{Ambjorn:1982bp}.  

To be concrete, let us consider QCD with $N_{\rm c}$ colors and $N_{\rm f}$ quark flavors in the presence of a covariantly-constant color electric field ${\bm{\mathcal E}} = \mathcal{E}^a T^a\,{\bf e}_\parallel$.  Here, $T^a$'s ($a = 1,2,\ldots, N_{\rm c}^2-1$) denote the generators of the color group SU($N_{\rm c}$) in the fundamental representation, normalized as ${\rm tr}\,T^aT^b = \delta^{ab}/2$.  The QCD generalizations of the Schwinger and Nikishov formulas for the
vacuum decay rate $w$ and the pair-production rate
$\Gamma$ then read
\begin{align}
	&\Gamma = \underbrace{ \frac{1}{T} \sum_{f} \sum_{i} \sum_{\rm spin} \int \frac{{\rm d}^3{\bm p}}{(2\pi)^3} f_{\rm quark} }_{=\,\Gamma_{\rm quark}} + \underbrace{ \frac{1}{T} \sum_{\alpha} \sum_{\rm pol.} \int \frac{{\rm d}^3{\bm p}}{(2\pi)^3} f_{\rm gluon} }_{=\,\Gamma_{\rm gluon}}
	= 2 \sum_{f} \sum_{i} \frac{(g\omega_i{\mathcal E})^2}{(2\pi)^3} \exp\left[ -\pi\frac{m_f^2}{|g\omega_i{\mathcal E}|} \right] + \frac{g^2 N_{\rm c}}{(2\pi)^3} \;, \label{eq:---10} \\
	&w = \underbrace{ \frac{1}{T} \sum_{f} \sum_{i} \sum_{\rm spin} \sum_{k=1}^\infty \int \frac{{\rm d}^3{\bm p}}{(2\pi)^3} \frac{f_{\rm quark}^{k}}{k} }_{=\,w_{\rm quark}} + \underbrace{ \frac{1}{T} \sum_{\alpha} \sum_{\rm pol.} \sum_{k=1}^\infty \int \frac{{\rm d}^3{\bm p}}{(2\pi)^3} (-1)^{k-1} \frac{f_{\rm gluon}^{k}}{k} 	}_{=\,w_{\rm gluon}}
	= 2 \sum_{f} \sum_{i} \frac{(g\omega_i{\mathcal E})^2}{(2\pi)^3} \sum_{k=1}^{\infty} \frac{1}{k^2} \exp\left[ -k\pi\frac{m_f^2}{|g\omega_i{\mathcal E}|} \right]  +   \frac{g^2 N_{\rm c}}{96\pi} \;, \nonumber 
\end{align}
where ${\mathcal E} := \sqrt{{\mathcal E}^a {\mathcal E}^a} = \sqrt{ 2\, {\rm tr}_{\rm c} {\bm {\mathcal E}} \cdot {\bm {\mathcal E}} }$ represents the strength of the color electric field.  The sum $\sum_f$ with $f=1,2,\ldots,N_{\rm f}$ runs over quark flavors, while $\sum_i$ and $\sum_\alpha$ with $i=1,2,\ldots,N_{\rm c}$ and $\alpha=1,2,\ldots,N_{\rm c}(N_{\rm c}-1)/2$\footnote{The total number of $\alpha$ is $N_{\rm c}(N_{\rm c}-1)/2$, which appears smaller than the total number of gluon colors, $N_{\rm c}^2-1$.  This is because $N_{\rm c}-1$ gluons are color neutral, i.e., they belong to the Cartan subalgebra of SU($N_{\rm c}$).  The remaining $(N_{\rm c}^2-1)-(N_{\rm c}-1)=N_{\rm c}(N_{\rm c}-1)$ gluons are colored and can be grouped into $N_{\rm c}(N_{\rm c}-1)/2$ pairs carrying opposite charges $(+\alpha,-\alpha)$.  These pairs of gluons are produced in the QCD Schwinger effect.  Mathematically, these can be formulated using the Cartan-Weyl basis of SU($N_{\rm c}$)~\cite{Gyulassy:1985oqt, Tanji:2011di, Taya:2016ovo, Taya:2017pdp}.}  account for the color degrees of freedom of quarks and gluons, respectively.  The sums $\sum_{\rm spin}$ and $\sum_{\rm pol.}$ run over the quark spin and the physical gluon polarization states, respectively.  The phase-space distribution functions of the produced quarks, $f_{\rm quark}$, and gluons, $f_{\rm gluon}$, take the same form as the electron distribution function $f$ in the QED case~(\ref{eq:7}), upon appropriate replacement of the
parameters as
\begin{align}
	f_{\rm quark} = f|_{m \to m_{f},\, eE \to \omega_i g {\mathcal E}} = \exp\left[ - \pi \frac{m_f^2 + {\bm p}_\perp^2}{|\omega_i g {\mathcal E}|} \right]
	\ \ \ {\rm and}\ \ \ 
	f_{\rm gluon} = f|_{m \to 0,\, eE \to \omega_\alpha g {\mathcal E}} = \exp\left[ - \pi \frac{{\bm p}_\perp^2}{|\omega_\alpha g {\mathcal E}|} \right] \;, \label{eq::26}
\end{align}
where $m_f$ is the quark current mass and $g$ denotes the QCD coupling constant.  Instead of the bare quark and gluon masses, one may also use effective masses, such as the constituent quark mass and infrared gluon mass, in the replacement~(\ref{eq::26}), which is sometimes used in phenomenological applications of the Schwinger effect in QCD, e.g., the flux-tube model to reproduce actual hadron multiplicities and strangeness suppression~\cite{Casher:1979gw}.  The weight factors, $\omega_i$ for quarks and $\omega_\alpha$ for gluons, control the effective coupling strengths to the color field.  These are needed because the color field has an orientation in the color space [i.e., $n^a$ in Eq.~(\ref{eq:30})] and hence each color of quarks and gluons couples differently to the field.  The concrete values of $\omega_i$ and $\omega_\alpha$ are fixed by the algebra of the color group SU($N_{\rm c}$)~\cite{Gyulassy:1985oqt}:  
\begin{align}
	\omega_i = w^\ell (H^\ell)_{ii}
	\ \ {\rm and}\ \ 
	\omega_\alpha = w^\ell v^\ell_{\ \alpha} \;, \label{eqdsq:29}
\end{align}
where $H^\ell$ ($\ell=1,2,\cdots,N_{\rm c}-1$) are the elements of the Cartan subalgebra (i.e., the $N_{\rm c}-1$ diagonal generators out of $T^a$ and the subscript $ii$ indicates the $i$-th diagonal component) and $v^\ell_{\ \alpha}$ are the root vectors, which are the solutions to the eigenvalue equation $[H^\ell, E_{\pm \alpha}] = \pm v^\ell_{\ \alpha} E_{\pm \alpha}$.  The remaining $w^\ell$ has the information of the color orientation and are obtained by diagonalizing the covariantly-constant field~(\ref{eq:30}) as $U^\dagger {\mathcal F}_{\mu\nu} U = \bar{F}_{\mu\nu} U^\dagger n^a T^a U = \bar{F}_{\mu\nu} w^\ell H^\ell $, with $U$ being the diagonalization matrix.  In the realistic $N_{\rm c}=3$, the weight factors~(\ref{eqdsq:29}) can be expressed explicitly in terms of the Casimir gauge invariants for the color field~\cite{Nayak:2005yv, Nayak:2005pf} (see also Refs.~\cite{Tanji:2010eu, Hattori:2023egw}).

We emphasize that the QCD formulas~(\ref{eq:---10}) take essentially the same form as their QED counterparts~(\ref{eq:10}).  This is intuitively reasonable, since the underlying physics mechanism of the Schwinger effect (see Fig.~\ref{fig:1}) should be independent of the origin of the energy source (i.e., color vs electromagnetic fields) and of the species of virtual particles involved (i.e., quarks and gluons vs electrons).  From a theoretical viewpoint, however, it is crucial that this coincidence relies on the use of the covariantly-constant background field~(\ref{eq:29}).  We remind that such fields constitute a special class of non-Abelian fields that can be reduced to an Abelian form by an appropriate gauge transformation~(\ref{eq:30}).  As a consequence, the motion of colored particles in these backgrounds becomes essentially identical to that of electrons in QED (e.g., linear increase of the momentum in a uniform electric field), so does the work performed by the background field.  By contrast, there exist genuinely non-Abelian background fields that cannot be transformed into an Abelian form.  In such cases, the dynamics of colored particles can differ drastically from those in QED (e.g., precession in color space), and thus the Schwinger effect in QCD can be modified.  An early investigation along these lines was carried out by Brown and Weisberger~\cite{Brown:1979bv} for a constant (not covariantly-constant) color electromagnetic field.  It still remains an open problem to derive general formulas for the QCD Schwinger effect beyond covariantly-constant backgrounds, although numerical methods are available (e.g., quark production from glasma, which is not necessarily covariantly-constant; see Sec.~\ref{sec:4.3}).  

Finally, we mention the effects of the addition of a covariantly-constant color magnetic field~\cite{Tanji:2011di, Hattori:2020guh}.  Similarly to QED, the role of a color magnetic field is to modify the dispersion via the Landau quantization~(\ref{eq--14}) and thus
\begin{align}
	f_{\rm quark} \to \exp\left[ - \pi \frac{m_f^2 + (2n +1)|w_i g {\mathcal B}| - 2s_{\parallel} w_i g {\mathcal B}}{|w_i g {\mathcal E}|} \right] \ \ {\rm and}\ \ 
	f_{\rm gluon} \to \exp\left[ - \pi \frac{(2n +1)|w_\alpha g {\mathcal B}| - 2s_{\parallel} w_\alpha g {\mathcal B}}{|w_\alpha g {\mathcal E}|} \right] \;,
\end{align}
where we assumed the gyromagnetic ratio ${\mathfrak g}=2$ and the color magnetic field, ${\bm {\mathcal B}}$, is aligned in the same color- and polarization-directions to the electric field for simplicity, and defined ${\mathcal B} := \sqrt{{\mathcal B}^a {\mathcal B}^a}$.  What is remarkable is the appearance of the positive exponent in $f_{\rm gluon}$ at the lowest-lying mode, $n=0$ and $s_\parallel = {\rm sgn}(w_\alpha g {\mathcal B})$, for which $f_{\rm gluon} = {\rm e}^{+\pi |{\mathcal B}/{\mathcal E}|}$.  This can be understood as a reminiscent of the Nielsen-Olensen instability~\cite{Nielsen:1978rm}, which is caused by that the lowest-lying mode becomes tachyonic, $\epsilon^2 \sim -|w_\alpha g {\mathcal B}| < 0$, and hence grows up exponentially.  Thus, the Schwinger effect could be enhanced significantly in the presence of a color magnetic field.  This is a peculiar result of QCD, which contains massless, charged vector mode.  Meanwhile, we remind that the result so far is at the one-loop level.  Higher-order effects must be taken into account once the unstable fluctuations have grown up, inclusion of which is an open problem in the QCD Schwinger effect.

\section{Application to nuclear physics} \label{sec:4}

The Schwinger effect has a broad range of phenomenological applications, particularly in nuclear physics.  Here, we briefly highlight several examples.  For other applications, we refer the reader to Refs.~\cite{Ruffini:2009hg, Fedotov:2022ely, Hattori:2023egw}.  

\subsection{High $Z$ nucleus}

\newcommand{\coulombene}[2]{
	2*cos((#1/#2)*90)-1
}
\begin{figure}[t]
\centering
\begin{tikzpicture}[scale=1, samples=100]
	\pgfmathsetmacro{\myenergy}{\coulombene{210}{173}}
	\draw[-, green] (0,\myenergy) -- ({min(1/(1-\myenergy),4.7)},\myenergy);
	\pgfmathsetmacro{\myenergy}{\coulombene{210}{185}}
	\draw[-, green] (0,\myenergy) -- ({min(1/(1-\myenergy),4.7)},\myenergy);
	\pgfmathsetmacro{\myenergy}{\coulombene{210}{192}}
	\draw[-, green] (0,\myenergy) -- ({min(1/(1-\myenergy),4.7)},\myenergy);
	\pgfmathsetmacro{\myenergy}{\coulombene{210}{205}}
	\draw[-, green] (0,\myenergy) -- ({min(1/(1-\myenergy),4.7)},\myenergy);
	\pgfmathsetmacro{\myenergy}{\coulombene{210}{211}}
	\draw[-, green] (0,\myenergy) -- ({min(1/(1-\myenergy),4.7)},\myenergy);
	\pgfmathsetmacro{\myenergy}{\coulombene{210}{228}}
	\draw[-, green] (0,\myenergy) -- ({min(1/(1-\myenergy),4.7)},\myenergy);
	\pgfmathsetmacro{\myenergy}{\coulombene{210}{234}}
	\draw[-, green] (0,\myenergy) -- ({min(1/(1-\myenergy),4.7)},\myenergy);
	\pgfmathsetmacro{\myenergy}{\coulombene{210}{259}}
	\draw[-, green] (0,\myenergy) -- ({min(1/(1-\myenergy),4.7)},\myenergy);
	\pgfmathsetmacro{\myenergy}{\coulombene{210}{277}}
	\draw[-, green] (0,\myenergy) -- ({min(1/(1-\myenergy),4.7)},\myenergy);
	\pgfmathsetmacro{\myenergy}{\coulombene{210}{284}}
	\draw[-, green] (0,\myenergy) -- ({min(1/(1-\myenergy),4.7)},\myenergy);
	\pgfmathsetmacro{\myenergy}{\coulombene{210}{302}}
	\draw[-, green] (0,\myenergy) -- ({min(1/(1-\myenergy),4.7)},\myenergy);
	\pgfmathsetmacro{\myenergy}{\coulombene{210}{310}}
	\draw[-, green] (0,\myenergy) -- ({min(1/(1-\myenergy),4.7)},\myenergy);
	\draw (0.6,0.20) node[left, green]{\tiny $\bullet$};
	\draw (0.6,0.38) node[left, green]{\tiny $\bullet$};
	\draw (0.6,0.56) node[left, green]{\tiny $\bullet$};
	\draw[-{Latex}, thick] (0,-3)--(0,1.6) node[right]{$\epsilon$}; 
	\draw[-{Latex}, thick] (0,0)--(5,0) node[right]{$r$}; 
	\draw (0,0) node[left]{0};
	\draw[line width=1.5pt, domain=0.5:4.7] plot(\x,{-1-1/\x});
	\draw[line width=1.5pt, domain=0.25:4.7] plot(\x,{+1-1/\x});
	\draw[-, dotted] (4.7,1) to (0,1) node[left]{$+m$}; 
	\draw[-, dotted] (4.7,-1) to (0,-1) node[left]{$-m$}; 
%
	\draw (1.2,-2.9) node[left, blue, opacity=0.6]{\huge $\bullet$};
	\draw (1.6,-2.5) node[left, blue, opacity=0.6]{\huge $\bullet$};
	\draw (1.6,-2.9) node[left, blue, opacity=0.6]{\huge $\bullet$};
	\draw (2.0,-2.1) node[left, blue, opacity=0.6]{\huge $\bullet$};
	\draw (2.0,-2.5) node[left, blue, opacity=0.6]{\huge $\bullet$};
	\draw (2.0,-2.9) node[left, blue, opacity=0.6]{\huge $\bullet$};
	\draw (2.4,-2.1) node[left, blue, opacity=0.6]{\huge $\bullet$};
	\draw (2.4,-2.5) node[left, blue, opacity=0.6]{\huge $\bullet$};
	\draw (2.4,-2.9) node[left, blue, opacity=0.6]{\huge $\bullet$};
	\draw (2.8-0.04,-1.7+0.02) node[left, blue, opacity=0.6, font=\fontsize{10.2pt}{0}\selectfont]{$\bigcirc$};
	\draw (2.8,-2.1) node[left, blue, opacity=0.6]{\huge $\bullet$};
	\draw (2.8,-2.5) node[left, blue, opacity=0.6]{\huge $\bullet$};
	\draw (2.8,-2.9) node[left, blue, opacity=0.6]{\huge $\bullet$};
	\draw (3.2,-1.7) node[left, blue, opacity=0.6]{\huge $\bullet$};
	\draw (3.2,-2.1) node[left, blue, opacity=0.6]{\huge $\bullet$};
	\draw (3.2,-2.5) node[left, blue, opacity=0.6]{\huge $\bullet$};
	\draw (3.2,-2.9) node[left, blue, opacity=0.6]{\huge $\bullet$};
	\draw (3.6,-1.7) node[left, blue, opacity=0.6]{\huge $\bullet$};
	\draw (3.6,-2.1) node[left, blue, opacity=0.6]{\huge $\bullet$};
	\draw (3.6,-2.5) node[left, blue, opacity=0.6]{\huge $\bullet$};
	\draw (3.6,-2.9) node[left, blue, opacity=0.6]{\huge $\bullet$};
	\draw (4.0,-1.7) node[left, blue, opacity=0.6]{\huge $\bullet$};
	\draw (4.0,-2.1) node[left, blue, opacity=0.6]{\huge $\bullet$};
	\draw (4.0,-2.5) node[left, blue, opacity=0.6]{\huge $\bullet$};
	\draw (4.0,-2.9) node[left, blue, opacity=0.6]{\huge $\bullet$};
	\draw (4.4,-1.7) node[left, blue, opacity=0.6]{\huge $\bullet$};
	\draw (4.4,-2.1) node[left, blue, opacity=0.6]{\huge $\bullet$};
	\draw (4.4,-2.5) node[left, blue, opacity=0.6]{\huge $\bullet$};
	\draw (4.4,-2.9) node[left, blue, opacity=0.6]{\huge $\bullet$};
	\draw (4.8,-1.7) node[left, blue, opacity=0.6]{\huge $\bullet$};
	\draw (4.8,-2.1) node[left, blue, opacity=0.6]{\huge $\bullet$};
	\draw (4.8,-2.5) node[left, blue, opacity=0.6]{\huge $\bullet$};
	\draw (4.8,-2.9) node[left, blue, opacity=0.6]{\huge $\bullet$};
	\draw [decorate,decoration={brace,amplitude=5pt,mirror,raise=4ex}] (0.14,-1.7) -- (0.14,-1.02) node[midway,xshift=5.2em,yshift=0.4em]{\small ``dived" states};
	\draw [decorate,decoration={brace,amplitude=5pt,mirror,raise=4ex}] (4.4,-3.1) -- (4.4,-1.02) node[midway,xshift=5em,yshift=0.0em]{\small Dirac sea};
	\draw [decorate,decoration={brace,amplitude=5pt,mirror,raise=4ex}] (-0.2,1.9) -- (-0.2,-1.70) node[midway,xshift=-7em,yshift=0.0em]{\small positive energy states};
	\fill[white] (-1.0,1.71) rectangle (-0.7,2.00);
	\fill[white] (-1.0,1.63) rectangle (-0.7,1.67);
	\fill[white] (-1.0,1.55) rectangle (-0.7,1.59);
	\fill[white] (-1.0,1.47) rectangle (-0.7,1.51);
	\draw (0.48,-1.7) node[left, red, opacity=0.6]{\huge $\bullet$};
	\draw[-{Stealth}, gray, line width=1.5pt] (2.5,-1.67)--(0.20,-1.67); 
\end{tikzpicture}
\vspace*{3mm}
\caption{
A schematic illustration of the vacuum structure of QED around a high-$Z$ nucleus.  The positive- and negative-energy electron states lie above $\epsilon = +m + V(r)$ (upper thick black curve) and below $\epsilon = -m + V(r)$ (lower thick black curve), respectively, where $V(r) \propto Z$ is the Coulomb potential by the nucleus.  All negative-energy states are initially filled (the Dirac sea).  For sufficiently large $Z$, some of the positive-energy states can ``dive" below $\epsilon = -m$ (indicated as ``dived states").  An electron in the Dirac sea can then tunnel into the positive energy state, leaving a hole behind, i.e., pair production occurs.  }
\label{fig:4.1-1}
\end{figure}
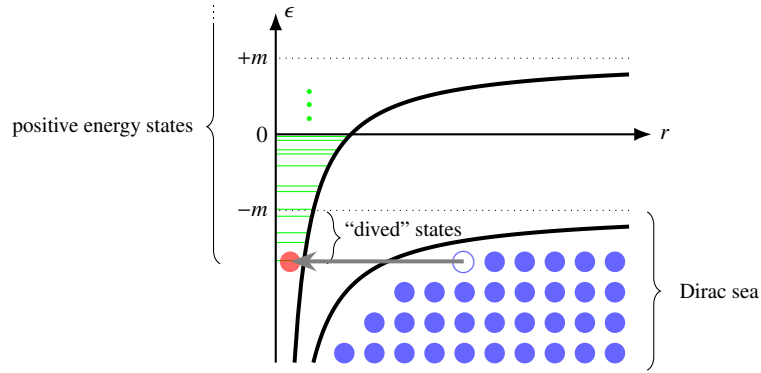

What happens when the Coulomb field of a nucleus becomes extremely strong as the atomic number $Z$ increases?  This question remains one of the oldest in QED, dating back to the seminal works of Gordon~\cite{Gordon1928} and Darwin~\cite{10.1098/rspa.1928.0076} in 1928.  They solved the Dirac equation in a point-like Coulomb potential $V(r) = -Z\alpha/r$, where $\alpha := e^{2}/4\pi \approx 1/137$ is the fine-structure constant of QED.  The corresponding bound-state electron energies are given by (known as the Sommerfeld formula~\cite{https://doi.org/10.1002/andp.19163561702})
\begin{align}
	\epsilon = m\left[ 1 + \left( \frac{Z\alpha}{ n - |\kappa| + \sqrt{\kappa^{2} - (Z\alpha)^{2}} } \right)^{2} \right]^{-1/2} \;, \label{eq-:32}
\end{align}
where $n = 1,2,3,\ldots$ is the principal quantum number and $\kappa = \pm 1, \pm 2, \ldots, \pm n$ is the Dirac angular quantum number that encodes the spin-orbit coupled angular momenta.  A striking feature of Eq.~(\ref{eq-:32}) is that the energy ceases to remain real once $Z\alpha > |\kappa|$.  For the lowest $1S_{1/2}$ state ($n=1$ and $|\kappa|=1$), this occurs when $Z \ge 1/\alpha \approx 137$.  This pathological behavior, often referred to as the {\it $Z=137$ catastrophe}, signals a fundamental departure from standard QED in the high-$Z$ regime.

The $Z=137$ catastrophe can be resolved by taking into account the finite extension of the nuclear
charge distribution, as proposed originally by Pomeranchuk and Smorodinsky in 1945~\cite{Pomeranchuk:1945cjv} and revisited by physicists from Moscow~\cite{1969JETP...30..358G, Popov:1970nz} and Frankfurt~\cite{Pieper:1969tbf} around 1970.  It is reasonable to consider the finiteness, since the Bohr radius $r \sim (Z\alpha m)^{-1}$ decreases with $Z$, and hence the electron will feel unrealistically large Coulomb potential $\propto 1/r$ at the short distances, if the nucleus is point-like.  Such a divergent behavior is regulated by the nucleus finiteness, rendering the energy level real for any $Z$.  Instead of developing an imaginary part, it turned out that the electron energy monotonically decreases with $Z$ and eventually ``dives" into the negative energy continuum $\epsilon \leq -m$ (the Dirac sea) at a certain critical value of $Z_{\rm cr}$ (see Fig.~\ref{fig:4.1-1}).  For the $1S_{1/2}$ state, it is estimated to be $Z_{\rm cr} \approx 173$, while the precise value changes slightly depending on the model, such as the charge-density profile and nucleus radius.  

The diving for $Z > Z_{\rm cr}$ implies that the vacuum becomes unstable and can produce an electron-positron pair, in close analogy with the Schwinger effect.  As illustrated in Fig.~\ref{fig:4.1-1}, the strong Coulomb potential induces a level crossing between a low-lying positive-energy bound state and the negative-energy continuum (the Dirac sea).  An electron originally occupying the Dirac sea can then tunnel into the dived bound state, leaving a hole (i.e., a positron) behind.  In this way, the strong Coulomb field of a high-$Z$ nucleus triggers vacuum pair production.  The produced electron stays at the bound state and screens the nuclear charge from $Ze$ to $(Z-1)e$, while the positron is accelerated outward by the Coulomb field and eventually escapes to infinity, where it can be detected.  Thus, spontaneous positron emission arises as a manifestation of vacuum decay in a supercritical Coulomb field (see Refs.~\cite{Popov2001, Rafelski:2016ixr, Voskresensky:2021okp} for reviews).  Note that a similar Dirac-sea picture can also be applied to other Schwinger-effect studies and often provides helpful intuition; e.g., the dynamically-assisted Schwinger effect~\cite{DiPiazza:2009py} and the Franz-Keldysh effect in strong-field QED~\cite{Taya:2018eng}.

Currently, low-energy heavy-ion collisions provide the only experimental opportunities to investigate the vacuum pair production induced by a strong Coulomb field~\cite{1969JETP...30..358G, Rafelski:1978xy, Reinhardt:1981zz}.  At sufficiently low collision energies, such as those near the Coulomb barrier, the ions may stick together after contact and form a transient quasi-molecular state with an effective atomic number $Z_{\rm eff} = Z_1 + Z_2$.  Although no individual nucleus reaches the critical charge (the heaviest known being Oganesson ${}_{118}{\rm Og}$~\cite{Giuliani:2019xgk}), the combined charge $Z_{\rm eff}$ can exceed it; for example, $Z_{\rm eff}=184$ in uranium-uranium collisions.  Experimental efforts began in 1976 at GSI and continued into the late 1990s, during which several measurements reported non-trivial positron peaks in the energy spectra that could be associated with the vacuum pair production~\cite{Cowan:1985cn}.  However, subsequent experiments reported a number of negative or contradictory results, such as coincident electrons~\cite{Cowan:1986fj}, positron lines in subcritical systems~\cite{Konig:1987qd}, and even the disappearance of the positron peaks~\cite{Ahmad:1999eu}.  To date, no universally accepted experimental evidence for the vacuum pair production in supercritical Coulomb fields has been established.

Continuous efforts are underway, both experimentally and theoretically, to study the vacuum pair production in low-energy heavy-ion collisions.  Upgraded experiments are planned at future facilities such as FAIR, NICA, and HIAF~\cite{GUMBERIDZE2009248, Ter-Akopian:2015sra, MA2017169}, which also attract interest in connection with the exploration of dense QCD matter~\cite{Taya:2024zpv, Ivanov:2026dwf}. 
It is also notable that an analog of the high-$Z$ nucleus has been realized in graphene~\cite{wang2013observing}, where a counterpart of the positron peak was already observed.  On the theory side, progress has been made in describing the dynamical evolution of heavy-ion reactions and the associated positron emission.  Recent numerical simulations, for example, go beyond the traditional ``monopole approximation," on which many earlier calculations rely and where only the spherically symmetric component of the two-center Coulomb potential is considered~\cite{Maltsev:2018nve, V:2023wvg}.

\subsection{String breaking and flux-tube model} \label{sec:4.2}

\setlength{\feynhandarrowsize}{6pt}
\newcommand{\getcrossing}[5]{
	\pgfmathsetmacro{\rx}{#1}
	\pgfmathsetmacro{\ry}{#2}
	\pgfmathsetmacro{\lx}{#3}
	\pgfmathsetmacro{\ly}{#4}
	\vertex (#5) at ({(\rx+\lx+\ry-\ly)/2}, {(\rx-\lx+\ry+\ly)/2});
}
\newcommand{\newcrossingT}[5]{
	\pgfmathsetmacro{\locx}{#1}
	\pgfmathsetmacro{\locy}{#2}
	\pgfmathsetmacro{\ang}{#3}
	\vertex (#5) at ({(\locy-#4*\locx)/(tan(\ang)-#4)},{tan(\ang)*(\locy-#4*\locx)/(tan(\ang)-#4)});
}
\newcommand{\newcrossingB}[6]{
	\pgfmathsetmacro{\locx}{#1}
	\pgfmathsetmacro{\locy}{#2}
	\pgfmathsetmacro{\locX}{#3}
	\pgfmathsetmacro{\locY}{#4}
	\vertex (#6) at ({((\locX+\locx)+(\locY-\locy)/#5)/2}, {((\locY+\locy)+#5*(\locX-\locx))/2});
}
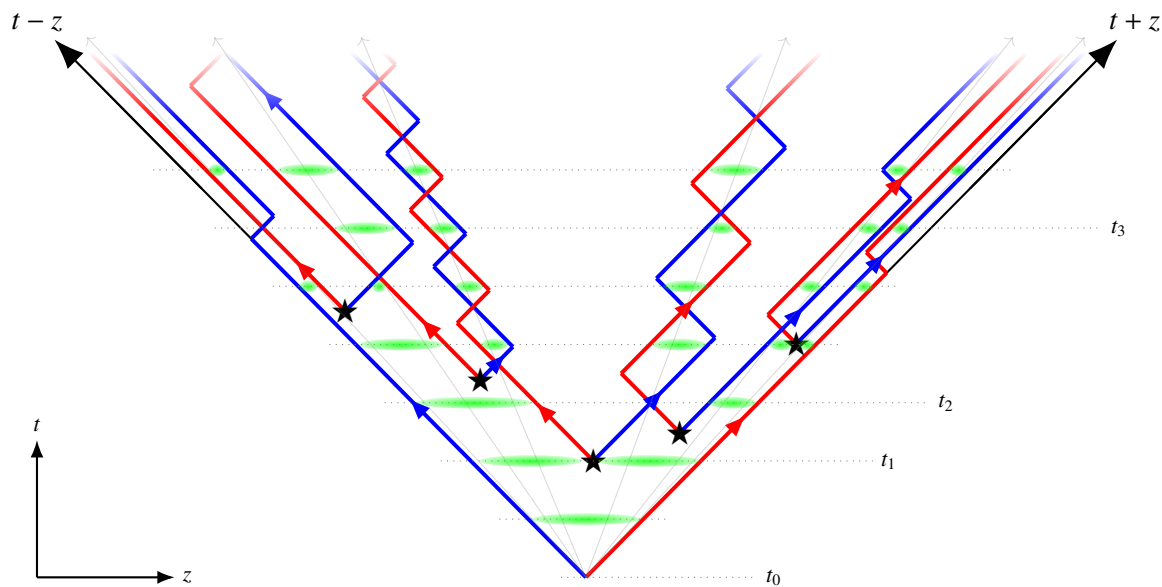
\begin{figure}[t]
\centering
\vspace*{-20mm}
\begin{tikzpicture}
\begin{feynhand}
	\pgfmathsetmacro{\offsetRx}{4};
	\pgfmathsetmacro{\offsetRy}{4};
	\pgfmathsetmacro{\offsetAx}{2.8};
	\pgfmathsetmacro{\offsetAy}{3.07};
	\pgfmathsetmacro{\offsetBx}{1.25};
	\pgfmathsetmacro{\offsetBy}{1.9};
	\pgfmathsetmacro{\offsetCx}{0.1};
	\pgfmathsetmacro{\offsetCy}{1.53};
	\pgfmathsetmacro{\offsetDx}{-1.4};
	\pgfmathsetmacro{\offsetDy}{2.6};
	\pgfmathsetmacro{\offsetEx}{-3.2};
	\pgfmathsetmacro{\offsetEy}{3.5};
	\pgfmathsetmacro{\offsetLx}{-4.45};
	\pgfmathsetmacro{\offsetLy}{4.45};
	\pgfmathsetmacro{\endy}{7.3};
	\pgfmathsetmacro{\endl}{7.1};
	\getcrossing{\offsetRx}{\offsetRy}{\offsetAx}{\offsetAy}{RA};
	\getcrossing{\offsetAx}{\offsetAy}{\offsetBx}{\offsetBy}{AB};
	\getcrossing{\offsetBx}{\offsetBy}{\offsetCx}{\offsetCy}{BC};
	\getcrossing{\offsetCx}{\offsetCy}{\offsetDx}{\offsetDy}{CD};
	\getcrossing{\offsetDx}{\offsetDy}{\offsetEx}{\offsetEy}{DE};
	\getcrossing{\offsetEx}{\offsetEy}{\offsetLx}{\offsetLy}{EL};
	\path (RA); \pgfgetlastxy{\RAx}{\RAy};
	\path (AB); \pgfgetlastxy{\ABx}{\ABy};
	\path (BC); \pgfgetlastxy{\BCx}{\BCy};
	\path (CD); \pgfgetlastxy{\CDx}{\CDy};
	\path (DE); \pgfgetlastxy{\DEx}{\DEy}	;
	\path (EL); \pgfgetlastxy{\ELx}{\ELy}	;
	\pgfmathparse{atan2(\RAy,\RAx)}\edef\formRAtheta{\pgfmathresult};
	\pgfmathparse{atan2(\ABy,\ABx)}\edef\formABtheta{\pgfmathresult};
	\pgfmathparse{atan2(\BCy,\BCx)}\edef\formBCtheta{\pgfmathresult};
	\pgfmathparse{atan2(\CDy,\CDx)}\edef\formCDtheta{\pgfmathresult};
	\pgfmathparse{atan2(\DEy,\DEx)}\edef\formDEtheta{\pgfmathresult};
	\pgfmathparse{atan2(\ELy,\ELx)}\edef\formELtheta{\pgfmathresult};
	\vertex (o) at (0, 0); 
	\vertex (R) at ({\endy}, {\endy}); 
	\vertex (L) at ({-\endy}, {\endy}); 
	\draw[-, black, line width=0.8pt] (o) to (R);
	\draw[-, black, line width=0.8pt] (o) to (L);
  	\vertex (R) at ({\offsetRx}, {\offsetRy}); 
	\draw[-, red, line width=1.5pt] (o) to (R);
	\propag[fer, red] (o) to (R);
	\vertex (RR) at ({2*\RAx-\offsetRx*28.45274pt}, {2*\RAy-\offsetRy*28.45274pt}); \path (RR); \pgfgetlastxy{\RRx}{\RRy};
	\draw[-, red, line width=1.5pt] (R) to (RR);
  	\newcrossingT{\RRx/28.45274}{\RRy/28.45274}{\formRAtheta}{+1}{RRR}; \path (RRR); \pgfgetlastxy{\RRRx}{\RRRy};
	\draw[-, red, line width=1.5pt] (RR) to (RRR);
	\vertex (A) at ({\offsetAx}, {\offsetAy}); 
	\vertex (AA) at ({2*\RAx-\offsetAx*28.45274pt}, {2*\RAy-\offsetAy*28.45274pt}); \path (AA); \pgfgetlastxy{\AAx}{\AAy};
	\draw[-, blue, line width=1.5pt] (A) to (AA);
	\propag[fer, blue] (A) to (AA);
  	\newcrossingB{\AAx/28.45274}{\AAy/28.45274}{\RRRx/28.45274}{\RRRy/28.45274}{+1}{AAA}; \path (AAA); \pgfgetlastxy{\AAAx}{\AAAy};
	\draw[-, blue, line width=1.5pt] (AA) to (AAA);
  	\vertex (A) at ({\offsetAx}, {\offsetAy}); 
	\vertex (AA) at ({2*\ABx-\offsetAx*28.45274pt}, {2*\ABy-\offsetAy*28.45274pt}); \path (AA); \pgfgetlastxy{\AAx}{\AAy};
	\draw[-, red, line width=1.5pt] (A) to (AA);
	\newcrossingT{\AAx/28.45274}{\AAy/28.45274}{\formABtheta}{+1}{AAA}; \path (AAA); \pgfgetlastxy{\AAAx}{\AAAy};
	\vertex (AAAA) at ({2*(\AAAx-\AAx)+\AAx}, {2*(\AAAy-\AAy)+\AAy}); \path (AAAA); \pgfgetlastxy{\AAAAx}{\AAAAy};
%
	\vertex (B) at ({\offsetBx}, {\offsetBy}); 
  	\newcrossingB{\ABx/28.45274}{\ABy/28.45274}{\AAAx/28.45274}{\AAAy/28.45274}{+1}{BB}; \path (BB); \pgfgetlastxy{\BBx}{\BBy};
	\draw[-, blue, line width=1.5pt] (B) to (BB);
	\propag[fer, blue] (B) to (BB);
	\vertex (BBB) at ({2*(\AAAx-\BBx)+\BBx}, {2*(\AAAy-\BBy)+\BBy}); \path (BBB); \pgfgetlastxy{\BBBx}{\BBBy};	
	\draw[-, blue, line width=1.5pt] (BB) to (BBB);
	\vertex (BBBB) at ({\BBBx+2*28.45274pt}, {\BBBy+2*28.45274pt}); \path (BBBB); \pgfgetlastxy{\BBBBx}{\BBBBy};	
	\draw[-, blue, line width=1.5pt] (BBB) to (BBBB);
	\draw[-, red, line width=1.5pt] (AA) to (AAAA);
	\propag[fer, red] (AA) to (AAAA);
  	\vertex (B) at ({\offsetBx}, {\offsetBy}); 
	\vertex (BB) at ({2*\BCx-\offsetBx*28.45274pt}, {2*\BCy-\offsetBy*28.45274pt}); \path (BB); \pgfgetlastxy{\BBx}{\BBy};
	\draw[-, red, line width=1.5pt] (B) to (BB);
	\newcrossingT{\BBx/28.45274}{\BBy/28.45274}{\formBCtheta}{+1}{BBB}; \path (BBB); \pgfgetlastxy{\BBBx}{\BBBy};
	\vertex (BBBB) at ({2*(\BBBx-\BBx)+\BBx}, {2*(\BBBy-\BBy)+\BBy}); \path (BBBB); \pgfgetlastxy{\BBBBx}{\BBBBy};
	\vertex (C) at ({\offsetCx}, {\offsetCy}); 
  	\newcrossingB{\BCx/28.45274}{\BCy/28.45274}{\BBBx/28.45274}{\BBBy/28.45274}{+1}{CC}; \path (CC); \pgfgetlastxy{\CCx}{\CCy};
	\draw[-, blue, line width=1.5pt] (C) to (CC);
	\propag[fer, blue] (C) to (CC);
	\vertex (CCC) at ({2*(\BBBx-\CCx)+\CCx}, {2*(\BBBy-\CCy)+\CCy}); \path (CCC); \pgfgetlastxy{\CCCx}{\CCCy};	
	\draw[-, blue, line width=1.5pt] (CC) to (CCC);
	\draw[-, red, line width=1.5pt] (BB) to (BBBB);
	\propag[fer, red] (BB) to (BBBB);
	\newcrossingT{\CCCx/28.45274}{\CCCy/28.45274}{\formBCtheta}{+1}{CCCC}; \path (CCCC); \pgfgetlastxy{\CCCCx}{\CCCCy};
	\vertex (CCCCC) at ({2*(\CCCCx-\CCCx)+\CCCx}, {2*(\CCCCy-\CCCy)+\CCCy}); \path (CCCCC); \pgfgetlastxy{\CCCCCx}{\CCCCCy};	
  	\newcrossingB{\CCCCx/28.45274}{\CCCCy/28.45274}{\BBBBx/28.45274}{\BBBBy/28.45274}{+1}{BBBBB}; \path (BBBBB); \pgfgetlastxy{\BBBBBx}{\BBBBBy};
	\draw[-, red, line width=1.5pt] (BBBB) to (BBBBB);
	\vertex (BBBBBB) at ({2*(\BBBBBx-\BBBBx)+\BBBBx}, {2*(\BBBBBy-\BBBBy)+\BBBBy}); \path (BBBBBB); \pgfgetlastxy{\BBBBBBx}{\BBBBBBy};	
	\draw[-, red, line width=1.5pt] (BBBBB) to (BBBBBB);
	\draw[-, blue, line width=1.5pt] (CCC) to (CCCCC);
	\newcrossingT{\BBBBBBx/28.45274}{\BBBBBBy/28.45274}{\formBCtheta}{+1}{BBBBBBB}; \path (BBBBBBB); \pgfgetlastxy{\BBBBBBBx}{\BBBBBBBy};
	\vertex (BBBBBBBB) at ({2*(\BBBBBBBx-\BBBBBBx)+\BBBBBBx}, {2*(\BBBBBBBy-\BBBBBBy)+\BBBBBBy}); \path (BBBBBBBB); \pgfgetlastxy{\BBBBBBBBx}{\BBBBBBBBy};
  	\newcrossingB{\CCCCCx/28.45274}{\CCCCCy/28.45274}{\BBBBBBBx/28.45274}{\BBBBBBBy/28.45274}{+1}{CCCCCC}; \path (CCCCCC); \pgfgetlastxy{\CCCCCCx}{\CCCCCCy};
	\vertex (CCCCCCC) at ({2*(\BBBBBBBx-\CCCCCCx)+\CCCCCCx}, {2*(\BBBBBBBy-\CCCCCCy)+\CCCCCCy}); \path (CCCCCCC); \pgfgetlastxy{\CCCCCCCx}{\CCCCCCCy};	
	\draw[-, blue, line width=1.5pt] (CCCCC) to (CCCCCCC);	
	\draw[-, red, line width=1.5pt] (BBBBBB) to (BBBBBBBB);
	\newcrossingT{\CCCCCCCx/28.45274}{\CCCCCCCy/28.45274}{\formBCtheta}{+1}{CCCCCCCC}; \path (CCCCCCCC); \pgfgetlastxy{\CCCCCCCCx}{\CCCCCCCCy};
	\vertex (CCCCCCCCC) at ({2*(\CCCCCCCCx-\CCCCCCCx)+\CCCCCCCx}, {2*(\CCCCCCCCy-\CCCCCCCy)+\CCCCCCCy}); \path (CCCCCCCCC); \pgfgetlastxy{\CCCCCCCCCx}{\CCCCCCCCCy};	
  	\newcrossingB{\CCCCCCCCx/28.45274}{\CCCCCCCCy/28.45274}{\BBBBBBBBx/28.45274}{\BBBBBBBBy/28.45274}{+1}{BBBBBBBBB}; \path (BBBBBBBBB); \pgfgetlastxy{\BBBBBBBBBx}{\BBBBBBBBBy};
  	\vertex (BBBBBBBBBB) at ({2*(\BBBBBBBBBx-\BBBBBBBBx)+\BBBBBBBBx}, {2*(\BBBBBBBBBy-\BBBBBBBBy)+\BBBBBBBBy}); \path (BBBBBBBBBB); \pgfgetlastxy{\BBBBBBBBBBx}{\BBBBBBBBBBy};
	\draw[-, red, line width=1.5pt] (BBBBBBBB) to (BBBBBBBBBB);	
	\draw[-, blue, line width=1.5pt] (CCCCCCC) to (CCCCCCCCC);	
	\newcrossingT{\BBBBBBBBBBx/28.45274}{\BBBBBBBBBBy/28.45274}{\formBCtheta}{+1}{BBBBBBBBBBB}; \path (BBBBBBBBBBB); \pgfgetlastxy{\BBBBBBBBBBBx}{\BBBBBBBBBBBy};
%
  	\vertex (C) at ({\offsetCx}, {\offsetCy}); 
  	\vertex (D) at ({\offsetDx}, {\offsetDy}); 
  	\vertex (DD) at ({2*\CDx-\offsetDx*28.45274pt}, {2*\CDy-\offsetDy*28.45274pt}); \path (DD); \pgfgetlastxy{\DDx}{\DDy};
	\newcrossingT{\DDx/28.45274}{\DDy/28.45274}{\formCDtheta}{-1}{DDD}; \path (DDD); \pgfgetlastxy{\DDDx}{\DDDy};
	\newcrossingB{\CDx/28.45274}{\CDy/28.45274}{\DDDx/28.45274}{\DDDy/28.45274}{-1}{CC}; \path (CC); \pgfgetlastxy{\CCx}{\CCy};
	\vertex (DDDD) at ({2*(\DDDx-\DDx)+\DDx}, {2*(\DDDy-\DDy)+\DDy}); \path (DDDD); \pgfgetlastxy{\DDDDx}{\DDDDy};	
	\draw[-, red, line width=1.5pt] (C) to (CD);
	\propag[fer, red] (C) to (CD);
	\draw[-, red, line width=1.5pt] (CD) to (CC);
	\draw[-, blue, line width=1.5pt] (D) to (DD);
	\propag[fer, blue] (D) to (DD);
	\draw[-, blue, line width=1.5pt] (DD) to (DDDD);
 	\vertex (CCC) at ({2*(\DDDx-\CCx)+\CCx}, {2*(\DDDy-\CCy)+\CCy}); \path (CCC); \pgfgetlastxy{\CCCx}{\CCCy};	
	\newcrossingT{\CCCx/28.45274}{\CCCy/28.45274}{\formCDtheta}{-1}{CCCC}; \path (CCCC); \pgfgetlastxy{\CCCCx}{\CCCCy};
	\vertex (CCCCC) at ({2*(\CCCCx-\CCCx)+\CCCx}, {2*(\CCCCy-\CCCy)+\CCCy}); \path (CCCCC); \pgfgetlastxy{\CCCCCx}{\CCCCCy};	
	\newcrossingB{\CCCCx/28.45274}{\CCCCy/28.45274}{\DDDDx/28.45274}{\DDDDy/28.45274}{-1}{DDDDD}; \path (DDDDD); \pgfgetlastxy{\DDDDDx}{\DDDDDy};
	\vertex (DDDDDD) at ({2*(\DDDDDx-\DDDDx)+\DDDDx}, {2*(\DDDDDy-\DDDDy)+\DDDDy}); \path (DDDDDD); \pgfgetlastxy{\DDDDDDx}{\DDDDDDy};	
	\draw[-, red, line width=1.5pt] (CC) to (CCC);
	\draw[-, red, line width=1.5pt] (CCC) to (CCCCC);
	\draw[-, blue, line width=1.5pt] (DDDD) to (DDDDDD);
	\newcrossingT{\DDDDDDx/28.45274}{\DDDDDDy/28.45274}{\formCDtheta}{-1}{DDDDDDD}; \path (DDDDDDD); \pgfgetlastxy{\DDDDDDDx}{\DDDDDDDy};
	\vertex (DDDDDDDD) at ({2*(\DDDDDDDx-\DDDDDDx)+\DDDDDDx}, {2*(\DDDDDDDy-\DDDDDDy)+\DDDDDDy}); \path (DDDDDDDD); \pgfgetlastxy{\DDDDDDDDx}{\DDDDDDDDy};	
	\newcrossingB{\DDDDDDDx/28.45274}{\DDDDDDDy/28.45274}{\CCCCCx/28.45274}{\CCCCCy/28.45274}{-1}{CCCCCC}; \path (CCCCCC); \pgfgetlastxy{\CCCCCCx}{\CCCCCCy};
	\vertex (CCCCCCC) at ({2*(\CCCCCCx-\CCCCCx)+\CCCCCx}, {2*(\CCCCCCy-\CCCCCy)+\CCCCCy}); \path (CCCCCCC); \pgfgetlastxy{\CCCCCCCx}{\CCCCCCCy};	
	\draw[-, blue, line width=1.5pt] (DDDDDD) to (DDDDDDDD);
	\draw[-, red, line width=1.5pt] (CCCCC) to (CCCCCCC);
	\newcrossingT{\CCCCCCCx/28.45274}{\CCCCCCCy/28.45274}{\formCDtheta}{-1}{CCCCCCCC}; \path (CCCCCCCC); \pgfgetlastxy{\CCCCCCCCx}{\CCCCCCCCy};
	\vertex (CCCCCCCCC) at ({2*(\CCCCCCCCx-\CCCCCCCx)+\CCCCCCCx}, {2*(\CCCCCCCCy-\CCCCCCCy)+\CCCCCCCy}); \path (CCCCCCCCC); \pgfgetlastxy{\CCCCCCCCCx}{\CCCCCCCCCy};
	\draw[-, red, line width=1.5pt] (CCCCCCC) to (CCCCCCCCC);
 	\vertex (DDDDDDDDD) at ({2*(\CCCCCCCCx-\DDDDDDDDx)+\DDDDDDDDx}, {2*(\CCCCCCCCy-\DDDDDDDDy)+\DDDDDDDDy}); \path (DDDDDDDDD); \pgfgetlastxy{\DDDDDDDDDx}{\DDDDDDDDDy};	
	\newcrossingT{\DDDDDDDDDx/28.45274}{\DDDDDDDDDy/28.45274}{\formCDtheta}{-1}{DDDDDDDDDD}; \path (DDDDDDDDDD); \pgfgetlastxy{\DDDDDDDDDDx}{\DDDDDDDDDDy};
 	\vertex (DDDDDDDDDDD) at ({2*(\DDDDDDDDDDx-\DDDDDDDDDx)+\DDDDDDDDDx}, {2*(\DDDDDDDDDDy-\DDDDDDDDDy)+\DDDDDDDDDy}); \path (DDDDDDDDDDD); \pgfgetlastxy{\DDDDDDDDDDDx}{\DDDDDDDDDDDy};	
	\draw[-, blue, line width=1.5pt] (DDDDDDDD) to (DDDDDDDDD);
	\draw[-, blue, line width=1.5pt] (DDDDDDDDD) to (DDDDDDDDDDD);
 	\vertex (CCCCCCCCCC) at ({2*(\DDDDDDDDDDx-\CCCCCCCCCx)+\CCCCCCCCCx}, {2*(\DDDDDDDDDDy-\CCCCCCCCCy)+\CCCCCCCCCy}); \path (CCCCCCCCCC); \pgfgetlastxy{\CCCCCCCCCCx}{\CCCCCCCCCCy};	
	\draw[-, red, line width=1.5pt] (CCCCCCCCC) to (CCCCCCCCCC);
 	\vertex (CCCCCCCCCCC) at ({\CCCCCCCCCCx-0.3*28.45274pt}, {\CCCCCCCCCCy+0.3*28.45274pt}); 	
	\draw[-, red, line width=1.5pt] (CCCCCCCCCC) to (CCCCCCCCCCC);
  	\vertex (D) at ({\offsetDx}, {\offsetDy}); 
  	\vertex (E) at ({\offsetEx}, {\offsetEy}); 
  	\vertex (EE) at ({2*\DEx-\offsetEx*28.45274pt}, {2*\DEy-\offsetEy*28.45274pt}); \path (EE); \pgfgetlastxy{\EEx}{\EEy};
	\newcrossingT{\EEx/28.45274}{\EEy/28.45274}{\formDEtheta}{-1}{EEE}; \path (EEE); \pgfgetlastxy{\EEEx}{\EEEy};
	\newcrossingB{\DEx/28.45274}{\DEy/28.45274}{\EEEx/28.45274}{\EEEy/28.45274}{-1}{DD}; \path (DD); \pgfgetlastxy{\DDx}{\DDy};
	\vertex (EEEE) at ({1.5*(\EEEx-\EEx)+\EEx}, {1.5*(\EEEy-\EEy)+\EEy}); \path (EEEE); \pgfgetlastxy{\EEEEx}{\EEEEy};	
	\draw[-, red, line width=1.5pt] (D) to (DE);
	\propag[fer, red] (D) to (DE);
	\draw[-, red, line width=1.5pt] (DE) to (DD);
	\draw[-, blue, line width=1.5pt] (E) to (EE);
	\draw[-, blue, line width=1.5pt] (EE) to (EEEE);
	\propag[fer, blue] (EE) to (EEEE);
 	\vertex (DDD) at ({2*(\EEEx-\DDx)+\DDx}, {2*(\EEEy-\DDy)+\DDy}); \path (DDD); \pgfgetlastxy{\DDDx}{\DDDy};	
	\newcrossingT{\DDDx/28.45274}{\DDDy/28.45274}{\formDEtheta}{-1}{DDDD}; \path (DDDD); \pgfgetlastxy{\DDDDx}{\DDDDy};
	\vertex (DDDDD) at ({1.2*(\DDDDx-\DDDx)+\DDDx}, {1.2*(\DDDDy-\DDDy)+\DDDy}); \path (DDDDD); \pgfgetlastxy{\DDDDDx}{\DDDDDy};	
	\newcrossingB{\DDDDx/28.45274}{\DDDDy/28.45274}{\EEEEx/28.45274}{\EEEEy/28.45274}{-1}{EEEEE}; \path (EEEEE); \pgfgetlastxy{\EEEEEx}{\EEEEEy};
	\vertex (EEEEEE) at ({2*(\EEEEEx-\EEEEx)+\EEEEx}, {2*(\EEEEEy-\EEEEy)+\EEEEy}); \path (EEEEEE); \pgfgetlastxy{\EEEEEEx}{\EEEEEEy};	
	\draw[-, red, line width=1.5pt] (DD) to (DDD);
%
  	\vertex (E) at ({\offsetEx}, {\offsetEy}); 
  	\vertex (L) at ({\offsetLx}, {\offsetLy}); 
  	\vertex (LL) at ({2*\ELx-\offsetLx*28.45274pt}, {2*\ELy-\offsetLy*28.45274pt}); \path (LL); \pgfgetlastxy{\LLx}{\LLy};
	\newcrossingT{\LLx/28.45274}{\LLy/28.45274}{\formELtheta}{-1}{LLL}; \path (LLL); \pgfgetlastxy{\LLLx}{\LLLy};
	\vertex (EE) at ({\offsetEx-4}, {\offsetEy+4}); 	
	\vertex (LLLL) at ({0.8*(\LLLx-\LLx)+\LLx}, {0.8*(\LLLy-\LLy)+\LLy}); \path (LLLL); \pgfgetlastxy{\LLLLx}{\LLLLy};	
	\draw[-, red, line width=1.5pt] (E) to (EL);
	\propag[fer, red] (E) to (EL);
	\draw[-, red, line width=1.5pt] (EL) to (EE);
	\draw[-, blue, line width=1.5pt] (o) to (L);
	\propag[fer, blue] (o) to (L);
	\draw[-, blue, line width=1.5pt] (L) to (LL);
	\draw[-, blue, line width=1.5pt] (LL) to (LLLL);
%
	\vertex[particle] (o) at ({\offsetAx}, {\offsetAy}) {\large $\bigstar$}; 
	\vertex[particle] (o) at ({\offsetBx}, {\offsetBy}) {\large $\bigstar$}; 
	\vertex[particle] (o) at ({\offsetCx}, {\offsetCy}) {\large $\bigstar$}; 
	\vertex[particle] (o) at ({\offsetDx}, {\offsetDy}) {\large $\bigstar$}; 
	\vertex[particle] (o) at ({\offsetEx}, {\offsetEy}) {\large $\bigstar$}; 
%
	\fill[white] (-8,6.9) rectangle (8,8.5);
 	\fill[white, path fading=fade top] (-8,6) rectangle (8,6.9);
	\vertex (o) at (0, 0); 
	\vertex (Rint) at ({0.7*\endy}, {0.7*\endy}); 
	\vertex[particle] (R) at ({\endy}, {\endy}) {\large $t+z$}; 
	\vertex (Lint) at ({-0.7*\endy}, {0.7*\endy}); 
	\vertex[particle] (L) at ({-\endy}, {\endy}) {\large $t-z$}; 
	\draw[-{Latex[length=4mm,width=3mm]}, black, line width=0.8pt] (Rint) to (R);
	\draw[-{Latex[length=4mm,width=3mm]}, black, line width=0.8pt] (Lint) to (L);
	\vertex[particle] (V) at ({-\endy+0}, +2) {$t$}; \vertex (Vo) at ({-\endy+0},0); \draw[-{Latex}, black, line width=0.8pt] (Vo) to (V);
	\vertex[particle] (H) at ({-\endy+2}, 0) {$z$}; \draw[-{Latex}, black, line width=0.8pt] (Vo) to (H);
	\vertex (f1) at ({\endl*cos(\formRAtheta)/sin(\formRAtheta)}, {\endl}); 
	\draw[->, gray, opacity=0.3] (o) to (f1);
	\vertex (f1) at ({\endl*cos(\formABtheta)/sin(\formABtheta)}, {\endl}); 
	\draw[->, gray, opacity=0.3] (o) to (f1);
	\vertex (f1) at ({\endl*cos(\formBCtheta)/sin(\formBCtheta)}, {\endl}); 
	\draw[->, gray, opacity=0.3] (o) to (f1);
	\vertex (f1) at ({\endl*cos(\formCDtheta)/sin(\formCDtheta)}, {\endl}); 
	\draw[->, gray, opacity=0.3] (o) to (f1);
	\vertex (f1) at ({\endl*cos(\formDEtheta)/sin(\formDEtheta)}, {\endl}); 
	\draw[->, gray, opacity=0.3] (o) to (f1);
	\vertex (f1) at ({\endl*cos(\formELtheta)/sin(\formELtheta)}, {\endl}); 
	\draw[->, gray, opacity=0.3] (o) to (f1);
	\vertex[particle] (o1) at (2.5, 0) {$\ t_0\ $}; \vertex (o2) at (-0.4, 0); \draw[-, gray, dotted] (o1) to (o2);
	\vertex[particle] (o1) at ({0.4+\offsetCy/2}, {\offsetCy/2}) {}; \vertex (o2) at ({-0.4-\offsetCy/2}, {\offsetCy/2}); \draw[-, gray, dotted] (o1) to (o2);
		\fill [green, path fading=fade out] (0,{\offsetCy/2}) circle [x radius={\offsetCy/2-0.05}, y radius={(\offsetCy/2)/10}];
	\vertex[particle] (o1) at ({2.5+\offsetCy}, {\offsetCy}) {$t_1$}; \vertex (o2) at ({-0.4-\offsetCy}, {\offsetCy}); \draw[-, gray, dotted] (o1) to (o2);
		\fill [green, path fading=fade out] ({-(\offsetCy-\offsetCx)/2-0.02},{\offsetCy}) circle [x radius={(\offsetCy-\offsetCx)/2-0.04}, y radius={(\offsetCy/2)/10}];
		\fill [green, path fading=fade out] ({(\offsetCy+\offsetCx)/2+0.02},{\offsetCy}) circle [x radius={(\offsetCy-\offsetCx)/2-0.04}, y radius={(\offsetCy/2)/10}];	
	\vertex[particle] (o1) at ({2.5+3*\offsetCy/2}, {3*\offsetCy/2}) {$\ t_2\ $}; \vertex (o2) at ({-0.4-3*\offsetCy/2}, {3*\offsetCy/2}); \draw[-, gray, dotted] (o1) to (o2);
		\fill [green, path fading=fade out] ({(\offsetCy+\offsetCx)/2-2.3},{3*\offsetCy/2}) circle [x radius={(\offsetCy-\offsetCx)/2+0.05}, y radius={(\offsetCy/2)/10}];
		\fill [green, path fading=fade out] ({3*\offsetCy/2-0.35},{3*\offsetCy/2}) circle [x radius={(\offsetCy-\offsetCx)/2-0.4}, y radius={(\offsetCy/2)/10}];
	\vertex[particle] (o1) at ({0.4+4*\offsetCy/2}, {4*\offsetCy/2}) {}; \vertex (o2) at ({-0.4-4*\offsetCy/2}, {4*\offsetCy/2}); \draw[-, gray, dotted] (o1) to (o2);
		\fill [green, path fading=fade out] ({-4*\offsetCy/2+0.6},{4*\offsetCy/2}) circle [x radius={(0.55}, y radius={(\offsetCy/2)/10}];
		\fill [green, path fading=fade out] ({-4*\offsetCy/2+1.84},{4*\offsetCy/2}) circle [x radius={(0.18}, y radius={(\offsetCy/2)/10}];
		\fill [green, path fading=fade out] ({-4*\offsetCy/2+4.3},{4*\offsetCy/2}) circle [x radius={(0.36}, y radius={(\offsetCy/2)/10}];
		\fill [green, path fading=fade out] ({-4*\offsetCy/2+5.65},{4*\offsetCy/2}) circle [x radius={(0.15}, y radius={(\offsetCy/2)/10}];
		\fill [green, path fading=fade out] ({-4*\offsetCy/2+6.0},{4*\offsetCy/2}) circle [x radius={(0.1}, y radius={(\offsetCy/2)/10}];
	\vertex[particle] (o1) at ({0.4+5*\offsetCy/2}, {5*\offsetCy/2}) {}; \vertex (o2) at ({-0.4-5*\offsetCy/2}, {5*\offsetCy/2}); \draw[-, gray, dotted] (o1) to (o2);
		\fill [green, path fading=fade out] ({-5*\offsetCy/2+0.13},{5*\offsetCy/2}) circle [x radius={(0.12}, y radius={(\offsetCy/2)/10}];
		\fill [green, path fading=fade out] ({-5*\offsetCy/2+1.07},{5*\offsetCy/2}) circle [x radius={(0.09}, y radius={(\offsetCy/2)/10}];	
		\fill [green, path fading=fade out] ({-5*\offsetCy/2+2.27},{5*\offsetCy/2}) circle [x radius={(0.183}, y radius={(\offsetCy/2)/10}];
		\fill [green, path fading=fade out] ({-5*\offsetCy/2+5.15},{5*\offsetCy/2}) circle [x radius={(0.3}, y radius={(\offsetCy/2)/10}];
		\fill [green, path fading=fade out] ({-5*\offsetCy/2+6.81},{5*\offsetCy/2}) circle [x radius={(0.15}, y radius={(\offsetCy/2)/10}];
		\fill [green, path fading=fade out] ({-5*\offsetCy/2+7.52},{5*\offsetCy/2}) circle [x radius={(0.12}, y radius={(\offsetCy/2)/10}];
	\vertex[particle] (o1) at ({2.5+6*\offsetCy/2}, {6*\offsetCy/2}) {$\ t_3\ $}; \vertex (o2) at ({-0.4-6*\offsetCy/2}, {6*\offsetCy/2}); \draw[-, gray, dotted] (o1) to (o2);
		\fill [green, path fading=fade out] ({-6*\offsetCy/2+1.66},{6*\offsetCy/2}) circle [x radius={(0.4}, y radius={(\offsetCy/2)/10}];
		\fill [green, path fading=fade out] ({-6*\offsetCy/2+2.7},{6*\offsetCy/2}) circle [x radius={(0.2}, y radius={(\offsetCy/2)/10}];	
		\fill [green, path fading=fade out] ({-6*\offsetCy/2+6.39},{6*\offsetCy/2}) circle [x radius={(0.18}, y radius={(\offsetCy/2)/10}];
		\fill [green, path fading=fade out] ({-6*\offsetCy/2+8.35},{6*\offsetCy/2}) circle [x radius={(0.15}, y radius={(\offsetCy/2)/10}];
		\fill [green, path fading=fade out] ({-6*\offsetCy/2+8.78},{6*\offsetCy/2}) circle [x radius={(0.12}, y radius={(\offsetCy/2)/10}];
	\vertex[particle] (o1) at ({0.4+7*\offsetCy/2}, {7*\offsetCy/2}) {}; \vertex (o2) at ({-0.4-7*\offsetCy/2}, {7*\offsetCy/2}); \draw[-, gray, dotted] (o1) to (o2);
		\fill [green, path fading=fade out] ({-7*\offsetCy/2+0.45},{7*\offsetCy/2}) circle [x radius={(0.12}, y radius={(\offsetCy/2)/10}];
		\fill [green, path fading=fade out] ({-7*\offsetCy/2+1.66},{7*\offsetCy/2}) circle [x radius={(0.4}, y radius={(\offsetCy/2)/10}];
		\fill [green, path fading=fade out] ({-7*\offsetCy/2+3.13},{7*\offsetCy/2}) circle [x radius={(0.2}, y radius={(\offsetCy/2)/10}];
		\fill [green, path fading=fade out] ({-7*\offsetCy/2+7.34},{7*\offsetCy/2}) circle [x radius={(0.35}, y radius={(\offsetCy/2)/10}];
		\fill [green, path fading=fade out] ({-7*\offsetCy/2+9.50},{7*\offsetCy/2}) circle [x radius={(0.15}, y radius={(\offsetCy/2)/10}];
		\fill [green, path fading=fade out] ({-7*\offsetCy/2+10.30},{7*\offsetCy/2}) circle [x radius={(0.12}, y radius={(\offsetCy/2)/10}];
\end{feynhand}
\end{tikzpicture}
\vspace*{3mm}
\caption{A schematic illustration of the spacetime evolution of string breaking.  An energetic quark (\red{\hspace*{0.15mm}\rule[0.4ex]{1.1em}{2pt}\hspace*{0.15mm}}) and anti-quark (\blue{\hspace*{0.15mm}\rule[0.4ex]{1.1em}{2pt}\hspace*{0.15mm}}) pair is produced by a high-energy reaction at $(t,z)=(t_0,0)$ and flies apart at the speed of light (in the massless-quark limit) along the $z$ direction in the center-of-mass frame.  The associated string (\green{\hspace*{0.15mm}\rule[0.4ex]{1.1em}{2pt}\hspace*{0.15mm}}) expands one-dimensionally in space.  At $t=t_1$, the string becomes sufficiently long and breaks into two shorter strings by producing a new quark and anti-quark pair, as indicated by $\bigstar$.  This process occurs repeatedly, producing many string fragments.  When the length of a string becomes sufficiently short such that the energy stored in the string falls below the pair-production threshold, it becomes stable against further pair production and forms up a hadron.  In this figure, the first hadron is produced at $t=t_2$ (where the red and blue lines cross for the first time), and subsequent hadron production proceeds in an inside-out manner: hadrons produced at later times ($t$ larger) are formed outer in space ($|z|$ larger).  The quark content of each hadron oscillates back and forth along the $z$ direction (so-called yo-yo motion) due to the attractive string force.  The final hadron production occurs at $t=t_3$, and six mesons are eventually produced in this figure.}
\label{fig:4.2-1}
\end{figure}

An isolated quark has never been observed, implying that quarks are confined within color-neutral hadrons.  This phenomenon is known as quark confinement in QCD.  A widely accepted explanation for quark confinement is that a color-electric flux-tube is formed between quarks and binds the quarks so strongly that they cannot be separated macroscopically, as originally proposed by Kogut and Susskind in 1974~\cite{Kogut:1974ag} (see also Refs.~\cite{Bali:2000gf, Greensite:2003bk, Alkofer:2006fu} for comprehensive reviews).  The flux tube is often called a {\it string}, because it extends essentially one-dimensionally in space, unlike the electric field in QED, which emanates radially from an electric charge.  

The string picture naturally explains the multi-particle production mechanism in high-energy reactions (see Fig.~\ref{fig:4.2-1}).  Suppose that an energetic quark and anti-quark pair is produced by a high-energy process such as electron-positron annihilation.  After the production, the quark and anti-quark recede from each other.  Accordingly, the associated string, which is sourced by one of the pair and sinks into the other, is stretched out.  The energy density of the string increases monotonically with its length because it requires work against the color-electric force.  However, since the energy cannot increase infinitely, the string should eventually break up into two shorter, less energetic strings.  This is achieved by producing a quark and anti-quark pair, which acts as new source and sink at some intermediate position along the parent string.  If the daughter strings are still long enough, they would break up again.  These processes repeat again and again, until the energy of the initial pair is fully converted into that of many fragments of stable short strings with color-neutral combinations of quarks at the ends, which are observed as hadrons by the detector.  

A few more characteristic features of the string-breaking picture of hadron production are in order (see Fig.~\ref{fig:4.2-1}):
\begin{itemize} 
\item A string that binds quarks extends spatially, which means that quarks that constitute a hadron are generated at spacelike-separated points.  This leads to an ``inside-out'' cascade of hadron production in spacetime, which is consistent with Lorentz covariance~\cite{Bjorken:1976mk, Andersson:1983ia}.  Namely, low-energy hadrons around mid-rapidity, $v = z/t \approx 0$, are produced first, followed by increasingly energetic hadrons formed at larger rapidities, and finally by hadrons composed of leading quarks produced near the light cone, $v \approx 1$.

\item It leads to a flat hadron distribution around mid-rapidity (``plateau"), whose extent increases with the initial energy of the primary quarks~\cite{Artru:1974hr}.  This is because a higher initial energy allows the string to become longer; consequently, hadron production takes place over a wider spatial region.  Since string breaking is a local process and hence its probability would be constant per unit length, the resulting hadron yield becomes approximately uniform along the string.

\item The final string composing a hadron is not static but exhibits a ``yo-yo" motion, i.e., the endpoint quarks oscillate back and forth in space as time goes~\cite{Artru:1974hr, Andersson:1978vj, Casher:1979gw}.  This can be regarded as an analog of the plasma oscillation in the QED backreaction problem (see Sec.~\ref{sec:2.3.4}).  The yo-yo motion appears in the following manner.  Since the endpoint quarks are always attracted to each other by the color-electric force of the string, the endpoint quarks are accelerated inward at early times, and then the string shrinks.  At a later time, the length of the string reaches zero, at which the endpoint quarks pass through each other and start to be decelerated by the color-electric force.  The string then expands.  This continues until the velocities of the quarks are decelerated to zero, corresponding to the moment when the string is maximally expanded.  Then, the string starts to accelerate the endpoint quarks inwards, i.e., we come back to the first step.  These steps repeat, leading to the oscillating yo-yo motion.  
\end{itemize}

Based on the string-breaking picture, Casher, Neuberger, and Nussinov developed in 1979 a phenomenological framework for hadron production in high-energy reactions, known as the {\it flux-tube model}~\cite{Casher:1978wy}.  They proposed that the Schwinger effect is a natural microscopic mechanism for string breaking and applied the Nikishov formula~(\ref{eq:7})\footnote{In the original paper by Casher {\it et al.} as well as follow-up studies at the time, the Schwinger formula for the vacuum-decay rate $w$ (\ref{eq:9}) was actually used.  It is more common recently (e.g., PYTHIA) to use the Nikishov formula~(\ref{eq:7}) and is appropriate to quantify the momentum spectrum of produced pairs (see also Ref.~\cite{Kluger:1998bm}).  } to estimate the differential probability ${\rm d}P$ for a breakup, 
\begin{align}
	{\rm d} P \propto \exp\left[ - \pi \frac{m^2 + {\bm p}_\perp^2}{\kappa} \right] {\rm d}^2{\bm p}_\perp {\rm d}^4x \; , \label{eq:7-}
\end{align}
where ${\bm p}_\perp$ is the transverse momentum of the produced pair and $\kappa \approx g{\mathcal E}$ is the string tension, i.e., the energy density of a string per unit length.  The string tension can be estimated from the Regge slope in hadron spectroscopy and lattice QCD, which yield $\kappa \sim 1\;{\rm GeV}/{\rm fm}$ (see, e.g., Ref.~\cite{Bali:2000gf}), or determined phenomenologically by fitting experimental data.  The flux-tube model~(\ref{eq:7-}) naturally explains some features of hadron yields in actual collider experiments, such as the suppression of heavy-flavor hadrons, according to the light-to-heavy-quark production ratio ${\rm d}P_{\rm heavy}/{\rm d}P_{\rm light} = \exp [ - \pi (m_{\rm heavy}^2 - m_{\rm light}^2 )/\kappa ]\ \Rightarrow\ u:d:s:c \sim 1:1:0.3:10^{-11}$ for the constituent quark masses and a Gaussian momentum-distribution, with width $\sigma^2 = \kappa/\pi \approx (0.25\;{\rm GeV})^2$ (which could be interpreted as an effective temperature $T=\sigma$~\cite{Bialas:1999zg, Kharzeev:2005iz}).  Note, however, that these are just qualitative, and more effects need to be included to quantitatively explain the data.  For example, the experimental value of the width is about $\sigma \approx 0.35\;{\rm GeV}$, which requires additional momentum-broadening sources such as soft gluon radiation~\cite{Bierlich:2022pfr}.  We also remind that the Nikishov formula~(\ref{eq:7}) is originally derived in QED under idealistic setups (see Sec.~\ref{sec:2.2}), meaning that the flux-tube model is built upon an Abelian approximation, where the non-Abelian features of the Schwinger effect such as the color-orientation dependence~(\ref{eq::26}) are dismissed, and also neglects realistic effects beyond the early treatments (see Sec.~\ref{sec:2.3}).  There exist a number of efforts to go beyond this; to name a few, backreaction~\cite{Glendenning:1983qq}, fluctuating string tension~\cite{Bialas:1999zg}, finite-time/size effect~\cite{Neuberger:1979tb, Wang:1988ct}, and time-dependent string tension~\cite{Hunt-Smith:2020lul}.  

The flux-tube model was extended by the Lund group in the 1980s (e.g., the ``popcorn" model for baryon production and inclusion of gluon as a kink of a string) and established as the {\it Lund string model}~\cite{Andersson:1983ia, Andersson:1997xwk}.  It was also implemented into numerical codes: originally released as JETSET, the code underwent extensive refinement and eventually culminated in PYTHIA---now one of the most widely used Monte-Carlo event generators for high-energy physics~\cite{Sjostrand:2019zhc, Bierlich:2022pfr}.  

Besides these phenomenological studies, first-principle numerical simulations have also been performed.  Numerical lattice-QCD simulations support the formation of strings in various multi-quark systems such as meson~\cite{Bali:1994de, Haymaker:1994fm, Yanagihara:2018qqg} and baryon~\cite{Bornyakov:2004uv}.  The breaking of a string has also been investigated by analyzing a static quark and anti-quark potential in lattice QCD~\cite{Bali:2005fu}, observing that the string breaking occurs at a critical distance of about $1\;{\rm fm}$, which is consistent with the typical size of hadrons.  Also, the realtime dynamics of the string breaking is now actively investigated, not only with classical-computing techniques such as realtime lattice method~\cite{Hebenstreit:2013baa, Spitz:2018eps} but also with quantum ones; see Ref.~\cite{Halimeh:2025vvp} and references therein.

\subsection{Early-time dynamics of relativistic heavy-ion collisions} \label{sec:4.3}

Relativistic heavy-ion collisions, performed at RHIC (Relativistic Heavy Ion Collider; $2000\!\sim$) and the LHC (Large Hadron Collider; $2008\!\sim$) provide the experimental means to create the hottest matter on Earth, reaching temperatures of a few trillion kelvin $={\mathcal O}(100\;\mathchar`-\;1000\;{\rm MeV})$.  Under such extreme conditions, QCD predicts that matter undergoes a phase transition in which quarks and gluons---normally confined inside hadrons---are liberated to form a plasma state, known as the {\it quark-gluon plasma} (QGP)~\cite{Yagi:2005yb}.  The study of QGP is not only crucial for deepening our understanding of QCD and the Standard Model of particle physics, but also for unveiling the ``origin" of the matter around us, since such an extraordinarily hot state should have existed in the early Universe according to the Big Bang theory, and all matter, including ourselves, evolved ultimately from QGP.  The physics of QGP and related topics are reviewed extensively in other chapters of this Encyclopedia, 
to which we refer the reader for more details.  For the purpose of this chapter, we recall that, after decades of intensive experimental efforts, it has been established that the QGP is actually created in heavy-ion collisions and that it exhibits remarkable properties, such as (nearly) perfect fluidity. 
At the same time, these experimental discoveries raised many theoretical challenges.  One of the most prominent is the early-time dynamics of relativistic heavy-ion collisions: it remains unclear how the QGP is actually formed from the colliding nuclei.  In particular, experimental results strongly suggest that the production of a vast number of quarks and gluons, followed by their thermalization (or ``hydrodynamization") into the QGP, occurs extremely rapidly, within a remarkably short time of $\lesssim 1\;{\rm fm}/c$ after a collision~\cite{Heinz:2001xi}.  This is known as the {\it early-thermalization problem}, and no fully established explanation exists, despite extensive theoretical investigations since the dawn of heavy-ion physics in the 1980s.  
Although this issue has already been reviewed in other chapters of this Encyclopedia, as well as in review papers~\cite{Fukushima:2016xgg, Schlichting:2019abc, Berges:2020fwq}, let us revisit it here from the viewpoint of the Schwinger effect.

The Schwinger effect has been applied to the early-time dynamics of heavy-ion collisions as a mechanism of the quark and gluon production (or entropy production).  This is essentially an extension of the flux-tube model reviewed in Sec.~\ref{sec:4.2}: a strong color electromagnetic field is formed immediately after a collision of heavy ions and decays into quark and gluon particles that constitute the QGP via the Schwinger effect.  An important difference lies in the field configuration.  The flux-tube model was originally developed for ``small" systems, such as lepton-lepton collisions and few-quark systems, where only a single string (or a few independent strings) is present at the initial time.  In heavy-ion collisions, by contrast, the system is ``large," in the sense that the transverse size of the system and the number of participating color charges are much greater and many strings can be formed.  The produced color electromagnetic field should therefore possess a far more complicated structure than that of a single string.  Accordingly, the pair-production dynamics is modified from the naive string-breaking picture.  

It was Biro, Nielsen, and Knoll in 1984 who first proposed a phenomenological extension of the flux-tube model to heavy-ion collisions~\cite{Biro:1984cf} (see also Refs.~\cite{Bialas:1984ye, Bialas:1985is, Kerman:1985tj}).  They proposed that a cluster of strings produced in heavy-ion collisions can fuse into a flux tube of higher color representation, called ``rope," hence the name {\it rope model}.  A rope has a stronger effective string tension $\kappa_{\rm rope}$ than a single string $\kappa_{\rm string}$, intuitively because of the superposition principle, i.e., sum of multiple strings can give a stronger field in total (of course, there are complications arising from the non-Abelian nature of QCD).  The resulting enhancement of the string tension scales approximately as $\kappa_{\rm rope} \propto \sqrt{N}\,\kappa_{\rm string}$, where $N={\mathcal O}(1\;\mathchar`-\;10) $ is the number of fused strings per cluster and thus $\kappa_{\rm rope} = {\mathcal O}({\rm a\ few\ GeV/fm})$~\cite{Bierlich:2022ned}.  An increased string tension enhances the pair-production rate, implying faster entropy production in heavy-ion collisions than would be expected from a simple extrapolation of elementary collisions~\cite{Bialas:1985is}.  It also favors the production of heavier quark flavors through ${\rm d}P_{\rm heavy}/{\rm d}P_{\rm light} = \exp [-\pi (m_{\rm heavy}^2 - m_{\rm light}^2)/\kappa ]$ [see Eq.~(\ref{eq:7-})].  In particular, it is able to explain strangeness enhancement, which is observed not only in heavy-ion collisions~\cite{Sorge:1992ej} but also in high-multiplicity proton-proton collisions~\cite{Bierlich:2014xba}.  The rope model is now implemented in modern event generators such as PYTHIA~\cite{Bierlich:2022pfr}, together with advanced treatments of string dynamics, e.g., string shoving~\cite{Bierlich:2017vhg}.

A more sophisticated and modern estimate of the color-field configuration is provided by the color-glass-condensate (CGC) picture of high-energy nuclei (see Refs.~\cite{Iancu:2003xm, Gelis:2010nm, Kovchegov:2012mbw} for reviews).  The resulting field configuration is commonly referred to as {\it glasma}, a term introduced by Lappi and McLerran in 2006~\cite{Lappi:2006fp} to indicate that it is a transient state between the initial color-{\it gla}ss-condensate and the final quark-gluon pla{\it sma}.  The central idea of the CGC framework is that a nucleus at very high energies becomes saturated with an enormous number of gluons.  This saturated, high-density gluonic state arises through a cascade of gluon emissions ${\rm g} \to {\rm gg}$, which is eventually balanced by the recombination process ${\rm gg} \to {\rm g}$ when the density becomes so large that gluons overlap and fuse with each other.  As the density is so large, the gluons can be treated as a coherent, classical color field, instead of a collection of incoherent, quantum particles.  A heavy-ion collision can, thus, be viewed as a collision of two classical color fields, whose subsequent evolution is governed by the classical Yang-Mills equation, 
\begin{align}
	{\mathcal D}_\nu {\mathcal F}^{\nu\mu} = {\mathcal J}^\mu \;, \label{eq:33}
\end{align}
where the source current ${\mathcal J}^\mu$ is determined by the color-charge distributions of the two colliding nuclei.  Notice that the Yang-Mills equation~(\ref{eq:33}) is conformal, so the physical scales of the produced field are determined solely by those contained in the color-charge distributions of the two colliding nuclei ${\mathcal J}^\mu$.  In the CGC framework, the only relevant scale arising from these distributions is the gluon density at saturation, known as the {\it saturation scale} $Q_{\rm s}$.  The saturation scale increases with the atomic number $A$ and decreases with Bjorken $x$ as $Q_{\rm s} \propto A^{1/3} x^{-\lambda}$, where $\lambda \approx 0.2\,\mathchar`-\,0.3$ is a phenomenological constant.  Typical values are $Q_{\rm s}\approx 1\;{\rm GeV}$ at RHIC and a few~GeV at the LHC.  Numerical solutions of Eq.~(\ref{eq:33}) have been obtained for boost-invariant setups~\cite{Lappi:2003bi, Lappi:2006fp, Lappi:2006hq, Fukushima:2011nq} (see also the IP-Glasma model for application to hydrodynamic initial conditions~\cite{Schenke:2012wb}) and, recently, for beyond-boost-invariant $(3+1)$-dimensional configurations~\cite{Gelfand:2016yho, Ipp:2017lho, Schlichting:2020wrv, McDonald:2023qwc, Ipp:2024ykh, Matsuda:2024moa, Matsuda:2024mmr}.  The resulting glasma field has the following characteristic features: 
\begin{itemize}
\item Glasma consists of a number of flux tubes.  These flux tubes are expanding in the beam direction and have typical transverse size of order $1/Q_{\rm s}$, reflecting the fact that the gluons are saturated on the incident ions with the length scale $1/Q_{\rm s}$.  

\item Glasma is not purely electric.  Each flux tube carries parallel color-electromagnetic fields oriented along the beam direction, ${\bm{\mathcal E}} \parallel {\bm{\mathcal B}}$.  The appearance of the longitudinal color-electric field ${\bm{\mathcal E}}$ is analogous to the formation of a longitudinal electric field between capacitor plates, with the two collided nuclei acting as color-charged analogs of such plates.  The longitudinal color-magnetic field ${\bm{\mathcal B}}$, in contrast, is a genuine consequence of the non-Abelian structure of QCD: the non-Abelian Gauss law allows a nonzero effective magnetic ``charge" even in the absence of
fundamental monopoles as ${\rm div}\,{\bm{\mathcal B}} = {\rm i}g\,[\,{\mathcal A}^i, {\mathcal B}^i\,] \neq 0$.  The presence of a longitudinal color-magnetic field is a distinctive feature of the glasma and is absent in the phenomenological flux-tube or rope models.  

\item Glasma is strong.  The typical color-field strength is determined by the saturation scale $Q_{\rm s}$.  Hence, $g{\mathcal E}, g{\mathcal B} \sim Q_{\rm s}^2$, which is much larger than the quark-mass scale.  

\item Glasma is topological.  As the electric and magnetic fields are parallel within each flux tube, the glasma carries local fluctuations of the topological Chern-Simons charge density,
\begin{align}
	\nu = -\frac{g^2}{16\pi^2}\,{\rm tr}_{\rm c}\!\left( {\mathcal F}_{\mu\nu}\tilde{\mathcal F}^{\mu\nu} \right)
		= \frac{g^2}{4\pi^2}\,{\rm tr}_{\rm c}\!\left( {\bm{\mathcal E}}\cdot{\bm{\mathcal B}} \right)
		\neq 0 \;. \label{eqdsqf:35}
\end{align}
Such topological fluctuations have important phenomenological implications via the chiral anomaly~\cite{Adler:1969gk, Bell:1969ts}, including the chiral magnetic effect~\cite{Fukushima:2008xe} (see Sec.~\ref{sec:4.4}). 
\end{itemize}

To describe the dynamics of the Schwinger effect in a non-trivial color-field configuration such as glasma, one must go beyond the early treatments of the Schwinger effect.  Numerous attempts have been made in this direction---often with techniques exchanged between QCD and QED studies---but the subject remains an active and cutting-edge research area.  Below, we briefly review several of these developments.

One of the earliest attempts was based on a kinetic approach, originally developed by Matsui and his collaborators~\cite{Kajantie:1985jh, Gatoff:1987uf} and, independently, by Bia\l{}as and Czy\.z~\cite{Bialas:1986mt, Bialas:1987en} in the late 1980s.  They considered an Abelian situation, in which the color electromagnetic field ${\mathcal F}_{\mu\nu}$ can be diagonalized as ${\mathcal F}_{\mu\nu} = \bar{F}_{\mu\nu} w^\ell H^\ell$, just as in the case of a covariantly-constant field~(\ref{eq:30}).  The problem then becomes essentially equivalent to QED, and the corresponding kinetic equations for quarks and gluons take the form of the Boltzmann equation,
\begin{align}
	p^{\mu} \left[ \frac{\partial}{\partial x^{\mu}} - q \bar{F}_{\mu\nu} \frac{\partial}{\partial p_{\nu}} \right] f
	= {\mathcal C} + {\mathcal S} \;, \label{eq::35}
\end{align}
where $f = f_{\rm quark}$ or $f_{\rm gluon}$ denotes the phase-space distribution functions for quarks and gluons, respectively, $q$ is the effective coupling [$q = g \omega_i$ for a quark of color $i$ and $q = g \omega_\alpha$ for a gluon of color $\alpha$; see Eq.~(\ref{eqdsq:29})], and ${\mathcal C}$ and ${\mathcal S}$ represent, respectively, the collision term among the produced particles and the source term describing the Schwinger effect.  In practice, the source term is commonly fixed by the locally-constant-field approximation (LCFA), i.e., use a naive extension of the Nikishov formula~(\ref{eq::26}) by promoting the fields spacetime-dependent by hand.  A more rigorous derivation based on quantum-field theory, assuming an intermediate particle picture (see Sec.~\ref{sec:2.3.3}), has been carried out first by Rau in 1994~\cite{Rau:1994ee} (see also Ref.~\cite{Tanji:2008ku}).  This derivation yields a highly non-Markovian source term, in contrast to the LCFA, yet numerical simulations show that both prescriptions lead to largely consistent spacetime evolutions once implemented in the kinetic equation~(\ref{eq::35})~\cite{Kluger:1991ib, Kluger:1992gb, Cooper:1992hw, Kluger:1998bm}.  The backreaction onto the color electromagnetic field ${\mathcal F}_{\mu\nu}$ can be incorporated by solving the Yang-Mills equation simultaneously with Eq.~(\ref{eq::35}) (within the Abelian approximation, this reduces to the Maxwell equation) with an additional source current induced by the Schwinger effect.  A full numerical simulation including the source term ${\mathcal S}$, the collision term ${\mathcal C}$ (within the relaxation-time approximation), and the backreaction was first performed by Banerjee {\it et al.} for a homogeneous color-electric field as an initial condition~\cite{Banerjee:1989by} in 1989.  A more sophisticated treatment of the collision term, with a two-to-two elastic scattering channel, was later carried out by Ruggieri {\it et al.}~\cite{Ruggieri:2015yea} in 2015.  They found that plasma oscillations present in the collisionless limit (see Sec.~\ref{sec:2.3.4}) become damped away and the initial color field decays more rapidly, as the interaction strength increases, and the system indeed relaxes toward thermal equilibrium due to the collision term.  The kinetic framework has also been used to study di-lepton emission from quarks produced in the early stages of heavy-ion collisions~\cite{Asakawa:1990se}.  

The kinetic approach can be extended to QCD by employing a non-Abelian kinetic equation, i.e., the Wong equation~\cite{Wong:1970fu}.  Such an extension was done by Nayak and Ravishankar in 1996~\cite{Nayak:1996ex, Nayak:1997kp}.  In the Wong framework, the color degree of freedom is incorporated as an additional phase-space variable, $f(x,p) \to f(x,p,Q)$, and the kinetic equation becomes
\begin{align}
  p^{\mu} \left[ \frac{\partial}{\partial x^{\mu}}
  - Q^a {\mathcal F}^a_{\mu\nu} \frac{\partial}{\partial p_{\nu}}
  - f^{abc} {\mathcal A}^b_\mu Q^c \frac{\partial}{\partial Q^a} \right] f
  = {\mathcal C} + {\mathcal S} \;, \label{eq:::35}
\end{align}
where $a = 1,2,\cdots, N_{\rm c}^2-1$ labels color and $f^{abc}$ are the anti-symmetric structure constants such that $[T^a,T^b] = i f^{abc} T^c$.  It was suggested that the resulting non-Abelian dynamics exhibits behaviors qualitatively different from those of the Abelianized system.

Quantum-field-theoretic approaches are required to go beyond the phenomenological kinetic treatment.  When radiative effects are neglected, the problem reduces to solving the Dirac equation in a glasma (or glasma-like) background to obtain the quark spectrum.  This line of research was pioneered by Gelis, Kajantie, and Lappi in 2004~\cite{Gelis:2004jp, Gelis:2005pb}, who ignored backreaction, and was extended by several authors in the late 2010s~\cite{Gelfand:2016prm, Taya:2016ovo, Tanji:2017xiw} to include the backreaction from the quark sector.  These studies consistently indicate that quark production proceeds rapidly.  This is indeed reasonable in light of the intuitive picture of the Schwinger effect (see Fig.~\ref{fig:1}): the characteristic tunneling time is $\tau \sim m/g{\mathcal E}$, implying that quark production should finish well before the QGP formation time $\approx 1\;{\rm fm}/c$ because $\tau \sim m/Q_{\rm s}^2 \ll 1/Q_{\rm s} \approx 0.1\;{\rm fm}/c$, where $Q_{\rm s}\approx{\rm a\ few\ GeV}$ is the typical saturation scale at RHIC and the LHC.  

Finally, we note that the early stage of relativistic heavy-ion collisions also provides a unique opportunity to study strong electromagnetic-field phenomena, in addition to the strong color-field physics reviewed above.  The Coulomb fields of the colliding ions are strong because of their large atomic numbers, $Z = {\mathcal O}(50\;\mathchar`-\;100)$, and are further enhanced by the strong Lorentz contraction, $\gamma \approx \sqrt{s_{NN}}/m_N = {\mathcal O}(100\;\mathchar`-\;1000)$, where $\sqrt{s_{NN}} = {\mathcal O}(100\;\mathchar`-\;1000\;{\rm GeV})$ is the center-of-mass energy per nucleon pair and $m_N \approx 1\;{\rm GeV}$ is the nucleon mass.  As these Coulomb fields overlap at the instant of the collision, an extremely strong electromagnetic field is generated.  The resulting field strength can reach $eE, eB = {\mathcal O}((100\;\mathchar`-\;1000\;{\rm MeV})^2)$, making it among the strongest known in the present Universe~\cite{Skokov:2009qp, Voronyuk:2011jd, Bzdak:2011yy, Deng:2012pc} (see Refs.~\cite{Hattori:2016emy, Shen:2025unr} for reviews).  This field has been exploited experimentally to investigate intriguing non-linear QED phenomena such as photon-photon scattering~\cite{ATLAS:2017fur} and the linear Breit-Wheeler process~\cite{STAR:2019wlg, Brandenburg:2022tna}, as well as various aspects of QGP physics, including electric conductivity~\cite{STAR:2016cio}, the search for the chiral magnetic effect~\cite{STAR:2021mii}, and charge-dependent directed flow~\cite{STAR:2023jdd}.  It has also stimulated a vast number of theoretical studies of QCD and hadron physics under strong electromagnetic fields; examples include hadron masses~\cite{Hidaka:2012mz, Taya:2014nha, Kawaguchi:2015gpt}, nuclear forces~\cite{Miura:2025flc}, the string tension~\cite{Bonati:2014ksa}, the QCD phase diagram~\cite{DElia:2021yvk, Endrodi:2023wwf}, the photon polarization tensor in QCD matter~\cite{Wang:2021eud, Fukushima:2024ete}, and relativistic magneto-hydrodynamics~\cite{Nakamura:2022wqr, Benoit:2025amn} (see also Refs.~\cite{Andersen:2014xxa, Miransky:2015ava, Iwasaki:2021nrz, Hattori:2023egw, Endrodi:2024cqn} and references therein for further topics).  It should be emphasized, however, that although the field is strong, it is extremely short-lived.  Since the field exists only during the instant of the collision, its typical lifetime may naively be estimated as $\tau = R/\gamma = {\mathcal O}(0.1\;\mathchar`-\;0.01\;{\rm fm}/c)$, where $R \approx 10\;{\rm fm}$ is the size of the colliding ions at rest (although it remains under discussion whether finite conductivity may prolong the field through the Faraday induction~\cite{Tuchin:2013ie, Huang:2015oca}).  This short lifetime significantly affects the non-perturbativity of the resulting physical processes, as reviewed in Sec.~\ref{sec:2.3..1}.  This point is particularly important for pair production (or the Schwinger effect)~\cite{Baur:2007zz, Baur:2008hn, Taya:2014taa}.  Indeed, the typical frequency scale is $\omega = 1/\tau = {\mathcal O}(1\;\mathchar`-\;10\;{\rm GeV})$, which is much larger than the quark-mass scale, so the perturbative mechanism becomes more effective than the non-perturbative one.  One possible way around this is to lower the collision energy, so that the colliding ions may stick together after the collision because of baryon stopping (the Landau picture~\cite{10.1143/ptp/5.4.570, Landau:1953gs}), rather than pass through each other as in the high-energy limit (the Bjorken picture~\cite{Bjorken:1982qr}).  The resulting electromagnetic field can then persist for a relatively long time, making the non-perturbative mechanism potentially relevant~\cite{Taya:2024wrm, Taya:2025utb}.

\subsection{Chiral anomaly} \label{sec:4.4}

In 1969, Adler, Bell, and Jackiw pointed out a phenomenon called the chiral anomaly: a quantum violation of chiral symmetry caused by a non-trivial gauge-field configuration~\cite{Adler:1969gk, Bell:1969ts}.  This quantum violation appears as a non-conservation of the axial-vector current, which is a conserved Noether current at the classical level in the massless limit.  In (3+1)-dimensional QED, it takes the form, 
\begin{align}
	\partial_\mu J_5^\mu = \frac{e^2}{2\pi^2} {\bm E} \cdot {\bm B}  +  2m P \;, \label{eq:::39}
\end{align}
where $J_5^\mu := \braket{\hat{\bar{\psi}}\gamma^\mu\gamma_5\hat{\psi}}$ is the expectation value of the axial-vector current.  The time component $J_5^0$ measures the density of the {\it chirality imbalance}, i.e., the difference between right- and left-handed chiralities.  The second term on the right-hand side, $2mP$, with $P := \braket{\hat{\bar{\psi}}{\rm i}\gamma_5\hat{\psi}}$ being the pseudo-scalar condensate, represents the explicit breaking of chiral symmetry by the mass.  The first term is the Chern-Simons charge density [denoted as $\nu$ in Eq.~(\ref{eqdsqf:35}) for QCD], which reflects the non-trivial gauge topology and is the origin of the chiral anomaly.  The anomaly relation~(\ref{eq:::39}) has rich phenomenological applications, including anomalous transport phenomena and the QCD phase structure, as reviewed in other chapters of this Encyclopedia.  

The Schwinger effect provides a natural microscopic mechanism for generating the chirality imbalance $J_5^0$~\cite{Fukushima:2018grm, Copinger:2020nyx}.  As an illustrative example, consider QED and apply a constant, parallel electromagnetic field.  We also adopt the idealized limits of massless and a strong magnetic field.  Because of the strong magnetic field, only particles in the lowest-Landau level with spin aligned with the magnetic field ($n=0$, $s_\parallel=+1/2$) can be produced via the Schwinger effect, along with anti-particles whose spin is anti-aligned ($n=0$, $s_\parallel=-1/2$).  After pair production, the produced particle and anti-particle are accelerated by the electric field: the particle moves in the same direction as the magnetic field, while the anti-particle moves in the opposite direction.  Thus, their momenta and spins are correlated as ${\bm p}\cdot{\bm s}>0$, meaning that both have right-handed helicity.  In the massless limit, helicity and chirality are tightly connected: a particle with right-handed helicity has right-handed chirality, whereas an anti-particle with right-handed helicity has left-handed chirality~\cite{Fukushima:2008xe}.  Consequently, the chirality imbalance is generated as 
\begin{align}
    \int {\rm d}^3{\bm x}\, J_5^0
    = N_{\rm right\ chirality}
     - N_{\rm left\ chirality}
     - \left(\bar N_{\rm right\ chirality}
     - \bar N_{\rm left\ chirality}\right)
    = N_{n=0, s_\parallel=+1/2}
     + \bar N_{n=0, s_\parallel=-1/2}
    = 2 N_{n=0, s_\parallel=+1/2}\;,
\end{align}
where $N$ and $\bar N$ denote the numbers of particles and anti-particles, respectively, in the specified states, and we used $N_{n=0, s_\parallel=+1/2} = \bar N_{n=0, s_\parallel=-1/2}$.  The number $N_{n=0,\, s_\parallel=+1/2}$ follows directly from
Eq.~(\ref{eq--14}), giving $N_{n=0, s_\parallel=+1/2} = T \times (e^2/4\pi^2) \int {\rm d}^3{\bm x}\,EB$ (for ${\mathfrak g}=2$).  Dropping the explicit-breaking term $2mP$ in the $m\to 0$ limit then reproduces the anomaly relation~(\ref{eq:::39}).  Note that this dropping is not trivial, as $P$ could compensate  the smallness of $m$~\cite{Schwinger:1951nm}.  A careful in-in treatment for the evaluation of the expectation value is necessary to justify $2mP \to 0$~\cite{Copinger:2018ftr}.  

This argument can be extended to more general situations, including finite mass, moderate magnetic fields, and inhomogeneous fields~\cite{Tanji:2010eu, Tanji:2018qws, Copinger:2018ftr, Taya:2020bcd, Aoi:2021azo, Fukushima:2023obj}.  Thus, the number of particles produced by the Schwinger effect is closely related to the resulting chirality imbalance.  Since the amount of pair production depends sensitively on the physical setup, the chirality imbalance also varies accordingly.  Indeed, the anomaly relation~(\ref{eq:::39}) is merely a {\it sum rule}: it states that the topological charge $\propto {\bm E}\cdot{\bm B}$ is saturated by the two contributions $J_5^\mu$ and $2mP$, but it does not specify how the saturation is divided between them~\cite{Fukushima:2023obj}.  To determine this ``anatomy'', one must compute either $J_5^\mu$ or $2mP$ explicitly.  For example, for a constant electromagnetic field, it was shown~\cite{Warringa:2012bq, Copinger:2018ftr}, 
\begin{align}
	\frac{e^2}{2\pi^2}{\bm E}\cdot{\bm B}
	= \underbrace{\frac{e^2}{2\pi^2}{\bm E}\cdot{\bm B} e^{-\pi m^2/|eE|}}_{=\,\partial_\mu J_5^\mu} + \underbrace{ \frac{e^2}{2\pi^2}{\bm E}\cdot{\bm B} \left(1 - e^{-\pi m^2/|eE|}\right)}_{=\,-2mP} \;. \label{eqwq:-41}
\end{align}
Thus, the contribution from the chirality imbalance becomes less significant when the field is weak ($m^2/|eE| \gtrsim 1$), where the Schwinger effect is suppressed, and becomes dominant for stronger fields.

The application of the Schwinger effect is also important for describing the realtime dynamics of the chiral-magnetic effect in heavy-ion collisions (see Ref.~\cite{Kharzeev:2024zzm} for a recent review).  The chiral-magnetic effect refers to the generation of an electric current along a magnetic field in the presence of a chirality imbalance ${\bm J} \propto J^0_5 {\bm B}$~\cite{Fukushima:2008xe}.  
In heavy-ion collisions, both the chirality imbalance and the quarks that carry the chiral-magnetic current are produced via the Schwinger effect from the glasma, and their evolution is strongly affected by the early-time dynamics.  Furthermore, the magnetic field is initially generated by the spectator ions and subsequently evolves dynamically, interacting with the quarks and gluons produced from the glasma.  Consequently, solving these complicated dynamics self-consistently is an extremely challenging task for theory.  For this reason, simplified but instructive setups have been explored, such as a homogeneous parallel color-electromagnetic field together with a perpendicular magnetic field~\cite{Fukushima:2015tza} (which can also be analyzed analytically to some extent using the Nikishov formula~\cite{Fukushima:2010vw, Warringa:2012bq}), and a magnetic field in the presence of a sphaleron background~\cite{Muller:2016jod, Mace:2016shq}.  These studies have numerically demonstrated that the chiral-magnetic current actually flows in the direction of the applied magnetic field.

\section{Take-home message} \label{sec:5}

We have reviewed the Schwinger effect in QED and QCD, as well as several of their applications in nuclear physics.  As a summary, we recapitulate here some of the key ideas of the Schwinger effect:
\begin{itemize}
\item A core idea of strong-field physics is that the vacuum is not empty but is filled with virtual fluctuations, which can leave observable effects through coupling to external fields.  When an external field becomes sufficiently strong to supply the energy needed to materialize these fluctuations, real pairs are produced and the vacuum decays---this is the Schwinger effect (see Fig.~\ref{fig:1}).  

\item Key formulas for the Schwinger effect in a constant electric field:
\begin{align}
	{\rm momentum\ spectrum}:\ 
	&f =\exp\left[ - \pi \frac{m^2 + {\bm p}_\perp^2}{|eE|} \right] \;, \tag{\ref{eq:8}} \\
\begin{split}
	{\rm pair\mathchar`-production\ rate}:\ 
	&\Gamma = \frac{1}{T} \sum_{\rm spin} \int \frac{{\rm d}^3{\bm p}}{(2\pi)^3} f = (2s+1)\frac{(eE)^2}{(2\pi)^3} \exp\left[ -\pi\frac{m^2}{|eE|} \right] \;, \\
	{\rm vacuum\mathchar`-decay\ rate}:\ 
	&w = \frac{1}{T} \sum_{\rm spin} \sum_{k=1}^\infty \int \frac{{\rm d}^3{\bm p}}{(2\pi)^3} \beta_k \frac{f^{k}}{k} = (2s+1) \frac{(eE)^2}{(2\pi)^3} \sum_{k=1}^{\infty} \frac{\beta_k}{k^2} \exp\left[ -k\pi\frac{m^2}{|eE|} \right] \;. 
\end{split} \tag{\ref{eq:10}} 
\end{align}
The non-perturbative, exponential dependence is the most characteristic feature of the Schwinger effect.  Magnetic-field effects can be included via modifying the dispersion via the Landau quantization (\ref{eq:11}).  The same formulas equally hold in QCD for covariantly-constant color electromagnetic fields~(\ref{eq:29}).  

\item Significant efforts have been devoted to extending the key formulas of the Schwinger effect, Eqs.~(\ref{eq:8}) and (\ref{eq:10}), by incorporating, for example, spacetime inhomogeneities, radiative corrections, realtime dynamics, and backreaction (see Sec.~\ref{sec:2.3}).

\item The Schwinger effect has many phenomenological applications, particularly in nuclear physics, including high-$Z$ nuclei, string breaking, relativistic heavy-ion collisions, and the chiral anomaly (see Sec.~\ref{sec:4}).

\item Although the Schwinger effect is an ``old" subject with more than 90 years of history, it remains an active research frontier, with many intriguing open questions!
\end{itemize}
For further study, we recommend the classic original papers~\cite{Sauter:1931zz, Sauter:1932gsa, Heisenberg:1936nmg, Schwinger:1951nm, Nikishov:1969tt} and the excellent review articles on the Schwinger effect~\cite{Dunne:2004nc, Ruffini:2009hg, Gelis:2015kya, Fedotov:2022ely, Hattori:2023egw}, as well as reviews of strong-field physics in a variety of contexts, including nuclear physics~\cite{Greiner:1985ce, Tuchin:2013ie, Hattori:2016emy, Fukushima:2018grm, Hattori:2023egw}, particle physics~\cite{Hartin:2018egj}, laser and plasma physics~\cite{Marklund:2006my, DiPiazza:2011tq, Gonoskov:2021hwf, Fedotov:2022ely}, solid-state physics~\cite{Oka:2011kf, Ghimire_2014, Cavaletto2025, Stammer:2025ekh}, and astrophysics~\cite{Harding:2006qn, Ruffini:2009hg, Turolla:2015mwa}.

\begin{ack}[Acknowledgments]

The author is supported by JSPS KAKENHI Grant No.~24K17058 and the RIKEN TRIP initiative (RIKEN Quantum).
\end{ack}


\bibliographystyle{Numbered-Style}
\bibliography{reference}

\end{document}